\documentclass[twocolumn]{aastex61}
\usepackage{amsmath,amssymb}

\newcommand\aastex{AAS\TeX}

\received{\today}
\revised{\***, 2017}
\accepted{\***,2017}
\submitjournal{ApJ}

\shorttitle{\aastex\ The TOP-SCOPE survey}
\shortauthors{Liu et al.}

\shorttitle{The TOP-SCOPE survey } \shortauthors{Liu et al.}

\begin{document}

\title{The TOP-SCOPE survey of Planck Galactic Cold Clumps: Survey overview and results of an exemplar source, PGCC G26.53+0.17}

\AuthorCollaborationLimit=200

\correspondingauthor{Tie Liu}
\email{liutiepku@gmail.com}

\author{Tie Liu}
\affiliation{Korea Astronomy and Space Science Institute, 776 Daedeokdaero, Yuseong-gu, Daejeon 34055, Republic of Korea}
\affiliation{East Asian Observatory, 660 N. A'ohoku Place, Hilo, HI 96720, USA}

\author{Kee-Tae Kim}
\affiliation{Korea Astronomy and Space Science Institute, 776 Daedeokdaero, Yuseong-gu, Daejeon 34055, Republic of Korea}

\author{Mika Juvela}
\affiliation{Department of Physics, P.O.Box 64, FI-00014, University of Helsinki, Finland}

\author{Ke Wang}
\affiliation{European Southern Observatory, Karl-Schwarzschild-Str.2, D-85748 Garching bei M\"{u}nchen, Germany}

\author{Ken'ichi Tatematsu}
\affiliation{National Astronomical Observatory of Japan, National Institutes of Natural Sciences, 2-21-1 Osawa, Mitaka, Tokyo 181-8588, Japan}

\author{James Di Francesco}
\affiliation{NRC Herzberg Astronomy and Astrophysics, 5071 West Saanich Rd, Victoria, BC V9E 2E7, Canada}
\affiliation{Department of Physics and Astronomy, University of Victoria, Victoria, BC V8P 1A1, Canada}

\author{Sheng-Yuan Liu}
\affiliation{Institute of Astronomy and Astrophysics, Academia Sinica. 11F of Astronomy-Mathematics Building, AS/NTU No.1, Sec. 4, Roosevelt Rd, Taipei 10617, Taiwan, R.O.C.}

\author{Yuefang Wu}
\affiliation{Department of Astronomy, Peking University, 100871, Beijing China}

\author{Mark Thompson}
\affiliation{Centre for Astrophysics Research, School of Physics Astronomy \& Mathematics, University of Hertfordshire, College Lane, Hatfield, AL10 9AB, UK}

\author{Gary Fuller}
\affiliation{UK ALMA Regional Centre Node, Jodrell Bank Centre for Astrophysics, School of Physics and Astronomy, The University of Manchester, Oxford Road, Manchester M13 9PL, UK}

% ...

\begin{abstract}

The low dust temperatures ($<$14 K) of Planck Galactic Cold Clumps (PGCCs) make them ideal targets to probe the initial conditions and very early phase of star formation. ``TOP-SCOPE" is a joint survey program targeting $\sim$2000 PGCCs in J=1-0 transitions of CO isotopologues and $\sim$1000 PGCCs in 850 $\micron$ continuum emisison. The objective of the ``TOP-SCOPE" survey and the joint surveys (SMT 10-m, KVN 21-m and NRO 45-m) is to statistically study the initial conditions occurring during star formation and the evolution of molecular clouds, across a wide range of environments. The observations, data analysis and example science cases for these surveys are introduced with an exemplar source, PGCC G26.53+0.17 (G26), which is a filamentary infrared dark cloud (IRDC). The total mass, the length and the mean line-mass (M/L) of the G26 filament are $\sim$6200 M$_{\sun}$, $\sim$12 pc and $\sim$500 M$_{\sun}$~pc$^{-1}$, respectively. Ten massive clumps including eight starless ones are found along the filament. The most massive Clump as a whole may be still in global collapse while its denser part seems to be undergoing expansion due to outflow feedback. The fragmentation in G26 filament from cloud scale to clump scale is in agreement with gravitational fragmentation of an isothermal, non-magnetized, and turbulent supported cylinder. A bimodal behavior in dust emissivity spectral index ($\beta$) distribution is found in G26, suggesting grain growth along the filament. The G26 filament may be formed due to large-scale compression flows evidenced by the temperature and velocity gradients across its natal cloud.

\end{abstract}

\keywords{surveys -- ISM: clouds -- ISM: kinematics and dynamics -- ISM: abundances -- stars: formation }

\section{Introduction}

In the current paradigm, stars form within cold and dense fragments in the clumpy and filamentary molecular clouds. Recent studies of nearby clouds by Herschel have revealed a ``universal" filamentary structure in the cold ISM \citep{and14}. A main filament surrounded by a network of perpendicular striations seems to be a very common pattern in molecular clouds. Within the filaments are compact (with sizes of 0.1~pc or less), cold (T$_{k}\leq$10 K), and dense (n(H$_{2}$)$>5\times10^{4}$ cm$^{-3}$) starless condensations, usually dubbed ``prestellar cores", which are centrally concentrated and largely thermally supported \citep{cas11}. However, the way that filaments form in the cold ISM is still far from being well characterized. Also, the properties of prestellar cores and how prestellar cores evolve to form stars are still not fully understood due to a lack of statistical studies toward a large sample.

The roles of turbulence, magnetic fields, gravity, and external compression in shaping molecular clouds and producing filaments can only be thoroughly understood by investigating an all-sky sample that contains a representative selection of molecular clouds in different environments. Observations by Herschel revealed that more than 70\% of the prestellar cores (and protostars) are embedded in the larger, parsec-scale filamentary structures in molecular clouds \citep{and10,and14,kon10}. The fact that the cores reside mostly within the densest filaments with column densities exceeding $\sim7\times10^{21}$ cm$^{-2}$ strongly suggests a column density threshold for core formation \citep{and14}. Such column density threshold for core formation was also suggested before Herschel \citep{john04,kirk06}. The (prestellar) core mass function being very similar in shape to the stellar initial mass function further suggests a connection to the underlying star formation process \citep[e.g.,][]{motte98,john00,alves07,and14,kon15}.

While significant progress has been made in recent years, past high-resolution continuum and molecular line surveys have mostly focused on the Gould belt clouds or the inner Galactic plane. There are hence still fundamental aspects of the initial conditions for star formation that remain unaddressed, which include but are not limited to:

$\bullet$ On smaller scales ($\sim$pc), how and where do prestellar cores (i.e., future star forming sites) form in abundance? Specifically, can prestellar cores form in less dense and high latitude clouds or short lived cloudlets? Is there really a ``universal" column density threshold for core/star formation?

$\bullet$ On larger scales ($\sim\times$10 pc), what controls the formation of hierarchical structure in molecular clouds? What is the interplay between turbulence, magnetic fields, gravity, kinematics and external pressure in molecular cloud formation and evolution in different environments (e.g., spiral arms, interarms, high latitude, expanding H{\sc ii} regions, supernova remnants)? How common are filaments in molecular clouds? What is the role of filaments in generating prestellar cores?

\subsection{Planck Galactic Cold Clumps}

\begin{figure}
\epsscale{1.3}
\plotone{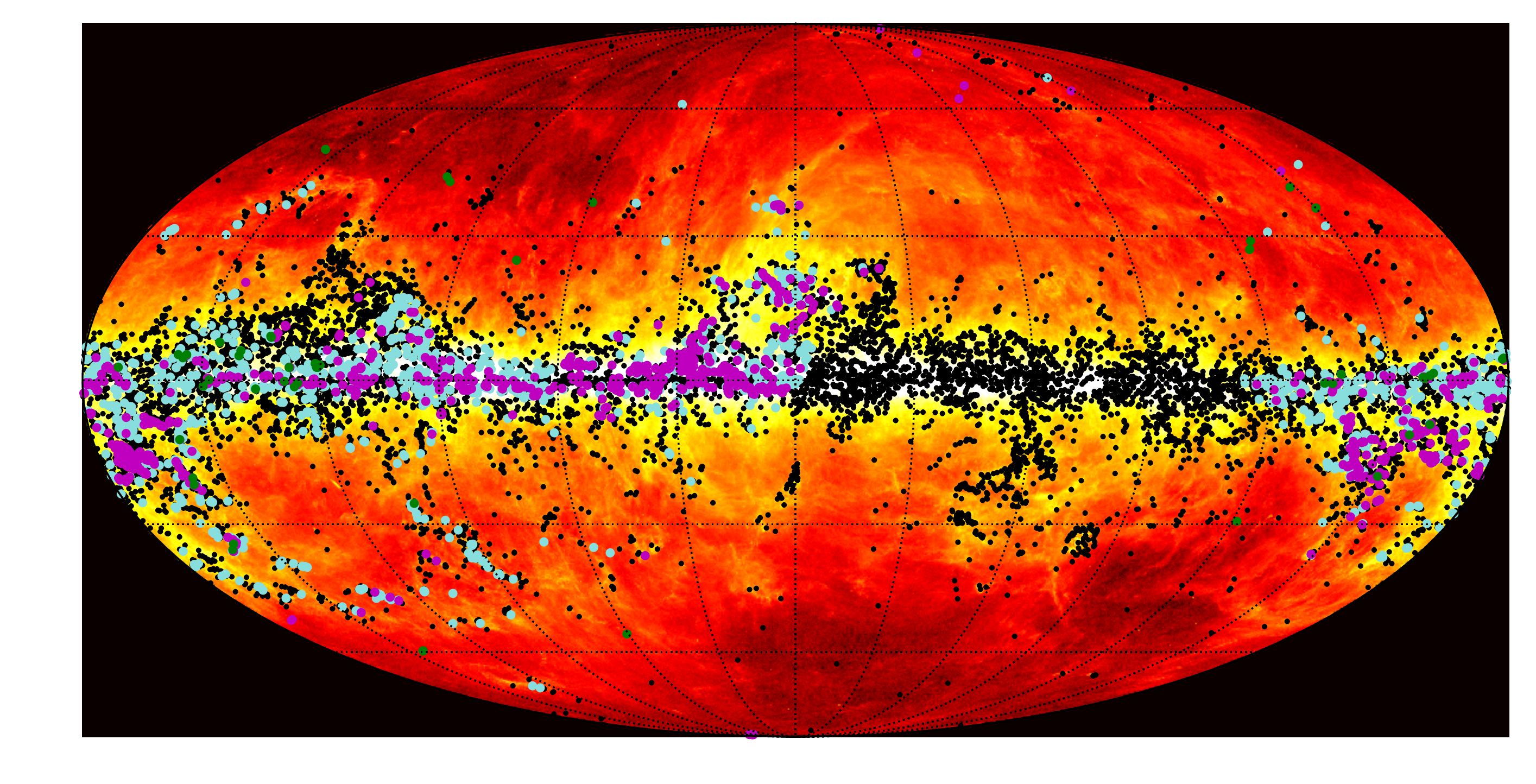}
\plotone{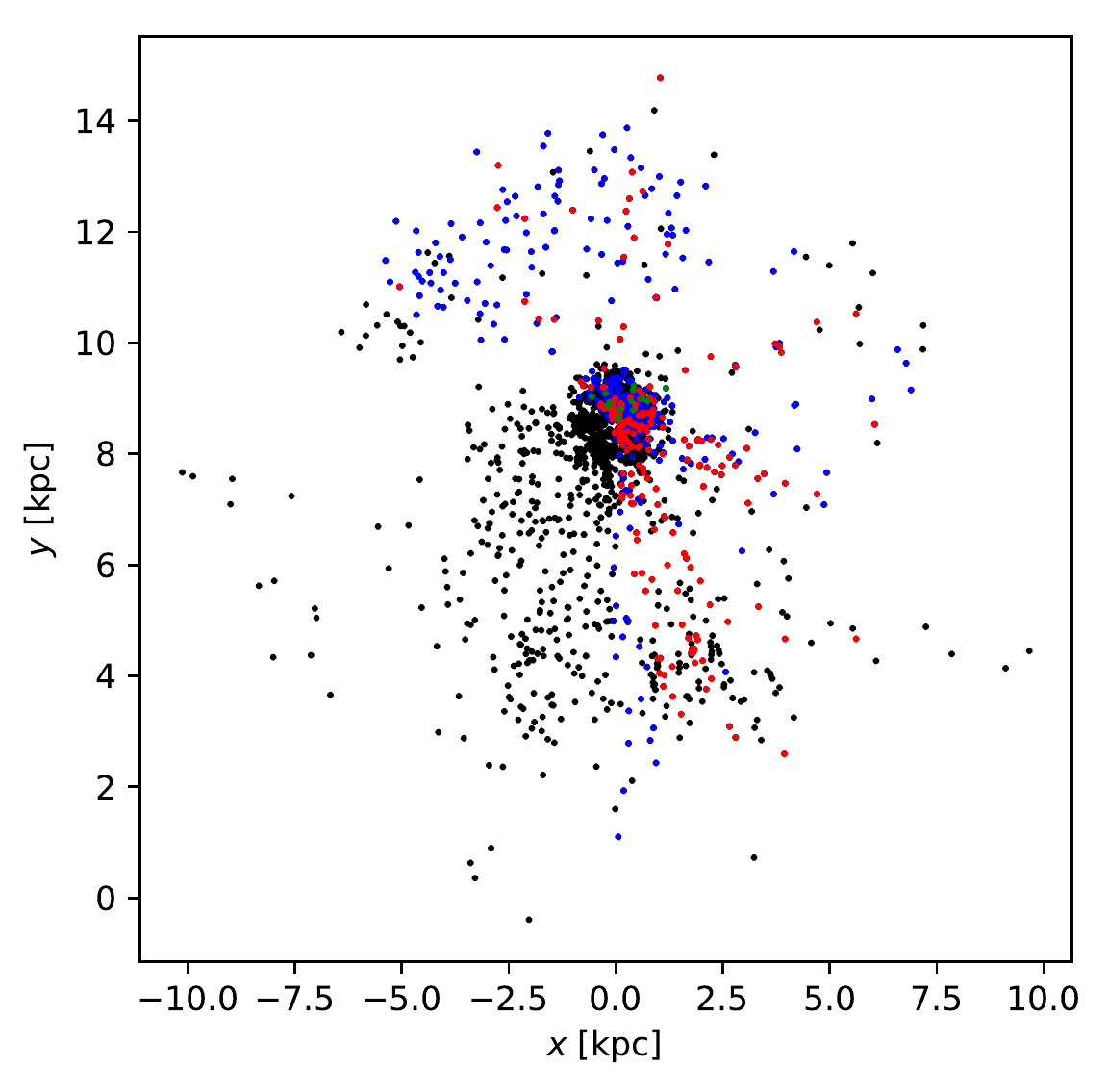}
\caption{\small Upper panel: All-sky distribution of the 13188 PGCC sources (black dots), the 2000 PGCC sources (blue dots) selected for TOP and 1000 PGCC sources (magenta dots) selected for SCOPE overlaid on the 857 GHz Planck map. Lower Panel: A face-on view of all-sky distribution of PGCCs with distances. The PGCCs in the initial PGCC sample, the TOP sample, and the SCOPE sample are shown in black dots, blue dots and red dots, respectively. The green dots represent PGCCs observed in the SCUBA-2 Ambitious Sky Survey (SASSy) at the JCMT, as part of pilot studies of SCOPE. \label{sky}}
\end{figure}

The Planck is the third-generation mission to measure the anisotropy of the cosmic microwave background radiation, and it observed the entire sky in nine frequency bands (between 30 and 857 GHz bands). The high frequency channels of Planck cover the peak thermal emission spectrum of dust colder than 14 K \citep{planck11,planck16}, indicating that Planck could probe the coldest parts of the ISM.
The Planck team has catalogued 13188 Planck galactic cold clumps (PGCCs), which are distributed across the whole sky, i.e., from the Galactic plane to high latitudes, following the spatial distribution of the main molecular cloud complexes. All 13188 PGCCs are overlaid on the Planck map in Figure \ref{sky}.

\begin{figure*}[tbh!]
\centering
\includegraphics[angle=0,scale=0.8]{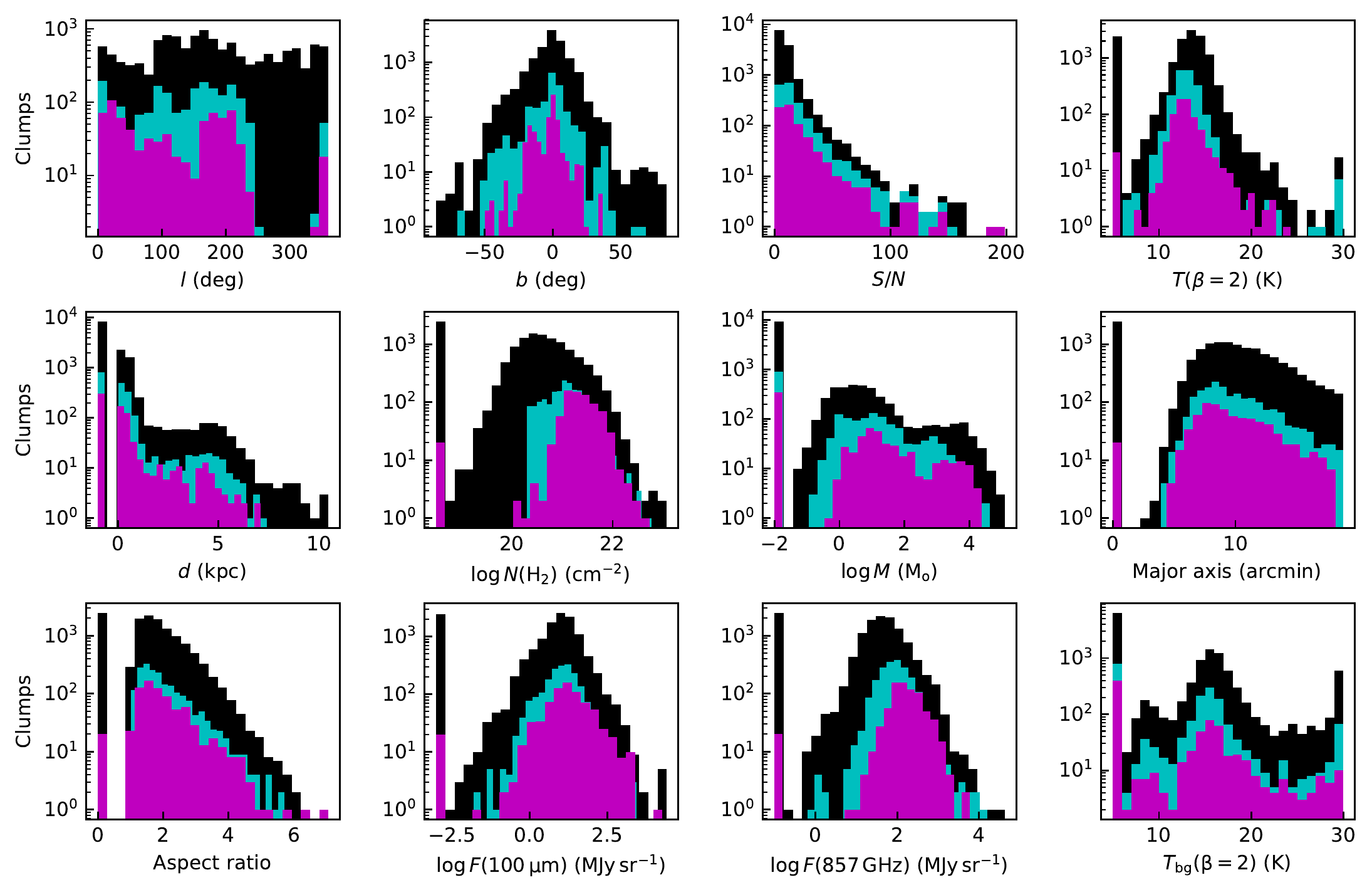}
\caption{Statistics of the initial PGCC sample (black), the TOP sample (blue) and the SCOPE sample (magenta). The y axis is number. From upper-left to lower-right: Longitude, Latitude, SNR, dust temperature, distance, column density, mass, Major axis, Aspect ratio, 100 $\micron$ flux density, 350 $\micron$ flux density, and background temperature. In the panels except for Longitude, Latitude, and SNR, the lowest bin corresponds to clumps for which the parameter value is unknown. \label{sources}  }
\end{figure*}

The catalogue of PGCCs was generated
using the CoCoCoDeT algorithm \citep{mont10}. The method uses
spectral information to locate sources of cold dust emission. Each of
the {\it Planck} 857, 545, and 353 GHz maps is compared to the IRAS
100\,$\mu$m data. The 100\,$\mu$m maps and the local average ratio of
the {\it Planck} and IRAS bands represent the distribution of warm
dust emission.
The spectrum of this extended component is estimated from an annulus
between 5$\arcmin$ and 15$\arcmin$ from the center position. This
also sets an upper limit to the angular size of the detected clumps.
Once the template of warm dust emission is subtracted from the {\it
Planck} data, the residual maps of cold dust emission can be used for
source detection. The initial catalogs were generated for the three
{\it Planck} bands separately and were then merged, requiring a
detection and a signal-to-noise ratio greater than 4 in all three
bands.

The properties of the PGCC catalogue are described in
\citet{planck16}. The source SEDs (based on the four bands
used in the source detection) give color temperatures 6-20\,K with a
median value of $\sim$14\,K. About 42\% of the sources have reliable
distance estimates that were derived with methods such as 3D
extinction mapping, kinematic distances, and association with known
cloud complexes. The distance distribution extends from 0.1\,kpc to
10\,kpc with a median of $\sim$0.4\,kpc. Because of the low angular
resolution of the data ($\sim 5\arcmin$) and the wide distribution of
distances, the PGCC catalog contains a heterogeneous set of objects
from nearby low-mass clumps to distant massive clouds. This is
reflected in the physical parameters where the object sizes range from
0.1\,pc to over 10\,pc and the masses from below 0.1\,$M_{\sun}$ to over
10$^4$\,$M_{\sun}$.

PGCC is the first homogeneous survey of cold and compact Galactic dust
clouds that extends over the whole sky. The main common feature of the
PGCC sources is the detection of a significant excess of cold dust
emission. This is suggestive of high column densities, a fact that has
also been corroborated by subsequent Herschel studies
\citep{planck11,planck16,GCC-IV}.
The catalog covers wide range of galactocentric distances (almost
0-15\,kpc) and a wide range of environments from quiescent
high-latitude clouds to active star-forming clouds and sites of
potential triggered star formation.
This makes the PGCC catalogue a good starting point for many studies of the star
formation process. Figure~\ref{sources} shows the
distributions of several parameters for the PGCC catalog sources.

A large fraction of PGCCs seems to be quiescent, not affected by on-going star-forming activity \citep{wu12,liu14}. Those sources are prime candidates for probing how prestellar cores form and evolve, and the initial stages of star formation across a wide variety of Galactic environments. The detection of gravitationally bound CO gas clumps \citep{liu12,meng13,zhang16} and dense molecular line tracers \citep{yuan16,tat17} inside PGCCs strongly suggests that many of PGCCs have the ability to form stars. Moreover, their low level of CO gas depletion indicates that the natal clouds of PGCCs are still in the early stages of molecular cloud evolution \citep{liu13}. A number of PGCCs were purposely followed-up with
observations with the Herschel satellite. These higher resolution data
revealed great variety in the morphology of the PGCCs
\citep{PlanckII, GCC-III, GCC-IV}. The clumps typically have significant
sub-structure and are often (but not always) associated with to cloud
filaments \citep{GCC-VII}. Further clues to the nature of the clumps is
provided by the dust emission properties, the PGCCs showing
particularly high values for submillimetre opacity and  the spectral
index $\beta$ \citep{GCC-V, GCC-VI}. Prestellar cores and extremely young Class 0 objects have been detected in PGCCs \citep{liu16,tat17}, indicating that some PGCCs can be used to trace the initial conditions when star formation commences.

Because of the uniqueness and importance of PGCC sources to understanding the earliest stages of star formation, we have been conducting a series of surveys to characterize the physical and dynamical state of PGCCs. These surveys are described in the following sections.

\section{The ``TOP-SCOPE" survey of PGCCs}

\subsection{TRAO Observations of PGCCs (TOP)}

``TOP" is a Key Science Program (KSP) of the Taeduk Radio Astronomy Observatory (TRAO). It is a survey of the J=1-0 transitions of $^{12}$CO and $^{13}$CO toward $\sim$2000 PGCCs. Some interesting PGCCs were also mapped in C$^{18}$O (1-0) line. The ``TOP" survey was initiated in December 2015. The main aims of ``TOP" are: to find CO dense clumps; to study the universality and ubiquity of filamentary structures in the cold ISM; to study cloud evolution in conjunction with HI surveys; to investigate how CO abundances change with the evolutionary status of the cloud, and across a range of different environments; to investigate the roles of turbulence, magnetic fields and gravity in structure formation; and to investigate the dynamical effect of stellar feedback and/or cloud-cloud collisions on star formation.\\

$\bullet$ \textbf{The TRAO 14-m telescope}

The Taeduk Radio Astronomy Observatory (TRAO\footnote{http://radio.kasi.re.kr/trao/main\_trao.php}) was established in October 1986 with the 14-meter radio telescope located on the campus of the Korea Astronomy and Space Science Institute (KASI) in Daejeon, South Korea. The surface accuracy of the primary reflector is $<$130 $\micron$ rms, and the telescope pointing accuracy is better than 10$\arcsec$ rms. The main FWHM beam sizes for the $^{12}$CO (1-0) and $^{13}$CO (1-0) lines are 45$\arcsec$ and 47$\arcsec$, respectively. The new receiver system, SEQUOIA-TRAO, is equipped with high-performance 16-pixel MMIC preamplifiers in a 4$\times$4 array, operating in the 85-115 GHz frequency range. The noise temperature of the receiver is 50-80 K over most of the band. The system temperature ranges from 150 K (86-110 GHz) to 450 K (115 GHz; $^{12}$CO). The 2nd IF modules with a narrow band, and the 8 channels with 4 FFT spectrometers, allow simultaneous observations at two frequencies within the 85-100 GHz or 100-115 GHz bands for all 16-pixels of the receiver. The backend system (a FFT spectrometer) provides 4096$\times$2 channels with fine velocity resolution of less than 0.1 km~s$^{-1}$ (15 kHz) per channel, and full spectral bandwidth of 60 MHz ($\sim$160 km~s$^{-1}$ for $^{12}$CO).

$\bullet$ \textbf{Source selection}

The TOP survey has a target list that probes the early stages of star formation across a wide range of Galactic environments. The starting point is the PGCC catalogue, which contains all cold dust clouds detected by Planck, in combination
with the IRAS 100 $\mu$m data. Given the very high sensitivity of the
Planck measurements, the PGCC catalogue is currently the most complete unbiased
all-sky catalog of cold clumps.

The TOP target list was constructed according to the following selection criteria. We excluded regions targeted as part of the previous Galactic Plane \citep{Moore15} or Gould
Belt Surveys \citep{ward07} at the James Clerk Maxwell Telescope (JCMT\footnote{http://www.eaobservatory.org/jcmt/about-jcmt/}) because these locations have been extensively observed by other telescopes (e.g., JCMT, NRO 45-m and PMO 14-m) both by molecular line and by dust continuum emission observations. The TOP survey is therefore highly complementary
to these previous surveys. The PGCCs in these excluded regions will also be investigated in our statistical studies using the archived data.

To ensure a representative study of the full PGCC population, the sources were divided into bins of
Galactic longitude (every 30$\arcdeg$), latitude (divided by $|b|$=0$\arcdeg$, 4$\arcdeg$, 10$\arcdeg$, and 90$\arcdeg$), and distance (divided by $d$=0 pc, 200 pc, 500 pc, 1000 pc, 2000 pc, 8000 pc and unknown), and
targets were selected from each bin.
Similarly, we cover the full range of source temperatures and column
densities in Planck measurements. The sampling is weighted towards the very coldest sources (Note that one third of the targets PGCC lists only an upper limit of the temperature). The sample includes 787 very cold sources for which PGCC catalog only gives a temperature
upper limit. To ensure a good detection rate, the sampling is weighted
towards higher column densities. Within the constraints listed above,
preference is given to regions covered previously by Herschel observations, which
can further confirm the presence of compact sources and give more
reliable estimates of their column densities. Figure \ref{sky} shows our target list of 2000 sources, sufficient to sample well the whole TRAO-visible sky and enable meaningful statistical studies of
targets with different physical characteristics and in
different environments.

About one third of the 2000 sources have Herschel data at 70/100-500
$\mu$m. A large fraction of these data come from the Herschel Galactic
Cold Cores survey (PI: M. Juvela) where the targets were also selected from the PGCC catalog,
employing a random sampling similar to the one outlined above. The TOP sample covers widely different Galactic environments as shown in Figure \ref{sky}.
Among the 2000 sources, 219 are located in the Galactic Plane with
$|b|<1^{\circ}$ and 154 at high latitudes with $|b|>30^{\circ}$.
Based on positional correlations, a sizable fraction of the targets
in the Plane are expected to be influenced by nearby H{\sc ii}
regions \citep{planck16}. For example, 50 PGCCs in the $\lambda$ Orionis Complex were included in the TOP sample. The $\lambda$ Orionis Complex containing the nearest giant H{\sc ii} region has moderately enhanced radiation field and the molecular clouds therein seem to be greatly affected by stellar feedback \citep{liu16,gold16}. Among the 1181 sources with known distances in the TOP sample, 997 are within 2
kpc and 99 are beyond 4 kpc. Out of the 2000 sources, 753 have axial ratios larger than 2,
suggestive of extended filamentary structures.

In Figure~\ref{sources}, we show the distributions of parameters for the TOP sample and the initial PGCC sample. In general, the TOP sample has similar distributions in longitude, latitude, and sizes as the initial PGCC sample. However, the TOP sample tends to have lower temperature and higher column density than the initial PGCC sample.\\

$\bullet$ \textbf{Observation strategy}

We firstly conduct single-pointing observations in $^{12}$CO (1-0) and $^{13}$CO (1-0) lines simultaneously, to determine the systemic velocity of each target and also to find suitable reference positions for mapping. The on and off source time in single-pointing survey is 30 seconds. For mapping observations, we applied the On-The-Fly (OTF) mode to map the PGCCs in the $^{12}$CO (1-0) and $^{13}$CO (1-0) lines simultaneously. A small fraction ($<$10\%) of interesting ``TOP" targets will also be mapped in C$^{18}$O (1-0) line. The map size is set to 15$\arcmin\times15\arcmin$. Since the PGCC sources have average angular sizes of 8$\arcmin$ \citep{planck16}, the map size is large enough to well cover most PGCCs. We did, however, obtain larger (e.g., 30$\arcmin$) OTF maps for some extended PGCCs. The FWHM  beam sizes ($\theta_{B}$) for the $^{12}$CO (1-0) and $^{13}$CO (1-0) lines are 45$\arcsec$ and 47$\arcsec$, respectively. The main beam efficiencies ($\eta_{B}$) for $^{12}$CO (1-0) and $^{13}$CO (1-0) lines are 54\% and 51\%, respectively. The typical system temperatures for $^{12}$CO (1-0) and $^{13}$CO (1-0) lines were $\sim$500 K and $\sim$250 K, respectively. We can achieve a typical sensitivity of $\sim$0.5 K and 0.2 K in T$_{A}^{*}$ for the $^{12}$CO (1-0) and $^{13}$CO (1-0) lines at a spectral resolution of 0.33 km~s$^{-1}$ over the mapping field in $\sim$40 minutes integration time. The OTF data are regridded into Class format files with a pixel size of 24$\arcsec$ with an OTFtool at TRAO. The OTF data were smoothed to 0.33 km~s$^{-1}$ and baseline removed with Gildas/Class with one or three-order polynomial. Then the Class format files were converted to FITS format files for analyzing with the MIRIAD and CASA packages.

\subsection{SCUBA-2 Continuum Observations of Pre-protostellar Evolution (SCOPE)}

The ``SCOPE"\footnote{SCOPE:  https://www.eaobservatory.org/jcmt/science/large-programs/scope/} is one of the eight large programs at the JCMT of the East Asia Observatory (EAO) in the 2015 call. It is a survey of $\sim$1000 PGCCs at 850 $\micron$ continuum. The ``SCOPE" project was launched in December 2015. The main aims of the ``SCOPE" survey are: to obtain a census of dense clumps distributed in widely different environments; to study the roles of filaments in dense core formation; to investigate dust properties; and to detect rare populations of dense clumps (e.g., first hydrostatic cores; massive starless cores; pre-/proto brown dwarfs; extremely cold cores). \\

$\bullet$ \textbf{The SCUBA-2 bolometer camera at the JCMT 15-m telescope}

With a diameter of 15-m, the JCMT is the largest single-dish astronomical telescope in the world designed specifically to operate in the submillimetre wavelength region of the spectrum. SCUBA-2 (Submillimetre Common-User Bolometer Array 2\footnote{http://www.eaobservatory.org/jcmt/instrumentation/\\continuum/scuba-2/}) is an array operating simultaneously at 450 $\micron$ and 850 $\micron$ with a total of 5120 bolometers per wavelength \citep{holl13}. The instrument has a field of view of 45 arcmin$^{2}$. The instrument is cooled by three pulse tube coolers and a dilution refrigerator. The dilution fridge has a base temperature of about 50 mK. The focal plane that is the heat bath for the bolometers is temperature controlled at 75 mK. The main FWHM beam size of SCUBA-2 is 7$\arcsec$.9 at 450 $\micron$ and 14$\arcsec$.1 at 850 $\micron$ \citep{demp13}. \\

$\bullet$ \textbf{Source selection}

\begin{figure}
\epsscale{1.1}
\plotone{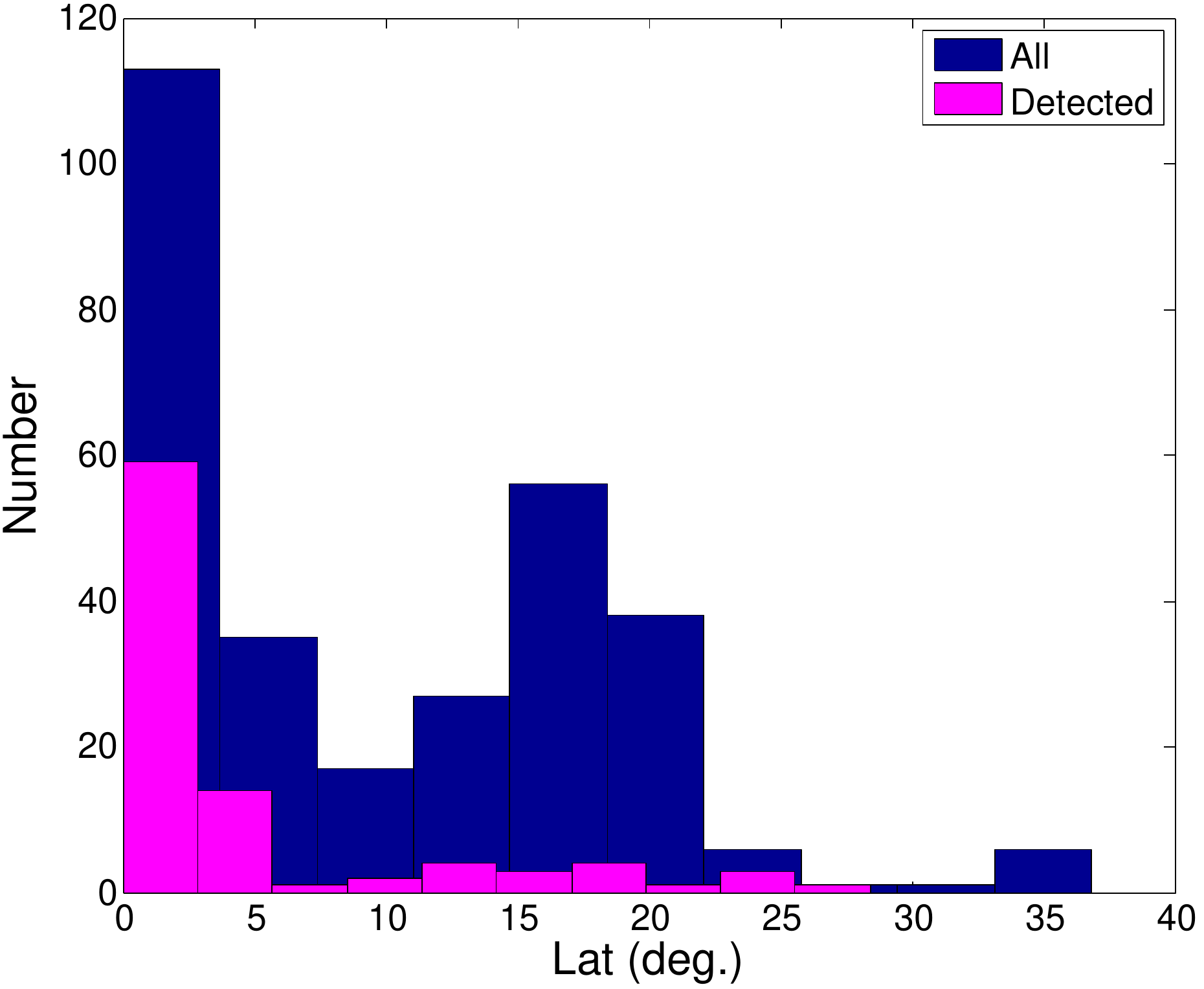}
\plotone{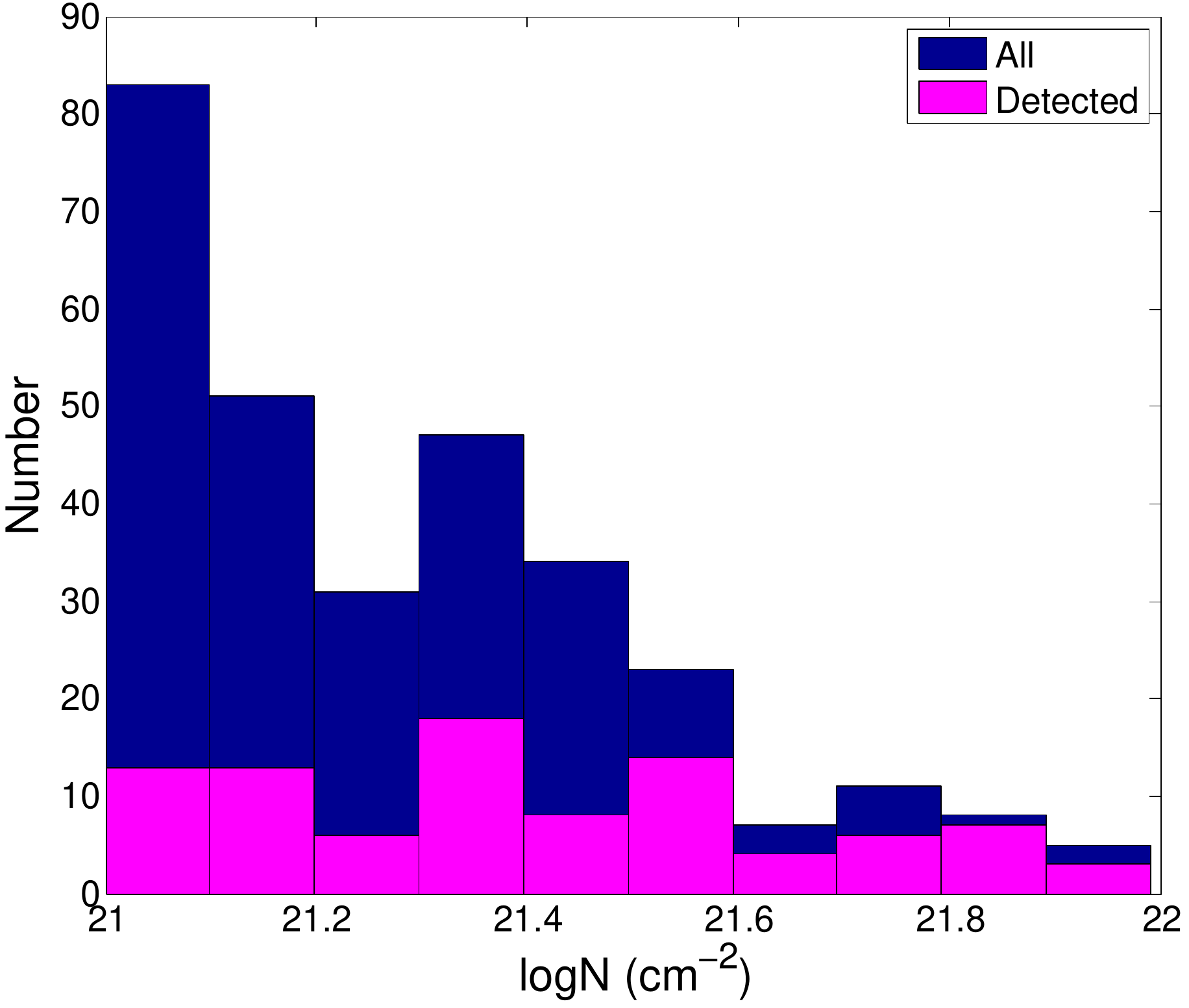}
\caption{\small SCUBA-2 detections as function of latitude (upper panel) and column density (lower panel) for PGCCs in the pilot study. The 300 PGCCs in pilot studies are shown in blue, while those with SCUBA-2 detections are shown in pink. \label{pilot}}
\end{figure}

The 1000 ``SCOPE" targets are selected from the 2000 PGCCs in the ``TOP" survey sample. Preference is given to regions covered by Herschel observations or high column density ($>1\times10^{21}$ cm$^{-2}$ in Planck measurements) clumps. From our pilot study of 300 PGCCs (see Figure \ref{pilot}), we found that the dense core detection rates in SCUBA-2 observations drastically drop toward PGCCs with higher latitude and lower column density. Therefore, the ``SCOPE" sample is further selected biased to high column density PGCCs in each parameter bin of the ``TOP" survey sample. However, to ensure a good representation of the Galactic distribution of the full PGCC population, we also included many lower column density ($>5\times10^{20}$ cm$^{-2}$ in Planck measurements) clumps at high latitudes as well. As shown in Figure \ref{sources}, the SCOPE sample has similar distributions in most parameters (except column density, mass and flux) as the TOP sample and the initial PGCC sample. The SCOPE sample tends to have larger column density and masses than the TOP sample and the initial PGCC sample. About half of SCOPE sources have distances within 2 kpc. The high resolution of SCUBA-2 at 850 $\micron$ can easily resolve dense cores (with sizes of $\sim$0.1 pc) inside these PGCCs with distances smaller than $\sim$2 kpc.\\

$\bullet$ \textbf{Observation strategy}

Since the PGCC sources have average angular sizes of 8$\arcmin$ \citep{planck16}, as noted above, the ``SCOPE" observations were conducted primarily using the CV Daisy mode\footnote{http://www.eaobservatory.org/jcmt/instrumentation/continuum/\\scuba-2/observing-modes/}. The CV Daisy is designed for small compact sources providing a deep 3$\arcmin$ (in diameter; the same as below) region in the centre of the map but coverage out to beyond 12$\arcmin$ \citep{bint14}.  All the SCUBA-2 850 $\micron$ continuum data were reduced using an iterative map-making technique \citep{chap13}. Specifically the data were all run with the same reduction tailored for compact sources, filtering out scales larger than 200$\arcsec$ on a 4$\arcsec$ pixel scale, for the first data release to the team. We also, however, tried different filtering and external masks in the data reduction for individual sources, which will be discussed below. The Flux Conversion Factor varies with time. A mean Flux Conversion Factor (FCF) of 554 Jy/pW/beam was used to convert data from pW to Jy/beam in the pipeline for the first data release. The FCF is higher than the canonical value derived by \cite{demp13}. This higher value reflects the impact of the data reduction technique and pixel size used by the us. The observations were conducted under grade 3/4 weather conditions with 225 GHz opacities between 0.1-0.15. With 16 minutes of integration time per map we reach an rms noise of $\sim$6-10 mJy~beam$^{-1}$ in the central 3$\arcmin$ region. The rms noise increases to 10-30 mJy~beam$^{-1}$ out to 12$\arcmin$, which is better than the sensitivity (50-70 mJy~beam$^{-1}$) in the 870 $\micron$ continuum ATLASGAL survey \citep{cont13}.

\section{Other joint surveys and follow-up observations}

Besides the ``TOP-SCOPE" survey, we are also conducting joint surveys and follow-up observations with other telescopes \citep[e.g., SMT 10-m, KVN 21-m, NRO 45-m, \& SMA; see][]{liu16,tat17}.

\subsection{SMT ``All-sky" Mapping of PLanck Interstellar Nebulae in the Galaxy (SAMPLING)}

``SAMPLING\footnote{http://sky-sampling.github.io}" is an ESO public survey to map up to 600 PGCCs in the J=2-1 transition of $^{12}$CO and $^{13}$CO using the SMT 10-m telescope (Wang et al. 2017, in preparation). The ``SAMPLING" project was also launched in December 2015. The typical map sizes in the ``SAMPLING" survey is 5$\arcmin\times5\arcmin$ with an rms level of $\sim$0.2 K at a spectral resolution of 0.34 km~s$^{-1}$. The beam size and main beam efficiency are 36$\arcsec$ and 0.7, respectively. In conjunction with the ``TOP-SCOPE" survey, the ``SAMPLING" survey aims to resolve the clump structure, to derive the internal variations of column density, turbulent and chemical properties, and also to make a connection to Galactic structure. We observed the exemplar source, PGCC G26.53+0.17, in $^{13}$CO (2-1) line with the SMT telescope. The observational parameters are summarized in Table \ref{obsG26} in Appendix A.

\subsection{KVN survey of SCUBA-2 dense clumps}

The Korean VLBI Network (KVN) is a three-element Very Long Baseline Interferometry (VLBI) network working at millimeter wavelengths. Three 21-m radio telescopes are located in Seoul, Ulsan, and Jeju island, Korea. We use the single-dish mode of the three 21-m radio telescopes to observe the dense clumps detected in the ``SCOPE" survey. The antenna pointing accuracy of the KVN telescopes is better than 5 arcsec. The KVN telescopes can operate at four frequency bands (i.e., 22, 44, 86, and 129 GHz) simultaneously. The beam sizes at the four bands are $\sim126\arcsec$, $\sim63\arcsec$, $\sim32\arcsec$, and $\sim23\arcsec$ at 22 GHz, 44GHz, 86 GHz and 129 GHz, respectively. The main beam efficiencies are $\sim$50\% at 22 GHz and 44 GHz and $\sim$40\% at 86 GHz and 129 GHz.

A key science proposal for single-pointing molecular line observations of $\sim$1000 ``SCOPE" dense clumps with KVN telescopes will be submitted in 2017. The main target lines are the 22 GHz water maser, 44 GHz Class I methanol maser, and other dense molecular lines, which are listed in Table \ref{obsG26} in Appendix A. The 22 GHz water maser and 44 GHz Class I methanol maser are indicators of outflow shocks. SiO thermal lines are also good tracers for outflow shocks or shocks induced by cloud-cloud collisions. Dense gas tracers (e.g., J=1-0 transitions of N$_{2}$H$^{+}$, H$^{13}$CO$^{+}$, HN$^{13}$C) can be used to determine the systemic velocities and the amount of turbulence in dense clumps. Optically thick lines (e.g., J=1-0 transitions of HCN and HCO$^{+}$) can be used to trace infall and outflow motions. H$_{2}$CO $2_{1,2}-1_{1,1}$ is a dense gas tracer and can also be used to reveal infall and outflow motions \citep{liu16}. The deuteration of H$_{2}$CO will be determined from observations of H$_{2}$CO (2$_{1,2}$-1$_{1,1}$) and HDCO (2$_{0,2}$-1$_{0,1}$) lines. The [HDCO]/[H$_2$CO] ratios will be used to trace the early phase of dense core evolution \citep{kang15}. Pilot surveys of $\sim$200 ``SCOPE" dense clumps with the KVN telescopes were conducted in 2016 \citep[e.g., Yi et al. 2017, in preparation; Kang et al. 2017, in preparation; ][]{liu16}. In the single-pointing molecular line observations, it takes about 20 minutes (on+off) to achieve an rms level of $<$0.1 K in brightness temperature with dual polarization at a spectral resolution of $\sim$0.2 km~s$^{-1}$ under normal weather conditions. The KVN observations of the exemplar source, PGCC G26.53+0.17, are summarized in Table \ref{obsG26} in Appendix A.

The KVN data are also reduced with the Gildas/Class package. All the scans were averaged to get the final averaged spectra and the baselines of the averaged spectra were removed with a linear fit.

\subsection{NRO 45-m follow-up survey}

By using the 45 m telescope of Nobeyama Radio Observatory (NRO), we plan to
carry out a comprehensive study of cores selected from the JCMT SCOPE
survey to scrutinize the initial conditions of star formation in widely
different environments including massive star forming regions.
We will observe 100 cores in various environments in
DNC, HN$^{13}$C,  N$_2$D$^+$, and cyclic-C$_3$H$_2$ with receiver T70 in single-pointing mode. Among the 100 cores, 35 cores will also be mapped in $^{12}$CO, $^{13}$CO ,C$^{18}$O, N$_2$H$^+$, HC$_{3}$N and CCS with the FOREST receiver. The full width at half maximum (FWHM) beam size of NRO 45-m telescope at 86 GHz \textbf{is} $\sim18\arcsec.8$ and the main beam efficiency is
$\sim$54\%. Adding 115 Orion cores already observed in single-pointing mode, we will have
215 single-pointing positions data.
By using DNC and N$_2$D$^+$ intensities and CEF for DNC/HN$^{13}$C,
we will select the best targets of 35 cores for OTF mapping with the
four-beam 2SB 2-polarization receiver FOREST.
By applying Chemical Evolution Factor (CEF) we developed \citep{tat17}, we should be
able to identify cores on the verge of massive star
formation. From deuterium fractionation, we can identify the earliest protostellar phase. We will confirm/revise CEF by observing Orion cores having similar distances. The chemical nature of cores including deuterium fractionation will be investigated. Core stability will be investigated to see how star formation starts. Dynamics of parent filaments will be investigated to see if there is accretion onto cores. Coefficient of the linewidth-size relation will be compared
among different environments to see its meaning in star formation. Specific angular momentum will be compared among regions.

The large proposal requesting 350 hrs over two years since December 2017 was accepted for
such follow-up observations. The NRO 45-m observations of the exemplar source, PGCC G26.53+0.17, have not yet been conducted. Therefore, no results from NRO 45-m observations on PGCC G26.53+0.17 will be presented in this paper. Some pilot studies with NRO 45-m telescope toward SCUBA-2 dense cores in PGCCs were presented in \cite{tat17}.

\section{Example science with an exemplar source, PGCC G26.53+0.17}

\begin{figure}
\centering
\includegraphics[angle=0,scale=0.3]{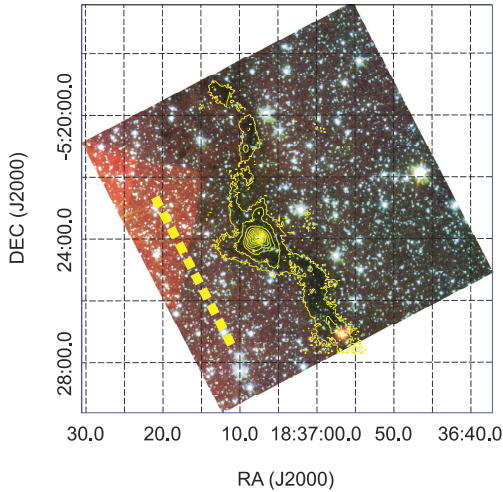}
\caption{Spitzer/IRAC three color (3.6 $\micron$ in blue, 4.5 $\micron$ in green, and 8 $\micron$ in red) composite image of PGCC G26.53+0.71. The yellow contours represent the SCUBA-2 850 $\micron$ continuum emission with filtering out scales of $>$200$\arcsec$. The contour levels are [0.03, 0.05, 0.1, 0.2, 0.4, 0.6, 0.8]$\times$1.38 Jy~beam$^{-1}$. The yellow dashed line shows the direction of the Galactic Plane. \label{IRDC} }
\end{figure}

Through the TOP-SCOPE survey and follow-up observations with other telescopes (e.g., SMT, KVN and NRO 45-m), we aim to statistically investigate the physical and chemical properties of thousands of dense clumps in widely different environments. Those studies will help answer the questions raised at the beginning of this paper. The full scientific exploitation of the TOP-SCOPE survey data will only be
possible upon the completion of the survey. In this paper, we introduce the survey data, data analysis, and example science cases of the above surveys with an exemplar source, PGCC G26.53+0.17 (hereafter denoted as G26). Located at a distance of 4.2$\pm$0.3 kpc, G26 has a mass of $\sim$5200 M$_{\sun}$ and a very low luminosity-to-mass ratio of $\sim$1.4 L$_{\sun}/$M$_{\sun}$ \citep{planck16}. The distance is estimated using the near-infrared extinction \citep{planck16}. The kinematic distance is 3.2-3.4 kpc \citep[Peretto et al., in preparation;][]{planck16}. In this paper, we adopt the distance of 4.2 kpc to be consistent with \cite{planck16}. The spatial resolution of SCUBA-2 at 850 $\micron$ is $\sim0.3$ pc at this distance, which is high enough to resolve massive clumps with sizes of $\sim$1 pc. As shown in Figure \ref{IRDC}, G26 is a filamentary infrared dark cloud (IRDC; \cite{pere09}) with a length of $\sim$10$\arcmin$, corresponding to $\sim$12 pc at a distance of 4.2 kpc. G26 was observed as part of the ``TOP,"``SCOPE,"``SAMPLING," and KVN surveys. The observations of G26 are summarized in Table \ref{obsG26} in Appendix A. The data reduction of JCMT/SCUBA-2 data will be presented in Section 6.2 in Appendix A. The methods used for G26 JCMT/SCUBA-2 data analysis in this work will also be applied to other targets in the ``TOP-SCOPE" survey.

\subsection{SEDs using filtered Herschel and SCUBA-2 maps}

\begin{figure*}
\centering
\includegraphics[angle=90,scale=0.6]{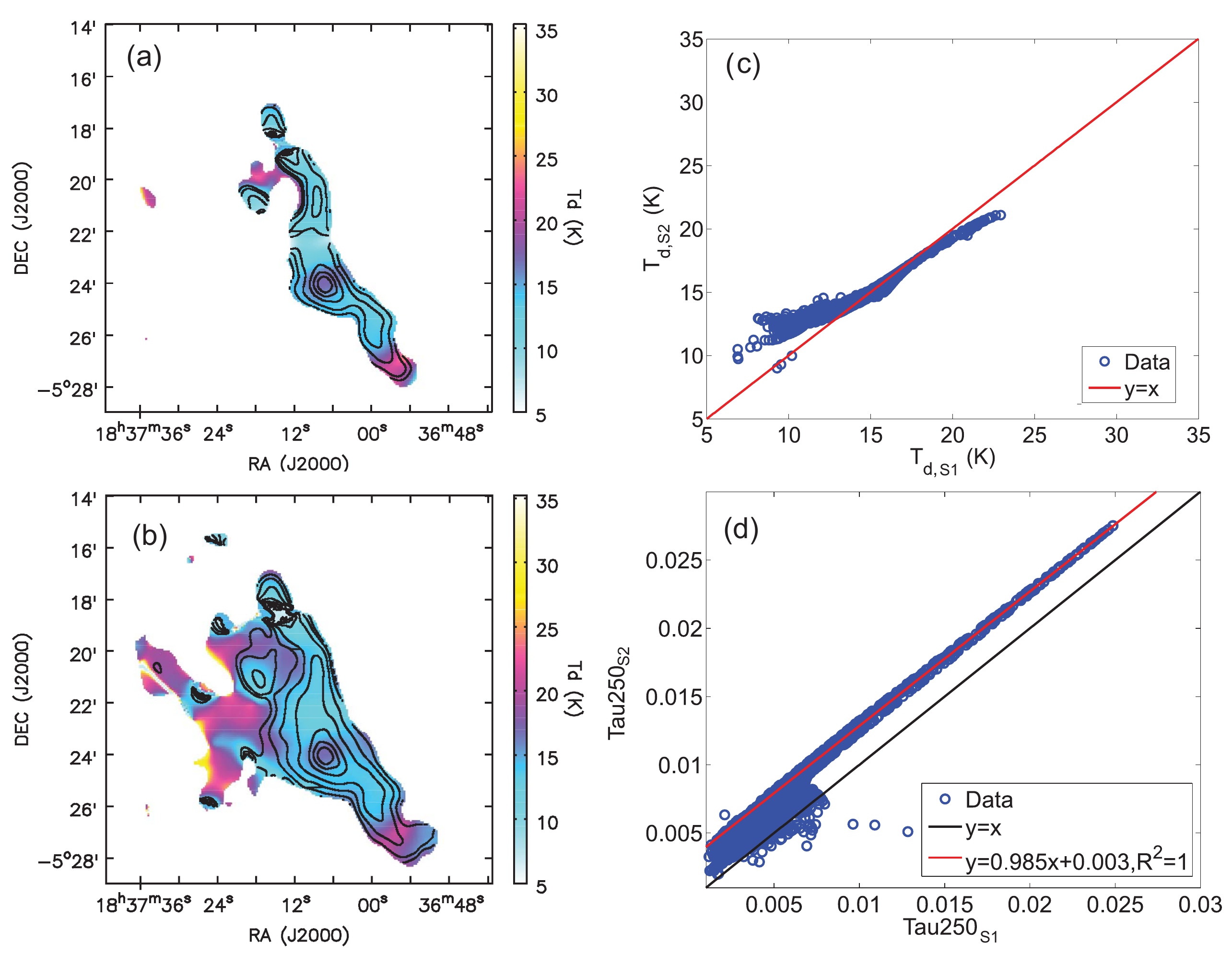}
\caption{(a). The dust temperature map in the \textit{\textbf{S1}} SED fit is shown in color-scale image. The 250 $\micron$ optical depth is shown in contours. The contour levels are [0.03, 0.05, 0.1, 0.2, 0.4, 0.6, 0.8]$\times$0.0270 (b). The dust temperature map in the \textit{\textbf{S2}} SED fit is shown in color-scale image. The 250 $\micron$ optical depth is shown in contours. The contour levels are [0.03, 0.05, 0.1, 0.2, 0.4, 0.6, 0.8]$\times$0.0275 (c). Comparison of the dust temperature values in \textit{\textbf{S1}} and \textit{\textbf{S2}}. Only data points with S/N$>$5 in SCUBA-2 map are considered. (d). Comparison of the 250 $\micron$ optical depth in \textit{\textbf{S1}} and \textit{\textbf{S2}}. Only data points with S/N$>$5 in SCUBA-2 map are considered. The red line shows the linear fit toward the data points with 250 $\micron$ optical depth larger than 0.01 in \textit{\textbf{S4}}.\label{HerSCUB} }
\end{figure*}

The high-resolution SCUBA-2 850 $\micron$ data filtered out large-scale extended emission and thus are sensitive to the denser structures (filaments, cores, or clumps) in molecular clouds. The 850 $\micron$ data are also important in providing high-resolution information about the dust emission at a wavelength that lies in the regime between the far-infrared part (e.g., Herschel)
and the millimetre part of the dust emission spectrum. One of the main goals of the ``SCOPE" survey is to provide high-quality 850 $\micron$ data to constrain better the dust emission spectrum of dense structures in molecular clouds.

In this section, we fit the SEDs pixel-by-pixel with a modified blackbody function using four or five bands, combining SCUBA-2
measurements and fluxes extracted from Herschel maps that have been run through the SCUBA-2 pipeline. The algorithm of the SED fit is the same as in section 6.1 in Appendix A. The Herschel and SCUBA-2 data with the subtraction of the local
background were fitted with modified blackbody spectra
with $\beta=1.8$ to estimate colour correction factors and to derive
estimates of the dust optical depth at a common spatial resolution of
40$\arcsec$. In Section 4.3, we fit the SEDs with free $\beta$ to investigate the dust grain
properties. The details of the SED fits using filtered Herschel and SCUBA-2 maps will be discussed in \cite{juvela17}. In this paper, we performed four individual SCUBA-2 data reductions labeled \textit{\textbf{R1, R2, R3, and R4}} with different external masks and spatial filters (see section 6.2 in Appendix A for details). \textit{\textbf{R1}} with a spatial filter of 200$\arcsec$ is very efficient to identify dense clumps/cores and is used for creating dense core/clump catalogues in our first data release to the team. However, here we fit the SEDs with datasets obtained from two other data reduction methods (\textit{\textbf{R2, R4}}), which use larger spatial filters (300 $\arcsec$ and 600 $\arcsec$) and thus can recover more flux. We applied two sets of SED fits (\textit{\textbf{S1}} and \textit{\textbf{S2}}) using different datasets.

\textit{\textbf{(1) S1:}} Herschel 160-500 $\micron$ data and SCUBA-2 850 $\micron$ data in the \textit{\textbf{R2}} data reduction (see section 6.2 in Appendix A) were used. The extended emission larger than 300$\arcsec$ in the Herschel and SCUBA-2 maps was filtered out.

\textit{\textbf{(2) S2:}} Herschel 160-500 $\micron$ data and SCUBA-2 850 $\micron$ data in the \textit{\textbf{R4}} data reduction (see section 6.2 in Appendix A) were used. The extended emission larger than 600$\arcsec$ in the Herschel and SCUBA-2 maps was filtered out.

Figure \ref{HerSCUB} a-b show the dust temperature (T$_{d}$) and 250 $\micron$ optical depth ($\tau_{250}$) maps from the two SED fits. The T$_{d}$ values across the filament ridge are around 10-14 K except for the dense clumps located in the central region ($\sim$17 K) and the southern end ($\sim$19 K), where protostars have formed and heated their local surroundings. The less dense region to the east of the filament ridge has higher T$_{d}$ than the dense filament, indicating that the filament is externally heated from its eastern side.

The T$_{d}$ and $\tau_{250}$ maps are quite different from SED fits with different effective spatial filters. In Figure \ref{HerSCUB} c-d, we compare the T$_{d}$ and $\tau_{250}$ values from SED fits \textit{\textbf{S1}} and \textit{\textbf{S2}} for pixels with S/N$>$5 in the SCUBA-2 map from the \textit{\textbf{R2}} imaging scheme. The mean dust temperature from \textit{\textbf{S1}} and \textit{\textbf{S2}} is 13.7 and 14.3 K, respectively. In comparison, the mean $\tau_{250}$ (or column density N) in \textit{\textbf{S1}} and \textit{\textbf{S2}} is 6.5$\times10^{-3}$ (or N$\sim1.0\times10^{22}$ cm$^{-2}$) and 8.3$\times10^{-3}$ (or N$\sim1.3\times10^{22}$ cm$^{-2}$), respectively. In contrast to \textit{\textbf{S1}}, \textit{\textbf{S2}} increases the T$_{d}$ in cold regions (with T$_{d}<15$ K). While in warmer regions (with T$_{d}>15$ K), the T$_{d}$ determined in \textit{\textbf{S1}} and \textit{\textbf{S2}} does not vary too much. \textit{\textbf{S2}} significantly increases $\tau_{250}$ by 0.003 (or N$\sim4.6\times10^{21}$ cm$^{-2}$) than \textit{\textbf{S1}} in dense regions (with $\tau_{250}\sim$0.01 or N$\sim1.5\times10^{22}$ cm$^{-2}$ in \textit{\textbf{S2}} map). The peak column densities in \textit{\textbf{S1}} and \textit{\textbf{S2}} are 3.8$\times10^{22}$ cm$^{-2}$ and 4.2$\times10^{22}$ cm$^{-2}$, respectively. In less dense regions, \textit{\textbf{S2}} also significantly increases $\tau_{250}$ but not as much as in dense regions.

SED fits with large effective spatial filters (600$\arcsec$; Figure \ref{HerSCUB}b) recovered more extended structure and thus more mass. The total mass of the G26 filament revealed in \textit{\textbf{S2}} is $\sim$6200 M$_{\sun}$. Given the length ($L\sim$12 pc) of the filament, the mean line-mass (M/L) of the filament is $\sim$500 M$_{\sun}$~pc$^{-1}$. For comparison, both the length and line-mass of G26 are comparable to those ($L\sim$8 pc; $M/L\sim$400 M$_{\sun}$~pc$^{-1}$) of the integral shaped filament in the Orion A cloud \citep{bally87,kai17}.

\subsection{Dense clumps}

\begin{deluxetable*}{ccccccccccccccc}[h!]
\centering
\tablecolumns{15} \tablewidth{0pc}\setlength{\tabcolsep}{0.05in}
\tablecaption{Parameters of dense clumps \label{clumppara}}\tablehead{\colhead{Clump} & \colhead{RA} & \colhead{DEC} & \colhead{$\theta_{maj}$} & \colhead{$\theta_{min}$} & \colhead{P.A.} & \colhead{R$_{eff}$} & \colhead{F$_{peak}$} & \colhead{F$_{int}$} & \colhead{SNR} & \colhead{T$_{d}$\tablenotemark{a}} & \colhead{M} & \colhead{n} & \colhead{N} & \colhead{M$_{Jeans}$}\\
\colhead{No.}  & \colhead{(J2000)} & \colhead{(J2000)} & \colhead{($\arcsec$)} & \colhead{($\arcsec$)}  & \colhead{($\arcdeg$)}& \colhead{(pc)} & \colhead{(Jy~beam$^{-1}$)} & \colhead{(Jy)} & \colhead{} & \colhead{(K)} & \colhead{(M$_{\sun}$)}& \colhead{(10$^{4}$ cm$^{-3}$)}& \colhead{(10$^{22}$ cm$^{-2}$)}& \colhead{(M$_{\sun}$)} }
\startdata
\multicolumn{15}{c}{\textit{\textbf{R1}}}\\
\cline{1-15}
1    &18:37:12.00 & -05:19:12.0   & 15	& 11	& 142	& $<$0.26   & 0.10(0.01)  &  0.23(0.01) & 7  & 10  & 89   & $>$2.1 &$>$2.2	& $<$2  \\
2    &18:37:08.40 & -05:20:27.6   & 19	& 9	  & 108	& $<$0.27   & 0.14(0.02)  &  0.35(0.02) & 11 & 13  & 82   & $>$1.7 &$>$1.9	& $<$3  \\
3    &18:37:09.36 & -05:21:03.6   & 14	& 13	& 108	& $<$0.27   & 0.17(0.02)  &  0.30(0.02) & 13 & 13  & 70   & $>$1.5 &$>$1.6	& $<$3  \\
4    &18:37:09.12 & -05:21:54.0   & 12	& 10	& 191	& $<$0.22   & 0.07(0.01)  &  0.10(0.01) & 5  & 12  & 27   & $>$1.1 &$>$1.0	& $<$4  \\
5    &18:37:07.44 & -05:23:09.6   & 10	& 8	  & 242	& $<$0.18   & 0.25(0.03)  &  0.32(0.02) & 20 & 13  & 75   & $>$5.3 &$>$3.9	& $<$2  \\
6    &18:37:07.68 & -05:23:56.4   & 29	& 20	& 133	&    0.40   & 1.35(0.14)  &  4.38(0.22) & 103& 17  & 654  & 4.2 &7.0	& 3  \\
7    &18:37:02.64 & -05:24:39.6   & 10	& 7	  & 167	& $<$0.17   & 0.11(0.01)  &  0.16(0.01) & 9  & 13  & 37   & $>$3.2 &$>$2.2	& $<$2  \\
9    &18:37:00.48 & -05:25:40.8   & 27	& 13	& 90	&   0.25    & 0.14(0.01)  &  0.70(0.04) & 11 & 14  & 144  & 3.8 &3.9	& 2  \\
10   &18:36:57.84 & -05:26:56.4   & 21	& 12	& 125	&   0.15    & 0.14(0.01)  &  0.46(0.02) & 11 & 19  & 58   & 7.1 &4.4	& 3  \\
\cline{1-15}
\multicolumn{15}{c}{\textit{\textbf{R2}}}\\
\cline{1-15}
1    &18:37:12.00 & -05:19:15.6   & 24	& 16	& 211	&    0.28      & 0.12(0.01)  &  0.44(0.02) & 8  & 10  & 170  & 3.2	& 3.7	 &2     \\
2    &18:37:08.64 & -05:20:20.4   & 22	& 11	& 112	&    0.13      & 0.16(0.02)  &  0.51(0.03) & 10 & 13  & 119  & 22.5	& 12.0 &1   \\
3    &18:37:09.36 & -05:21:03.6   & 22	& 14	& 143	&    0.21      & 0.20(0.03)  &  0.53(0.03) & 13 & 13  & 124  & 5.6	& 4.8	 &2     \\
4    &18:37:09.12 & -05:21:54.0   & 25	& 14	& 155	&    0.25      & 0.11(0.01)  &  0.47(0.02) & 7  & 12  & 127  & 3.4	& 3.5	 &2     \\
5    &18:37:07.44 & -05:23:09.6   & 15	& 11	& 254	& $<$0.26      & 0.31(0.04)  &  0.60(0.03) & 20 & 13  & 140  & $>$3.3	& $>$3.5	 &$<$2     \\
6    &18:37:07.68 & -05:23:56.4   & 39	& 27	& 147	&    0.60      & 1.40(0.12)  &  6.41(0.32) & 88 & 17  & 957  & 1.8	& 4.5	 &5     \\
7    &18:37:02.64 & -05:24:39.6   & 19	& 15	& 126	&    0.19      & 0.17(0.01)  &  0.71(0.04) & 11 & 13  & 166  & 10.0	& 7.8	 &1     \\
9    &18:37:00.48 & -05:25:40.8   & 22	& 17	& 247	&    0.27      & 0.17(0.02)  &  0.93(0.05) & 10 & 14  & 191  & 4.0	& 4.5	 &2     \\
10   &18:36:57.60 & -05:26:56.4   & 24	& 17	& 142	&    0.29      & 0.18(0.01)  &  0.91(0.05) & 11 & 19  & 115  & 1.9	& 2.3	 &5     \\
\cline{1-15}
\multicolumn{15}{c}{\textit{\textbf{R3}}}\\
\cline{1-15}
3    &18:37:09.36 & -05:21:03.6   & 26	& 18	& 266	&    0.33     & 0.23(0.03)   &  1.20(0.06) & 7  & 14  & 247  & 2.8	& 3.9	 & 3     \\
5    &18:37:07.44 & -05:23:09.6   & 22	& 13	& 230	&    0.19     & 0.38(0.05)   &  1.12(0.06) & 12 & 14  & 230  & 13.9	& 10.9 & 1   \\
6    &18:37:07.68 & -05:23:56.4   & 57	& 42	& 170	&    0.95     & 1.47(0.12)   & 11.44(0.57) & 47 & 17  & 1708 & 0.8	& 3.2	 & 7     \\
7    &18:37:02.64 & -05:24:39.6   & 15	& 10	& 213	& $<$0.25     & 0.21(0.02)   &  0.52(0.03) & 7  & 14  & 107  & $>$2.8	& $>$2.9	 & $<$3     \\
8    &18:37:00.72 & -05:25:01.2   & 32	& 26	& 209	&    0.51     & 0.22(0.02)   &  2.20(0.11) & 7  & 14  & 452  & 1.4	& 3.0	 & 4     \\
10   &18:36:57.60 & -05:26:56.4   & 41	& 32	& 179	&    0.68     & 0.24(0.01)   &  2.42(0.12) & 8  & 19  & 305  & 0.4	& 1.1	 & 11    \\
\cline{1-15}
\multicolumn{15}{c}{\textit{\textbf{R4}}}\\
\cline{1-15}
1   &18:37:12.24 & -05:19:12.0   & 23	& 21	& 160	  &    0.34    & 0.16(0.01)   &  1.04(0.05) & 5 & 12 & 281  & 3.0	 &4.2	& 2   \\
3   &18:37:09.36 & -05:21:03.6   & 31	& 23	& 103	  &    0.46    & 0.26(0.03)   &  1.96(0.10) & 9 & 14 & 403  & 1.7	 &3.2	& 4   \\
4   &18:37:09.12 & -05:21:57.6   & 23	& 15	& 210	  &    0.25    & 0.18(0.02)   &  0.96(0.05) & 6 & 13 & 225  & 6.0	 &6.1	& 2   \\
5   &18:37:07.44 & -05:23:09.6   & 25	& 14	& 223	  &    0.25    & 0.40(0.05)   &  1.32(0.07) & 13& 14 & 271  & 7.2	 &7.4	& 2   \\
6   &18:37:07.68 & -05:23:56.4   & 69	& 49	& 145	  &    1.15    & 1.48(0.12)   & 13.87(0.69) & 49& 17 & 2071 & 0.6	 &2.7	& 8   \\
7   &18:37:02.64 & -05:24:39.6   & 17	& 9	  & 210	  & $<$0.25    & 0.21(0.02)   &  0.55(0.03) & 7 & 14 & 113  & $>$3.0	 &$>$3.1	& $<$3   \\
8   &18:37:00.72 & -05:25:01.2   & 38	& 27	& 217	  &    0.59    & 0.23(0.02)   &  2.43(0.12) & 8 & 14 & 499  & 1.0	 &2.4	& 5   \\
10  &18:36:57.60 & -05:26:56.4   & 49	& 36	& 178	  &    0.81    & 0.25(0.01)   &  3.60(0.18) & 8 & 19 & 453  & 0.4	 &1.2	& 12  \\
\enddata
\tablenotetext{a}{T$_{d}$ in \textit{\textbf{R1}} \& \textit{\textbf{R2}} is derived from SED fit \textit{\textbf{S1}}. T$_{d}$ in \textit{\textbf{R3}} \& \textit{\textbf{R4}} is derived from SED fit \textit{\textbf{S2}}.}
\end{deluxetable*}

\begin{figure}
\centering
\includegraphics[angle=0,scale=0.6]{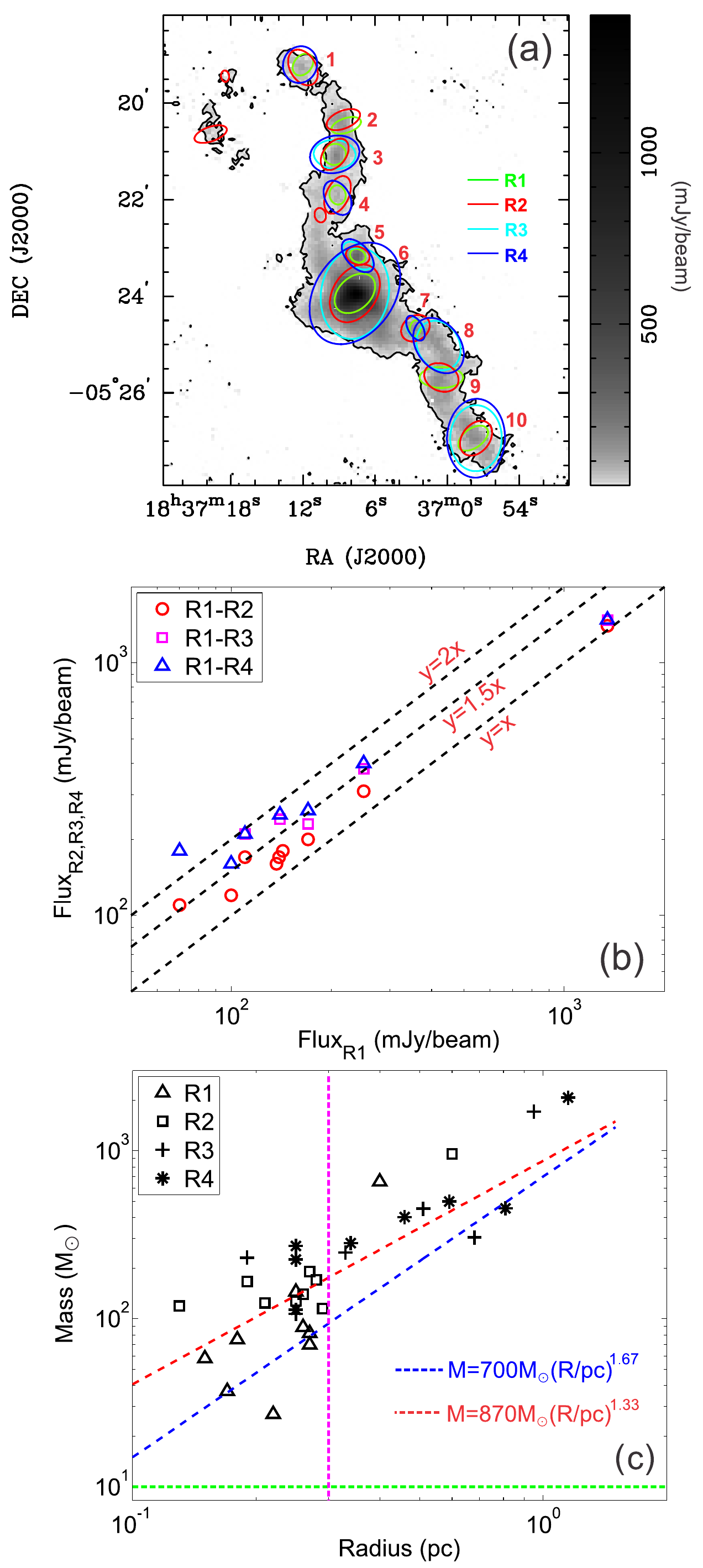}
\caption{(a). The 850 $\micron$ continuum emission in the \textit{\textbf{R2}} reduction is shown in gray-scale and black contours. The contour levels are the same as in the lower panel of Figure \ref{SCUBA-2}. The green, red, cyan, and blue ellipses represent the dense clumps identified in the 850 $\micron$ continuum emission maps from the \textit{\textbf{R1}}, \textit{\textbf{R2}}, \textit{\textbf{R3}} and \textit{\textbf{R4}} reductions, respectively. (b). Comparison of the peak fluxes of dense clumps in different data reductions. (c). Mass-Radius relation of the dense clumps. The red dashed line shows a threshold to
form high-mass protostars \citep{kauff10}. The blue dashed line shows the radius-mass relation for the clumps which undergo quasi-isolated gravitational
collapse in a turbulent medium \citep{Li17}. Here we remind readers that the spatial resolution (beam size; magenta dashed line) of SCUBA-2 at 850 $\micron$ is $\sim$0.3 pc and the 3$\sigma$ mass sensitivity (black dashed line) is $\sim$10 M$_{\sun}$ (assuming T$_{d}$=10 K) at the distance of 4.2 kpc. \label{core} }
\end{figure}

The ``SCOPE" survey aims to obtain a census of ``all-sky" distributed dense clumps and cores. Since PGCC sources trace some of the coldest ISM in the Galaxy, an extensive survey of PGCC sources is able to provide us with a number of candidates of clumps and cores at their earliest evolutionary phases. In nearby clouds, we expect to discover a population of prestellar core candidates or extremely young (e.g., Class 0) protostellar objects \citep{liu16,tat17}. In Galactic Plane PGCCs, we are particularly interested in searching for massive clumps or cores that may represent the initial conditions of high-mass star formation. Below, we demonstrate the source extraction process based on the G26 SCUBA-2 images. The properties of clumps identified in G26 will also be investigated.

Extraction of the dense clumps was done
using the FELLWALKER \citep{berry15} source-extraction algorithm,
part of the Starlink CUPID package \citep{berry07}. The core of the FellWalker algorithm is a gradient-tracing scheme consisting of following
many different paths of steepest ascent in order to reach a significant
summit, each of which is associated with a clump \citep{berry07}. FellWalker is less dependent on specific parameter settings than
CLUMPFIND \citep{berry07}. The source-extraction process with FellWalker in the SCOPE survey is the same as that used by the JCMT Plane Survey and details can be found in \cite{Moore15,Eden17}. A mask constructed above a threshold of 3 $\sigma$ (i.e.,
three times the pixel-to-pixel noise) in the SNR map is applied to the intensity map as input for the
task CUPID:EXTRACTCLUMPS, which extracts the peak and integrated
flux-density values of the clumps. A further
threshold for CUPID:FINDCLUMPS was the minimum number of
contiguous pixels, which was set at 12 corresponding to the number of pixels expected to be found in an unresolved
source with a peak SNR of 5 $\sigma$, given a 14-arcsec beam and 4-arcsec pixels. The cores close to the map edges which are associated with artificial structures were further removed manually.

In total, ten clumps were identified from the SCUBA-2 images. Those ten clumps can be identified in at least two of the SCUBA-2 data reductions (\textit{\textbf{R1}}, \textit{\textbf{R2}}, \textit{\textbf{R3}} and \textit{\textbf{R4}}; see section 6.2 in Appendix A). Figure \ref{core}a presents the distributions of these ten clumps on the 850 $\micron$ image from the \textit{\textbf{R2}} reduction. From checking Spitzer (see Figure \ref{IRDC}) and the Herschel/PACS 70 $\micron$ data, we found that clumps ``6" and ``10" are associated with young stellar objects, while the others are likely starless. The parameters of these clumps are summarized in Table \ref{clumppara}. The effective radius is defined as $R_{eff}=\sqrt{ab}$, where a and b are the deconvolved FWHM sizes of the clump major and minor axes. The clump masses (M) are derived with Equation A1 in Appendix A with total fluxes of 850 $\micron$ continuum emission from the SCUBA-2. The radii and masses are listed in the 7th and 12th columns of Table \ref{clumppara}, respectively. The particle number density (n) and column density (N) were calculated as $n=\frac{M}{\frac{4}{3}\pi R_{eff}^3\mu m_{H}}$ and $N=\frac{M}{\pi R_{eff}^2\mu m_{H}}$, respectively, where $\mu=2.37$ is the mean molecular weight per ``free particle" (H$_{2}$ and He) and $m_{H}$ is the atomic hydrogen mass.

Figure \ref{core}b compares the peak fluxes of the dense clumps in different imaging schemes, respectively. In general, applying larger spatial filters increases the peak fluxes particularly for less dense clumps. The peak flux of the most massive clump ``6" does not change too much ($<$10\%) in different imaging schemes. In contrast to \textit{\textbf{R1}}, however, the peak fluxes of other less dense clumps increased by $\sim$20\%-50\% in \textit{\textbf{R2}}, and by 50\% to 100\% in \textit{\textbf{R3}} and \textit{\textbf{R4}}. Imaging scheme \textit{\textbf{R1}} is used in the first data release of the SCOPE survey. \textit{\textbf{R1}} is very efficient to detect dense clumps but cannot well recover flux from extended structures. Therefore, we suggest to use external masks and larger spatial filtering in SCUBA-2 data reduction to recover more flux if Herschel data are available.

Figure \ref{core}c presents the mass-radius relation for the dense clumps. In general, larger scales kept by spatial filtering lead to larger masses and radii. The red dashed line shows a density threshold for high-mass star formation \citep{kauff10}. Except for clump ``6", all the other clumps in \textit{\textbf{R1}} are below this threshold. Most clumps, however, will move above the empirical threshold when larger scales are kept by the spatial filtering, as in \textit{\textbf{R2}}, \textit{\textbf{R3}}, and \textit{\textbf{R4}}. Most clumps (revealed in \textit{\textbf{R2}}, \textit{\textbf{R3}}, and \textit{\textbf{R4}}) are above the predicted mass-radius relation for clumps undergoing quasi-isolated gravitational collapse in a turbulent medium \citep{Li17}. The clumps in G26 especially the dense and massive ones are very likely bounded by gravity and have ability to form high-mass stars.

We derived the thermal Jeans masses $M_{J}=0.877M_{\sun}(\frac{T}{10 K})^{3/2}(\frac{n}{10^5 cm^{-3}})^{-1/2}$ following \cite{wang14}. The resulting n, N and $M_{J}$ are listed in the last three columns in Table \ref{clumppara}. The clump masses are also about one order of magnitude larger than the corresponding thermal Jeans masses.
The clumps are all located along the filament with a mean separation of $\sim$1.3 pc, which is much larger than the thermal Jeans length ($\sim$ 0.1 pc for T=14 K and the mean volume density n=$6.2\times10^4$ cm$^{-3}$ for clumps in \textit{\textbf{R2}}), indicating that fragmentation in G26 is not governed by thermal instability but determined by other factors like turbulence or magnetic fields, as revealed in other IRDCs \citep[e.g.,][]{wang11,wang14,cont16}.

The so-called ``sausage instability" of a gas cylinder proposed by \cite{Chand53} and promoted by many others \citep[e.g.,][]{stod63,ostr64,inut92,fiege00,fisch12} has been often applied for studies of the fragmentation in filamentary clouds \citep[e.g.,][]{jack10,wang11,wang14,liu17}. In this scenario, an isothermal, hydrostatic, non-magnetised, infinite filament becomes unstable and fragments into a chain of equally spaced fragments if its mass per unit length along the cylinder (or linear mass density) is close enough to the critical value so that perturbations in the filament have time to grow. In G26, the mean spacing and mass of clumps in G26 are $\sim$1.3 pc and $\sim$540 M$_{\sun}$, respectively, which were derived for clumps in \textit{\textbf{R4}} reduction. The fragment mass in cylindrical fragmentation can be estimated as $M_\mathrm{cl} = 575.3 M_\odot
\left( \frac{\sigma}{1 ~ \rm km \, s^{-1}} \right)^3
\left( \frac{n_c}{10^5 ~ \rm cm^{-3}} \right)^{-1/2}$; The typical spacing in cylindrical fragmentation is $\lambda_\mathrm{cl} = 1.24\, {\rm pc}
\left( \frac{\sigma}{1 ~ \rm km \, s^{-1}} \right)
\left( \frac{n_c}{10^5 ~ \rm cm^{-3}} \right)^{-1/2}$ \citep{wang14}. $\sigma$ is the 1-D velocity dispersion and n$_{c}$ is the volume density at the centre
of the cylinder. Here we take n$_{c}$ as the mean number density ($6.2\times10^4$ cm$^{-3}$) for clumps identified in \textit{\textbf{R2}} reduction because \textit{\textbf{R2}} filtered more extended emission than \textit{\textbf{R4}} and thus the mean number density for clumps in \textit{\textbf{R2}} is much closer to the volume density at the centre
of the cylinder. If we only consider thermal pressure (assuming T=14 K), the $\sigma$, $M_\mathrm{cl}$ and $\lambda_\mathrm{cl}$ are 0.22 km~s$^{-1}$, 7.8 M$_{\sun}$, and 0.35 pc. The $M_\mathrm{cl}$ and $\lambda_\mathrm{cl}$ are much smaller than the observed values in G26, indicating that thermal pressure alone cannot govern the fragmentation. However, if we also consider turbulent support, the $\sigma$, $M_\mathrm{cl}$ and $\lambda_\mathrm{cl}$ will become $\sim$0.94 km~s$^{-1}$ (estimated from C$^{18}$O line width; see Table \ref{linepara}), 600 M$_{\sun}$ and 1.5 pc. The $M_\mathrm{cl}$ and $\lambda_\mathrm{cl}$ derived from turbulence-dominated cylindrical fragmentation are very consistent with the observed values ($\sim$540 M$_{\sun}$ and $\sim$1.3 pc), suggesting that the fragmentation in G26 from cloud (at $\sim$10 pc scale) to clumps (at $\sim$1 pc scale) is very likely dominated by turbulence in the absence of magnetic fields. Limited by the spatial resolution ($\sim$0.3 pc at the distance of 4.2 kpc) and mass sensitivity (3$\sigma$ mass sensitivity is $\sim$10 M$_{\sun}$), we cannot study fragmentation from clump scale to core scale (at $\sim$0.1 pc scale) with present data. Further fragmentation below the completeness and resolution limit of the present data is very likely to take place as seen in other IRDCs \citep{wang11,wang14}. The hierarchical fragmentation in G26 at all spatial scales can only be studied from higher spatial resolution interferometric observations.

\subsection{Evidence for grain growth}

\begin{figure*}
\centering
\includegraphics[angle=90,scale=0.5]{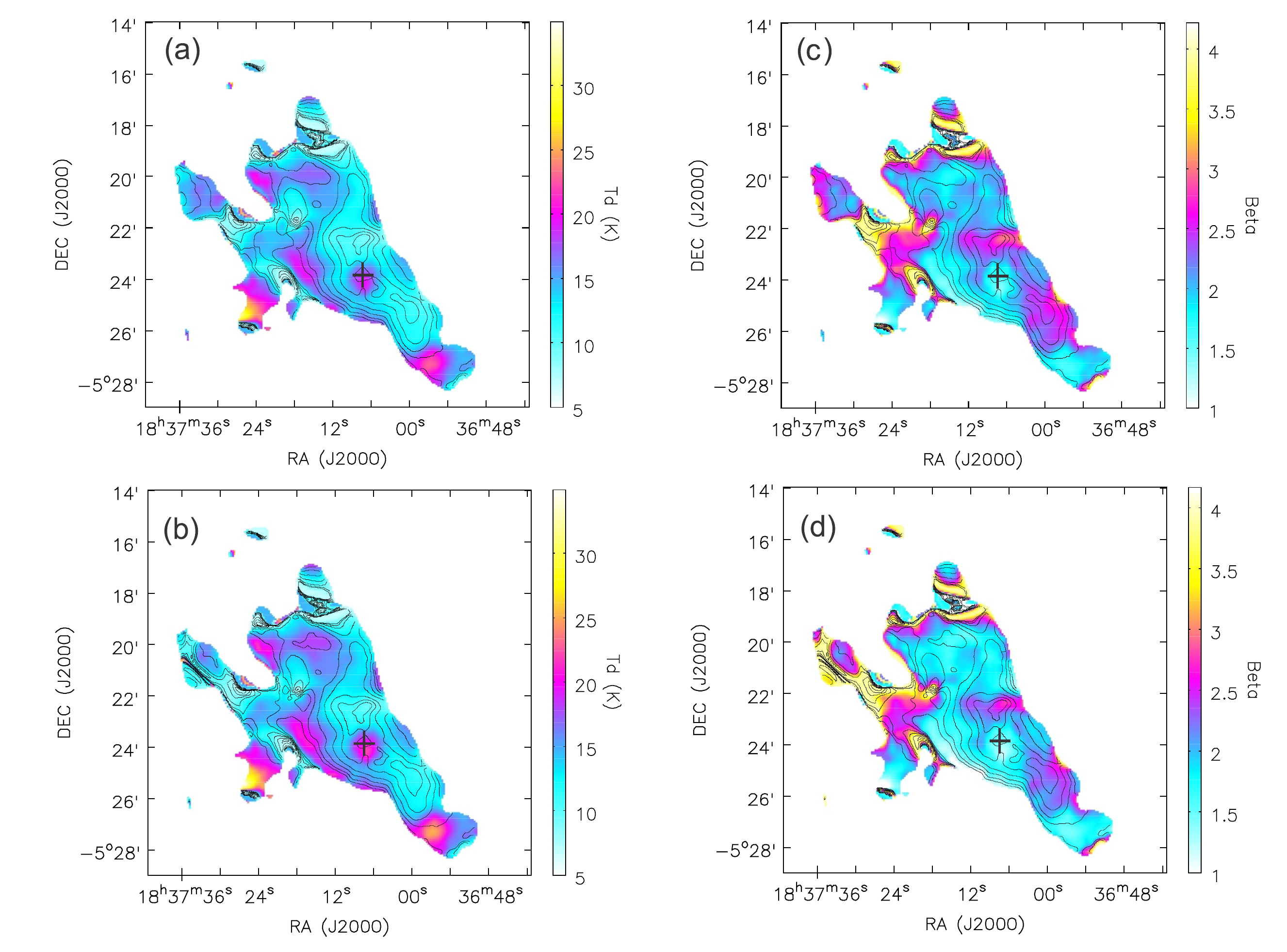}
\caption{(a). The dust temperature map in the SED fit with free $\beta$ using the Herschel data only is shown in color image. The 250 $\micron$ optical depth is shown in contours. The contour levels are [0.03, 0.05, 0.1, 0.2, 0.4, 0.6, 0.8]$\times$0.0282 (b). The dust temperature map in the  SED fit with free $\beta$ using both the Herschel and SCUBA-2 data is shown in color image. The 250 $\micron$ optical depth is shown in contours. The contour levels are [0.03, 0.05, 0.1, 0.2, 0.4, 0.6, 0.8]$\times$0.0211 (c). The $\beta$ map in the SED fit with free $\beta$ using the Herschel data only is shown in color image. The 250 $\micron$ optical depth is shown in contours. The contour levels are the same as in panel a. (d). The $\beta$ map in the SED fit with free $\beta$ using both the Herschel and SCUBA-2 data is shown in color image. The 250 $\micron$ optical depth is shown in contours. The contour levels are the same as in panel b. The crosses in the four panels mark the position of the most massive clump.\label{freebeta} }
\end{figure*}

\begin{figure}
\centering
\includegraphics[angle=0,scale=0.5]{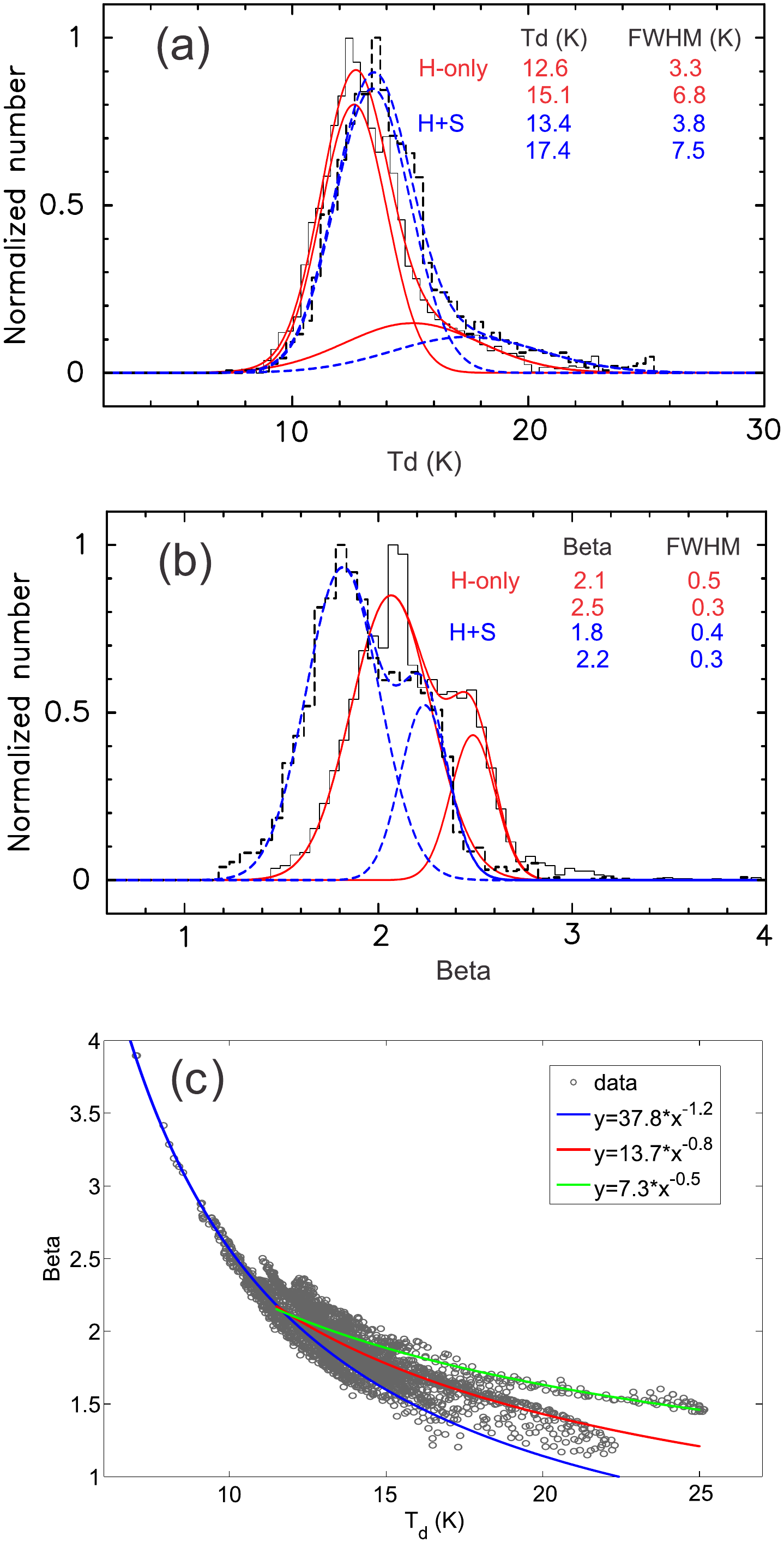}
\caption{(a). Histograms of T$_{d}$. The T$_{d}$ in the SED fit with free $\beta$ using the Herschel data only is shown in solid lines, while the T$_{d}$ in the SED fit with free $\beta$ using both the Herschel and SCUBA-2 data is shown in dashed lines. The red and blue color lines show the normal distribution fits. (b). Histograms of $\beta$. The $\beta$ in the SED fit with free $\beta$ using the Herschel data only is shown in solid lines, while the $\beta$ in the SED fit with free $\beta$ using both the Herschel and SCUBA-2 data is shown in dashed lines. The red and blue color lines show the normal distribution fits. (c). Spectral index $\beta$ vs. dust temperature T$_{d}$ in the SED fit with free $\beta$ using both the Herschel and SCUBA-2 data. The blue line shows the power-law fit toward the data with T$_{d}<$10 K. The red line shows the power-law fit toward the data with T$_{d}>$10 K. The green line outlines the trend for the data with T$_{d}>$15 K and high $\beta$.\label{Tempbeta}  }
\end{figure}

Observationally, dust properties (e.g., opacity) greatly affect the determination of density and temperature profiles of dense cores, and therefore influence the estimation of core masses. The dust grain size distributions in molecular clouds and cores can change with time because of grain sputtering, shattering,
coagulation, and growth \citep{Ormel09}.
Grain growth in the densest and coldest regions of interstellar clouds has been evidenced by an increase in the far-infrared opacity \citep{GCC-VI}. Indeed, the dust spectral index ($\beta$) and its temperature (T$_{d}$) dependence can provide additional information on the chemical composition, structure, and the size distribution of interstellar dust grains \citep{juve11}. This dust emissivity change has been attributed to the formation of fluffy aggregates in the dense medium, resulting from grain coagulation.

Recently, \cite{GCC-VI} studied dust optical depths by comparing measurements of submillimetre dust emission and the reddening of the light of background stars in the near-infrared (NIR) for a sample of 116 PGCCs in Herschel fields. In their studies, there are indications that $\beta$ increases towards the coldest regions and decreases strongly near internal heating sources. Although NIR reddening is an independent and reliable measure of column density, NIR extinction could not, however, trace the densest portion of the cloud due to low resolution. Additionally, the Herschel 250 $\mu$m band is less sensitive to very cold dust and may completely miss the highest column density peak \citep[see Figure 2 in][]{paga15}. Herschel 500 $\mu$m data could trace cold dust well but have relatively low angular resolution (35$^{\prime\prime}$). Adding high resolution photometric data at longer wavelengths (e.g. 850 $\mu$m from SCUBA-2) could help more accurately derive the dust spectral index, dust temperature, and column density (N$_{H_{2}}$) simultaneously \citep{sad13,chen16}.

For PGCC sources having Herschel data in the ``SCOPE" survey, we will determine the dust emissivity spectral index ($\beta$), dust temperature (T$_{d}$), and column density (N$_{H_{2}}$) simultaneously by combining 850 $\mu$m data with spatially--filtered Herschel photometric data. Since the PGCC sources are at different stages of cloud evolution from starless clumps to protostellar cores and are located in different Galactic environments, we will search for variations of $\beta$ that might be attributed to the evolutionary stage of the sources or to environmental factors, including the location within the Galaxy and stellar feedback. A pilot study of dust properties toward $\sim$100 PGCCs with Herschel data and SCUBA-2 data will be presented in \cite{juvela17}. Below we investigate the dust properties of G26 as an example.

In Figure \ref{freebeta}, we present the T$_{d}$, $\beta$, and $\tau_{250}$ maps of G26 from SED fits with free $\beta$ and T$_{d}$. The extended emission larger than 600$\arcsec$ in the Herschel and SCUBA-2 maps (from imaging scheme \textit{\textbf{R4}}; see section 6.2 in Appendix A) was filtered out. In general, the $\beta$ map is anti-correlated with the T$_{d}$ map. The most massive core, marked with a black cross, clearly shows higher temperature and lower $\beta$ than its surroundings. The low $\beta$ may indicate grain growth during core or star formation. Figure \ref{Tempbeta} a-b show histograms of T$_{d}$ and $\beta$. Only the data points with S/N$>$10 in the SCUBA-2 map, corresponding to the ridge region of the filament, are included in the statistics. In general, including 850 $\micron$ data tends to increase the T$_{d}$ by $\sim$ 1 K and lower $\beta$ by $\sim$0.3. The dust temperature histograms show a high temperature tail, which is caused by on-going star formation. Interestingly, the $\beta$ histograms can be well fitted with two normal distributions regardless of whether or not 850 $\micron$ data were included in the SED fits. Such a bimodal behavior in $\beta$ distribution may suggest grain growth along the filament.

Figure \ref{Tempbeta}c presents the correlation between T$_{d}$ and $\beta$. $\beta$ is generally anti-correlated with T$_{d}$. Interestingly, the $\beta$ vs. T$_{d}$ correlations can be roughly depicted with three power-laws, which again indicates the existence of different kinds of dust grains in the filament. Such an anti-correlation between T$_{d}$ and $\beta$ was also found in all Perseus
clumps \citep{chen16}. \cite{chen16} also found that these anti-correlations cannot be solely
accounted for by anti-correlated T$_{d}$ and $\beta$ uncertainties
associated with SED fitting. The anti-correlation may be partially explained by the dust grains' intrinsic
$\beta$ dependency on temperature but more likely by the sublimation
of surface ice mantles, which can increase $\beta$ when present on a dust grain \citep{chen16}. A thorough investigation of the correlation between T$_{d}$ and $\beta$ is beyond the scope of this paper. We will systematically study this issue in the future with more ``SCOPE" objects.

\subsection{Cloud structure}

\begin{figure*}
\centering
\includegraphics[angle=-90,scale=0.6]{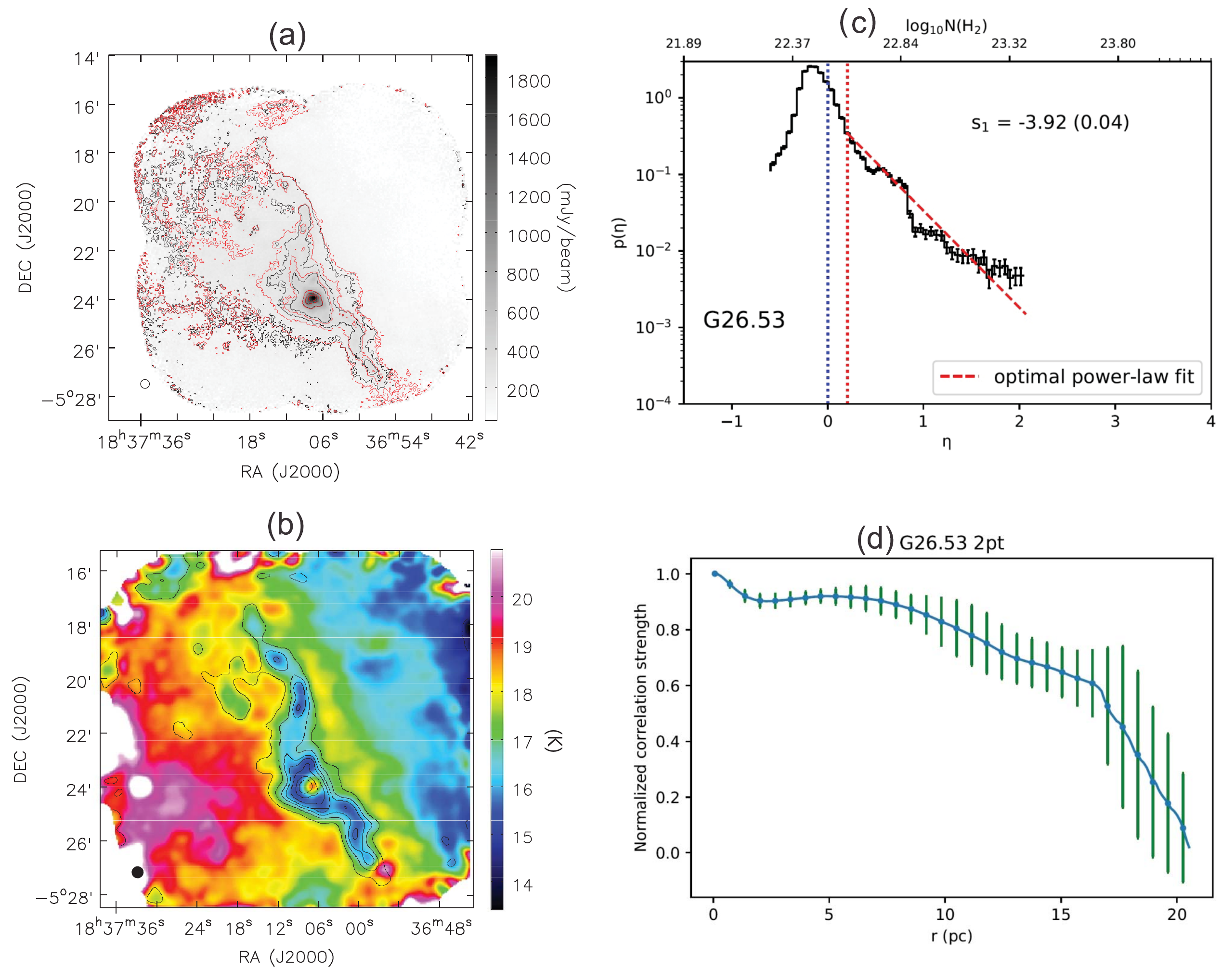}
\caption{(a). The gray-scale image and black contours show the 850 $\micron$ continuum emission from combined Planck data and SCUBA-2 data in \textit{\textbf{R2}}. The contour levels are [0.1, 0.15, 0.2, 0.4, 0.6, 0.8]$\times$1.92 Jy~beam$^{-1}$.  The red contours show the 850 $\micron$ continuum emission from combined Planck data and SCUBA-2 data in \textit{\textbf{R4}}. The contour levels are [0.1, 0.15, 0.2, 0.4, 0.6, 0.8]$\times$1.97 Jy~beam$^{-1}$. (b). The dust temperature map from SED fits with Herschel and combined Planck and SCUBA-2 850 $\micron$ data in \textit{\textbf{R2}} is shown in color image. The column density map is shown in contours. The contour levels are [0.15, 0.2, 0.4, 0.6, 0.8]$\times$2.76$\times10^{23}$ cm$^{-2}$. (c). Column density probability distribution function (N-PDF). The blue vertical dashed line marks the mean column density of $\sim3.3\times10^{22}$ cm$^{-2}$. The red vertical dashed line shows the starting column density ($\sim4\times10^{22}$ cm$^{-2}$) of the power law fit.  (d). Two-point correlation functions of the column density distribution.\label{planck}}
\end{figure*}

To understand better the star or core formation process, it is crucially important to study their parent molecular cloud properties. SCUBA-2 has better resolution and longer wavelengths than Herschel/SPIRE bands, and thus is more suitable to trace the cold ISM. Due to variations of the atmosphere which mimic emission from extended astronomical objects, SCUBA-2 data are not well suited to capture the extended structures in molecular clouds. The loss of the filtered emission by SCUBA-2, however, can be somewhat corrected based on the Planck data. The combined \textit{Planck} 353 GHz data and the SCUBA-2 850 $\micron$ data will be used to investigate the density
distribution and structures at various cloud scales. Below we use G26 to demonstrate the combination of the Planck and SCUBA-2 data.

We combined the \textit{Planck} 353 GHz image with the SCUBA-2 850 $\micron$ images in the \textit{\textbf{R2}} and \textit{\textbf{R4}} reductions following the same procedure as in \cite{lin16,lin17}. In principle, the data combination was performed in the Fourier domain, yielding high-resolution (14$\arcsec$) combined data that have little or no loss of extended structure \citep{lin16,lin17}. The combined images are shown in Figure \ref{planck}a. In general, the two images show very similar morphology and flux density. From these combined images, we find that the dense filament resides in a cloud with very smooth structures. The cloud boundary is well enclosed by the 10\% contours in Figure \ref{planck}a. Particularly, the dense filament is close to the cloud edge, indicating possible dynamical clues to its formation (e.g., external compression).

Figure \ref{planck}b presents the dust temperature and column density maps obtained from SED fits with Herschel/PACS 160 $\micron$, SPIRE 250/350 $\micron$ and combined Planck+SCUBA-2 850 $\micron$ data in \textit{\textbf{R2}} reduction. The final dust temperature and column density maps have an angular resolution of 25$\arcsec$. The temperature map clearly reveals a temperature gradient from south-east to north-west.

To quantify the dense gas distributions systematically, we perform analyses of the column density probability distribution functions (N-PDF) following \cite{lin16,lin17}. The natural logarithm of the ratio of column density and mean column density is
$\eta = ln(N_{\rm H_{2}}/\langle N_{\rm H_{2}}\rangle)$, and the normalization of the probability function is given by
$\int_{-\infty}^{+\infty}p(\eta)d\eta = \int_{0}^{+\infty}p(N_{\rm H_{2}})d(N_{\rm H_{2}})=1 $. Figure \ref{planck}c shows the column density probability distribution functions (N-PDF). The N-PDF in molecular clouds is usually found to consist of a lognormal like part at low column density and a power-law like part at high column density \citep{Klessen00,Kritsuk11,Fed13,Schneider12,Schneider15,lin16,lin17}. The N-PDF power-law tail is usually linked to the effects of gravity \citep{Klessen00,Kritsuk11,Fed13,lin16,lin17}. We tried to fit the tail with a power law for the N-PDF of G26. Although the power-law fit is poor, its slope (-3.9) is comparable to those of IRDCs G28.34+0.06 (-3.9) and G14.225-0.506 (-4.1) but smaller than those of protoclusters \citep{lin16,lin17}, indicating that G26 is still at a very young evolutionary stage and is not greatly affected by star formation activities as in high-mass protoclusters.

To diagnose the characteristic spatial scales in G26, we determine the two-point correlation functions (2PT) of gas column density following \cite{lin16,lin17}. Figure \ref{planck}d shows the two-point correlation (2PT) function of column densities in the observed field. The correlation strength at a separation scale (lag) of $l$ is calculated by
\begin{equation}
S_{tr}(l) = \frac{\left\langle\emph{X(\textbf{\emph{r}})X(\textbf{\emph{r+l}})}\right\rangle_{\textbf{\emph{l}}}}{\left\langle\emph{X(\textbf{\emph{r}})X(\textbf{\emph{r}})}\right\rangle_{\textbf{\emph{l}}}},
\end{equation}
where $\emph{X(\textbf{\emph{r}})}$ denotes the column density value at position \textbf{\emph{r}}, and the angle brackets are an average over all pairs of positions with a separation of $l$. The final form of the correlation function is normalized by the peak correlation strength to enable comparisons between different fields. For a more detailed description, see \citep{lin16,lin17}.

In general, the 2PT
function shows a smooth decay of correlation strengths over all spatial scales up to $\sim$15 pc. Such a flat profile in 2PT correlation is also witnessed towards IRDCs like G11.11-0.12 \citep[see Figure 5 in][]{lin17}, but is in contrast with most of the active OB-cluster-forming regions, which have a steep decay of correlation strength at small spatial scales \citep{lin16}. The flat 2PT correlation indicates that the column density distribution of G26 is very homogeneous over all spatial scales larger than 1 pc and the mass is less concentrated overall than in high-mass protoclusters.

The N-PDF and 2PT are powerful tools to investigate the cloud structure evolution and should be applied to other SCOPE fields.

\subsection{Dense gas fraction}

\begin{figure*}
\centering
\includegraphics[angle=-90,scale=0.6]{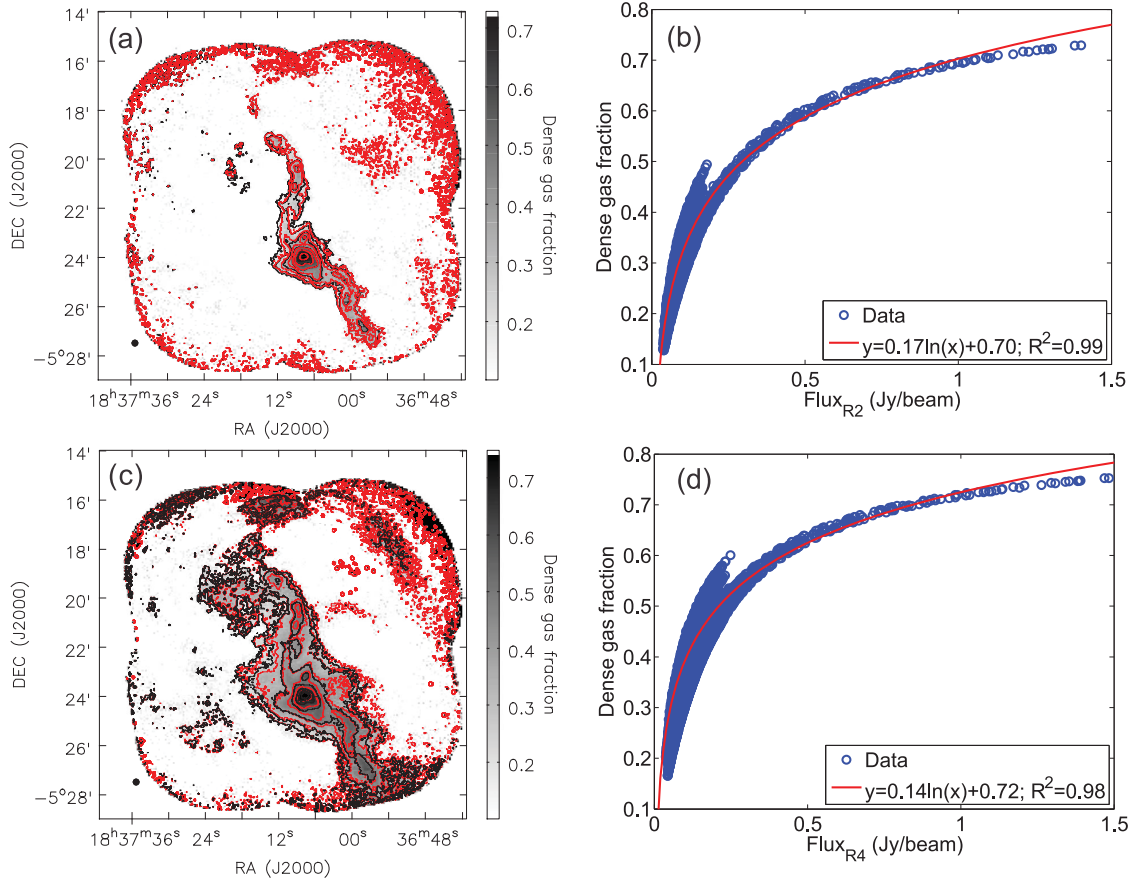}
\caption{(a). The dense gas fraction in \textit{\textbf{R2}} reduction is shown in gray-scale and red contours. The contour levels are [0.2, 0.3, 0.4, 0.5, 0.6, 0.7]. The 850 $\micron$ continuum emission in \textit{\textbf{R2}} reduction is shown in black contours. The contour levels are the same as in the lower panel of Figure \ref{SCUBA-2}. (b). The dense gas fraction vs. the 850 $\micron$ continuum flux density in the \textit{\textbf{R2}} reduction. The red line shows the logarithmic function fit toward the pixels with flux density larger than 200 mJy~beam$^{-1}$ in the SCUBA-2 image. (c). The dense gas fraction in \textit{\textbf{R4}} reduction is shown in gray-scale and red contours. The contour levels are [0.2, 0.3, 0.4, 0.5, 0.6, 0.7]. The 850 $\micron$ continuum emission in \textit{\textbf{R4}} reduction is shown in black contours. The contour levels are the same as in Figure \ref{SCUBA-Lab}b. (d). The dense gas fraction vs. the 850 $\micron$ continuum flux density in the \textit{\textbf{R4}} reduction. The red line shows the logarithmic function fit toward the pixels with flux density larger than 300 mJy~beam$^{-1}$ in the SCUBA-2 image.\label{DFG}}
\end{figure*}

The empirical power-law relations between the star formation rate (SFR), surface
density ($\Sigma_{SFR}$) and the surface density of cold gas ($\Sigma_{gas}$), pioneered in the works of \cite{sch59} and \cite{ken98}, the so-called Kennicutt--Schmidt (K-S) law, is of
great importance as an input for theoretical models of galaxy evolution. Recently, nearly linear correlations between star formation rates and line luminosities of dense molecular gas tracers (e.g., HCN and CS) have been found toward both Galactic dense clumps and galaxies \citep{gao04,wu05,wu10,lada10,zhang14,liu16c,step16}, strongly suggesting that star formation is mainly related to dense gas in molecular clouds. Therefore, it is important to evaluate the proportions of dense gas in molecular clouds. As discussed in \cite{liu16c}, the relation between star formation rates and clump masses (traced by filtered continuum maps) is as tight as correlations between star formation rates and line luminosities of dense gas tracers (e.g., HCN, CS). Particularly, filtered continuum maps have advantages to trace the total dense gas masses than dense gas tracers (e.g., HCN, HCO$^{+}$, CS), whose emissions are usually optically thick \citep{liu16c}. Since the filtered SCUBA-2 images intrinsically filter the large-scale diffuse gas and are most sensitive to the high volume densities ($\sim10^4$ cm$^{-3}$ for G26) in clouds, the flux ratio between the SCUBA-2 data and the Planck+SCUBA-2 combined data is well representative of the dense gas fraction in the cold dust (f$_{DG}$), as also suggested in \cite{cseng16}.

Figure \ref{DFG} a,c show the dense gas fraction maps for G26. In general, the dense gas fraction increases with column density. In panels (b) and (d), we investigate the relationship between the dense gas fraction and flux density for the pixels with SNR$>$5 in the SCUBA-2 images. The correlations can be well fitted with logarithmic functions toward high flux density points. The logarithmic function fits can also extend to low flux density points even though the low flux density points show much larger scatter. The f$_{DG}$ in image from \textit{\textbf{R2}} reduction range from 0.13 to 0.73 with a median value of 0.30. While the f$_{DG}$ in the image from \textit{\textbf{R4}} reduction range from 0.17 to 0.75 with a median value of 0.37. The dense gas fractions determined from images with different spatial filters are slightly different but only vary less than $\sim$10\%. It seems that only 30\%-40\% of the cloud gas in G26 is dense.

\subsection{On the origin of the dense filament}
\subsubsection{Evidence for large scale compression flows}

\begin{figure}
\centering
\includegraphics[angle=0,scale=0.6]{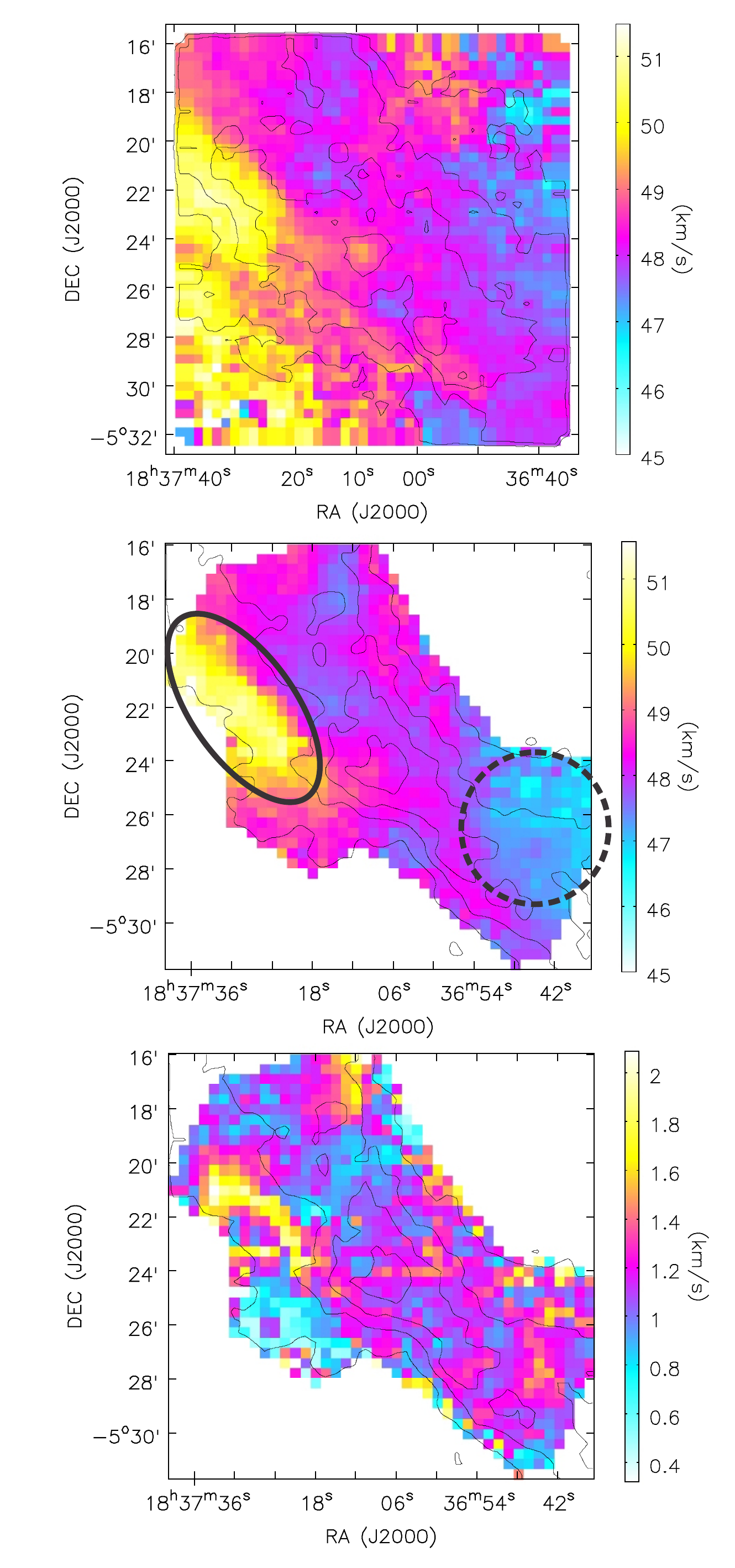}
\caption{Upper panel: $^{12}$CO (1-0) integrated intensity contours overlaid on its first-moment map. Middle panel: $^{13}$CO (1-0) integrated intensity contours overlaid on its first-moment map. The dashed circle and solid ellipse mark the regions that show large velocity gradients. Lower panel: $^{13}$CO (1-0) integrated intensity contours overlaid on its second-moment map. The contours for both $^{12}$CO (1-0) and $^{13}$CO (1-0) are from 20\% to 80\% in steps of 20\% of the peak values. The peak integrated intensities of $^{12}$CO (1-0) and $^{13}$CO (1-0) are 56 K~km~s$^{-1}$ and 20 K~km~s$^{-1}$, respectively.\label{TRAO-mom}  }
\end{figure}

\begin{figure}
\centering
\includegraphics[angle=-90,scale=0.4]{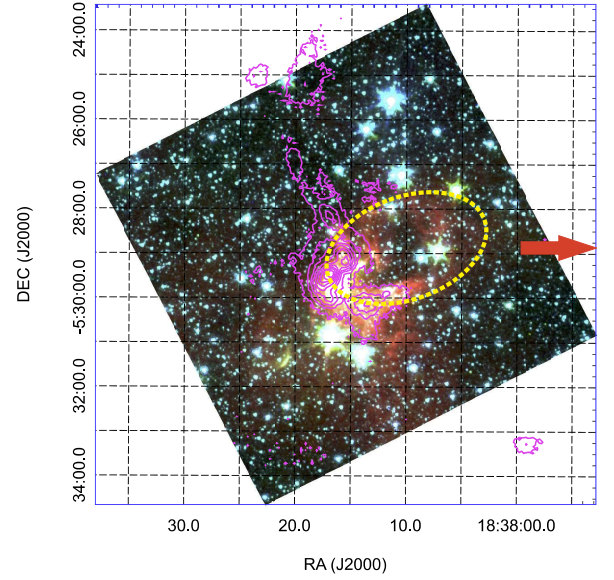}
\caption{Spitzer/IRAC three-color (3.6 $\micron$ in blue, 4.5 $\micron$ in green, and 8 $\micron$ in red) composite image of the G26 bubble region. The magenta contours represent the SCUBA-2 850 $\micron$ continuum emission. The contour levels are [0.1, 0.2, 0.3, 0.4, 0.5, 0.6, 0.7, 0.8, 0.9]$\times$958.6 mJy~beam$^{-1}$. The yellow dashed ellipse outlines the infrared bubble. The red arrow marks the direction of G26. \label{HII} }
\end{figure}

\begin{figure*}
\centering
\includegraphics[angle=-90,scale=0.65]{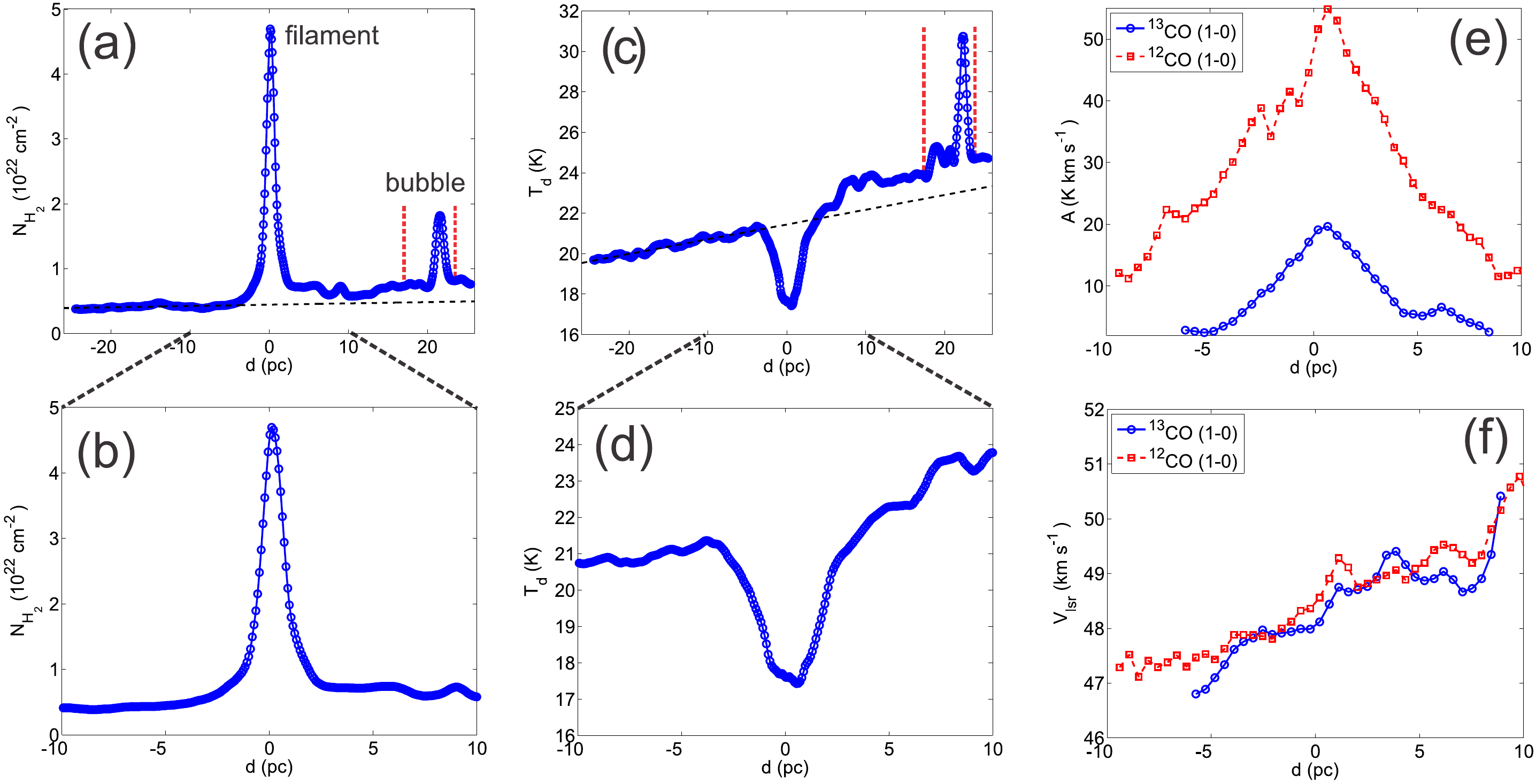}
\caption{The variation of column density (a, b), dust temperature (c,d), integrated intensity (e) and intensity weighted velocity (f) of $^{12}$CO (1-0) and $^{13}$CO (1-0), along the yellow dashed line across the filament seen in the upper panel of Figure \ref{Herschel}. The distances are calculated with respect to the peak of the filament. The black dashed lines in (a) and (c) are linear fits toward the positions (-25 pc to 5 pc) behind the filament. The red dashed lines in (a) and (c) mark the bubble region.\label{compress}   }
\end{figure*}

The origin of dense filaments in giant clouds still remains a puzzle to astronomers \citep{and14}. Filaments have been well predicted in simulations of supersonic turbulence in the absence of gravity, which can produce hierarchical structure with a lognormal density distribution seen in observations \citep{vaz94}. Filaments in strongly magnetized turbulent clouds \citep[e.g. B211/3, Musca;][]{pal13,cox10} are oriented preferentially perpendicular to the magnetic field lines, suggesting an important role of magnetic fields in filament formation. Filamentary structures in the Galaxy are preferentially aligned parallel to the Galactic mid-plane and therefore with the direction of large-scale Galactic magnetic field, suggesting a possible connection between large-scale Galactic dynamics and filament formation \citep{wang15,wang16,li16}.  The formation of filaments by self-gravitational fragmentation of sheet-like clouds has been seen in simulations of 1D compression \citep[e.g., by an expanding H{\sc ii} region, an old supernova remnant, or the collision of two clouds; ][]{Inu15,Fed16,LiP17}. For example, the asymmetric column density profiles of filaments in the Pipe Nebula are most likely the result of large-scale compressive flows generated by the winds of the nearby Sco OB2 association \citep{pere12}.

The ``SCOPE" survey aims to resolve Planck cold clumps and reveal their substructures. We will assess the detection rate of filaments in Planck cold clumps in different environments (e.g., spiral arms, interarms, high latitude, expanding H{\sc ii} regions, supernova remnants). In conjunction with molecular line data from the ``TOP" and ``SAMPLING" surveys, which provide kinematic information, we will be able to study the formation mechanisms of dense filaments in widely different environments. Below we present evidence of compression flows that may be responsible to the formation of the G26 filament.

As shown in the upper panel of Figure \ref{Herschel} and Figure \ref{planck}b, the temperature gradient on large scales is perpendicular to the dense filament G26. Figure \ref{TRAO-mom} a-b present maps of the first moment of $^{12}$CO (1-0) and $^{13}$CO (1-0), respectively. The maps reveal large-scale velocity gradients along the NW-SE direction across the whole map. Temperature gradient suggests an asymmetric heating source. Together with the velocity gradient, one can image compression flows from the left.

The origin of the large-scale compression flows is unclear. To the east of G26, we found an infrared bubble. Figure \ref{HII} shows a three-color composite Spitzer/IRAC image of the infrared bubble region overlaid with SCUBA-2 850 $\micron$ continuum emission in contours. As shown in Figure \ref{compress}, the locations between the bubble and the filament have higher column density ($\sim$1.6 times higher) and higher temperature ($\sim$3 K higher) than the locations west of the filament, suggesting a pressure gradient exists from the south-east to north-west. Also, the filament shows an asymmetric column density profile (panel a \& b), again indicating that it may be compressed by external pressure from its south-east side. Panel (e) and (f) present the variance of the integrated intensity and intensity weighted velocity of $^{12}$CO (1-0) and $^{13}$CO (1-0) line emission, respectively. Both $^{12}$CO (1-0) and $^{13}$CO (1-0) line emission reveal a large-scale velocity gradient ($\sim$0.16 km~s$^{-1}$~pc$^{-1}$) perpendicular to the filament, again suggesting the existence of compression flows.

Note that the infrared bubble only has an extent of $\sim$5 pc and is $\sim$15 pc away in projection from the filament G26 (see Figures \ref{Herschel}a and \ref{Pong}a in Appendix A). Hence the bubble may not have ability to generate the large-scale compression flows. Since the G26 filament is parallel to the Galactic Plane (see Figure \ref{IRDC}), the large-scale compression flow may originate from the ram pressure from the OB associations in the Galactic Plane. In any case, the higher pressure to the south-east of G26 can continuously sweep up the interstellar gas to feed the dense filament.

\subsubsection{Collisions of sub-filaments?}

\begin{figure*}
\centering
\includegraphics[angle=0,scale=0.8]{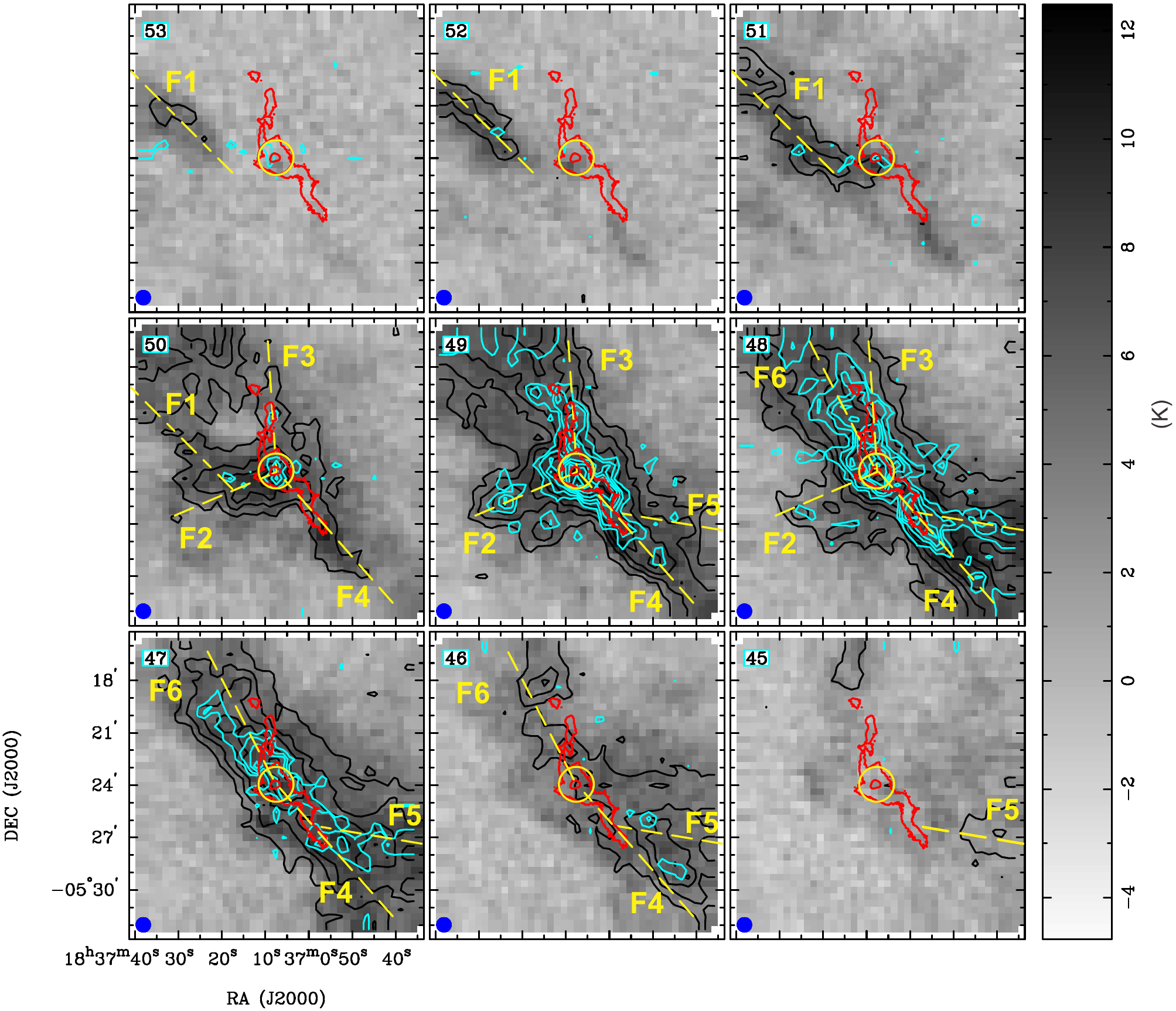}
\caption{Channel maps of J=1-0 transitions of $^{12}$CO (gray scale images), $^{13}$CO (black contours), and C$^{18}$O (blue contours). The contour levels for $^{13}$CO are from 10\% to 90\% in steps of 10\% of the peak intensity (8.2 K). The contour levels for $^{13}$CO are from 10\% to 90\% in steps of 10\% of the peak intensity (3.0 K). The yellow circles mark the central massive clump. The yellow lines outline the sub-filaments. The red contours show the SCUBA-2 850 $\micron$ continuum emission in the \textit{\textbf{R2}} data reduction. The contour levels are [0.05, 0.5]$\times$1.40 Jy~beam$^{-1}$.\label{TRAO-chan} }
\end{figure*}

\begin{figure*}
\centering
\includegraphics[angle=0,scale=0.8]{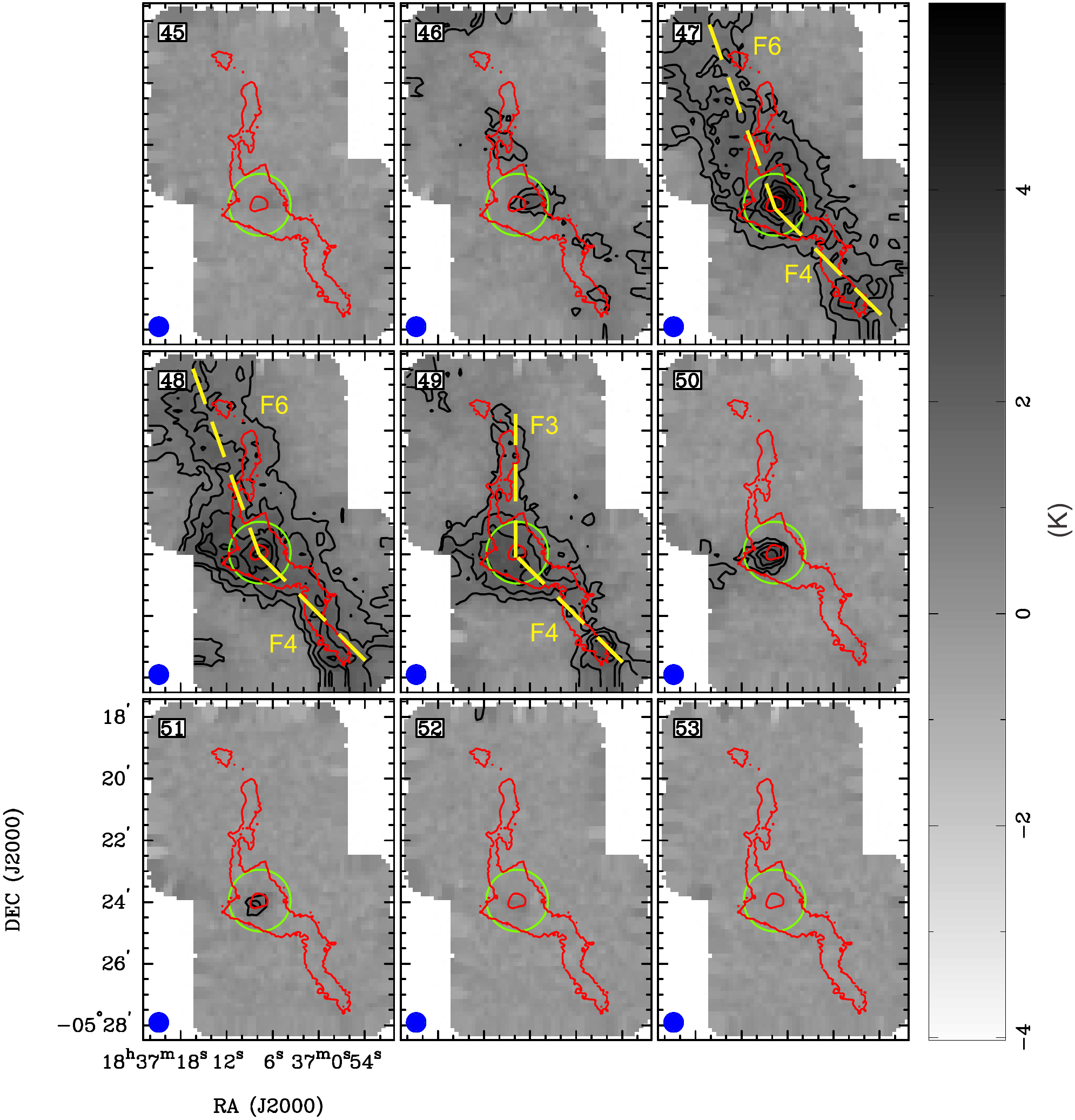}
\caption{ Channel maps of $^{13}$CO (2-1) line emission. The contour levels for $^{13}$CO are from 20\% to 90\% in steps of 10\% of the peak intensity (5.8 K). The yellow dashed lines outline the directions of Position-Velocity cuts in Figure \ref{PV}. The green circles mark the central massive clump. The red contours show the SCUBA-2 850 $\micron$ continuum emission in the \textit{\textbf{R2}} data reduction. The contour levels are [0.05, 0.5]$\times$1.40 Jy~beam$^{-1}$.\label{SMT-chan}}
\end{figure*}

Figure \ref{TRAO-chan} presents channel maps of the J=1-0 transitions of $^{12}$CO, $^{13}$CO and C$^{18}$O lines. In contrast to the 850 $\micron$ continuum emission, the molecular line emission reveals more complicated structures. Several velocity coherent subfilaments (F1 to F6) can be identified from the channel maps. The position angles of F1, F2, F3, F4, F5, and F6 are about 45$\arcdeg$, 115$\arcdeg$, 0$\arcdeg$, 44$\arcdeg$, 80$\arcdeg$ and 20$\arcdeg$, respectively. The sub-filaments F3, F4, and F6 are also seen in SMT $^{13}$CO (2-1) channel maps (see Figure \ref{SMT-chan}). Those sub-filaments have noticeable differences in velocity because they emerge in different velocity channels.

F4 and F6 form the main gas sub-filament, most of whose emission is in the 47-48 km~s$^{-1}$ channels. F1, which has the most redshited velocities among the sub-filaments, is located to the north-west and only appears in the 51-53 km~s$^{-1}$ channels. F1 seems to interact with the main filament (F6) as well as another sub-filament (F2). The interface between F1 and F6 shows much larger velocity dispersion, as revealed in the second moment map of $^{13}$CO (1-0) emission (see the lower panel of Figure \ref{TRAO-chan}). F5 has more blueshifted velocities than other sub-filaments. It is connected to F4. The G26 filament as revealed by SCUBA-2 850 $\micron$ continuum has a curved (or ``S" type) shape, suggesting that it may be dynamically interacting with its surroundings \citep{wang15,wang16}. The continuum filament is mainly associated with the F3 and F4 gas sub-filaments. Interestingly, F3 is clearly offset from the axis of the main gas filament (F6) and has redshifted velocities with respect to F6. F3 may be compressed due to a collision between F1 and F6. The collision may have reshaped the continuum filament into a curved shape. F5 may also collide with F4 and reshape the southern arm of the continuum filament. With the detection of several sub-filaments in G26 and because IRDCs are at the very early stages of star formation, we add further support to the idea that IRDCs are in a stage in which they are still being assembled \citep{jim10,sanh13}.

\subsubsection{Filament accretion}

\begin{figure}
\centering
\includegraphics[angle=0,scale=0.45]{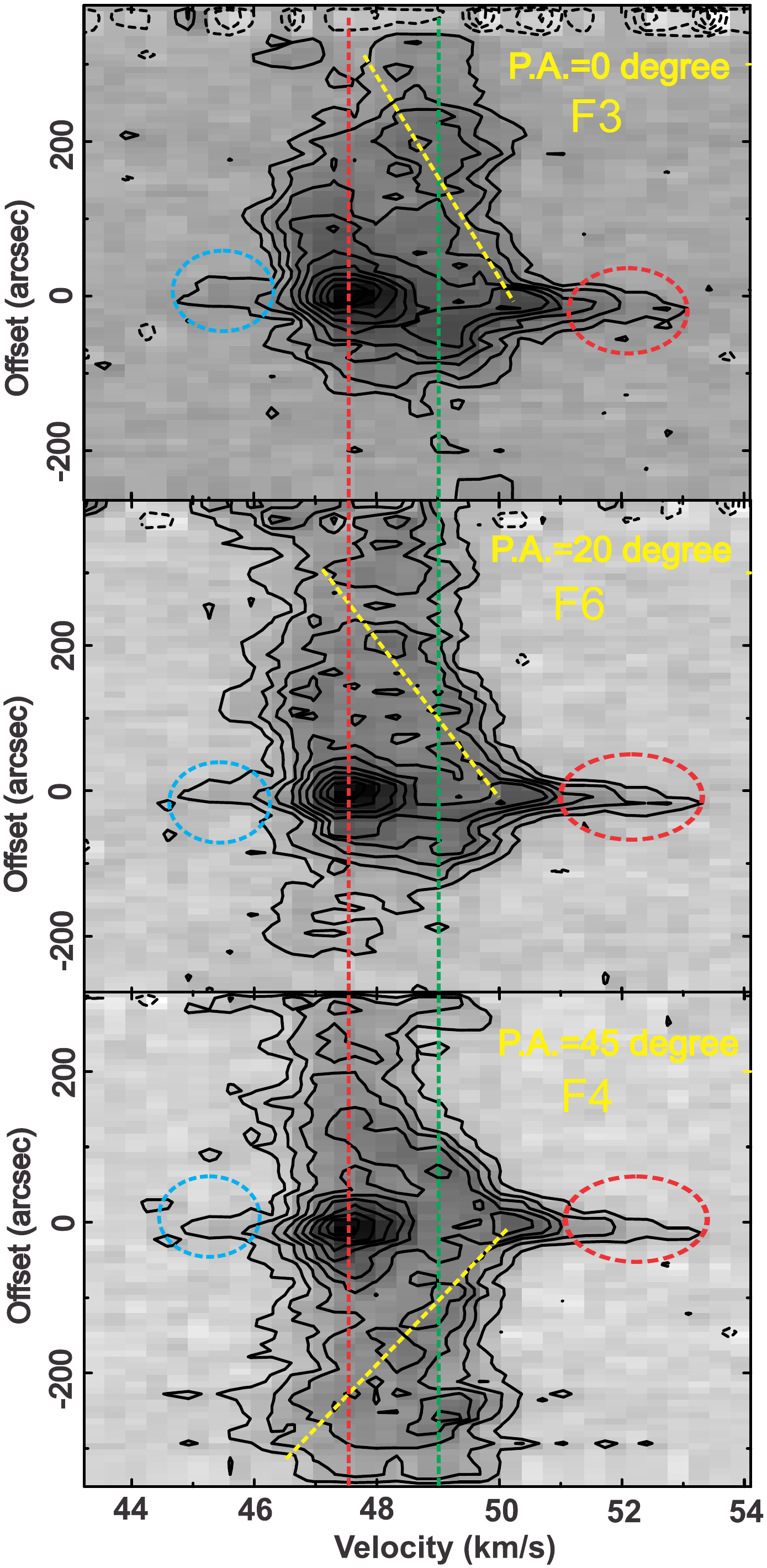}
\caption{Position-Velocity diagrams of $^{13}$CO (2-1) line emission. The contour levels are from 10\% to 90\% in steps of 10\% of the peak value (5.3 K). The green dashed vertical line marks the systemic velocity of 49 km~s$^{-1}$. The blueshifted and redshifted high-velocity emission is marked with blue and red dashed ellipses, respectively. The red vertical dashed line depicts the peak emission. The yellow dashed lines show the velocity gradients along the filaments. \label{PV}   }
\end{figure}

Figure \ref{PV} presents PV diagrams of $^{13}$CO (2-1) line emission along sub-filaments F3, F4, and F6. The peak emission of the clump 6 is significantly blueshifted with respect to the systemic velocity and the main sub-filament. As denoted by yellow dashed lines, the sub-filaments connected to the central clump show a clear velocity gradients with respect to the peak emission of the central clump. The velocity gradients are $\sim$0.4 km~s$^{-1}$~pc$^{-1}$, $\sim$0.6 km~s$^{-1}$~pc$^{-1}$, and 0.5 km~s$^{-1}$~pc$^{-1}$ for ``F3", ``F4" and ``F6", respectively, which are similar to the values found in other Galactic long filaments \citep{wang16}.

One interpretation of velocity gradients is that they are caused by inflows along the sub-filaments as also seen in other filamentary clouds \citep{kirk13,pere13,liu16b,yuan17}. The velocity differences ($\delta V$) between the sub-filaments and central clump become larger as they approach the central clump (i.e., $\delta V\propto \delta R^{-1}$, where $\delta R$ is the distance to the central clump), suggesting that the inflows along the sub-filaments are likely driven by the gravity of the central clump MM6.

We can estimate the mass inflow rate ($\dot{M}_{\parallel}$) along filaments following \cite{kirk13}:
\begin{equation}
\dot{M}_{\parallel} = \frac{\nabla V_{\parallel,obs} M}{tan(\alpha)}
\end{equation}
where $\nabla V_{\parallel,obs}$, M and $\alpha$ are the velocity gradient, mass and inclination angle with respect to the plane of the sky of a filament, respectively. Since most gas is bounded in dense clumps, only a small fraction of free gas may be accreted along the filament. We estimated the mass of the free gas as the difference ($\sim1900$ M$_{\sun}$) between the total filament mass ($\sim6200$ M$_{\sun}$) and the sum of the clump masses ($\sim$4300 M$_\sun$) from the \textit{\textbf{R4}} imaging scheme. We take a mean velocity gradient of $\sim\nabla V_{\parallel,obs}=0.5$ km~s$^{-1}$~pc$^{-1}$ to estimate the net mass inflow rate. Assuming $\alpha=45\arcdeg$, the inflow rate along the filaments to the central clump is $\sim1\times10^{-3}$ M$_{\sun}$~yr$^{-1}$.

\subsection{Chemical properties of filaments and dense clumps}

The CO molecular line data from the ``TOP" and ``SAMPLING" surveys will be used to study the gaseous CO properties (e.g., abundances, CO-to-H$_{2}$ conversion factors, CO depletion) of a large sample of PGCCs through joint analysis with the continuum emission data. The KVN survey is designed to observe dozens of dense gas lines toward dense clumps and cores detected in the ``SCOPE" survey and characterize their chemical properties. Below, we will investigate the CO depletion of the G26 filament and the dense gas abundances in clump 6.

\subsubsection{CO depletion}

Gaseous CO significantly freezes out onto grain surfaces when densities exceed
$\sim3\times10^{4}$ cm$^{-3}$ \citep{bac02}. As CO is a major destroyer of molecular ions, CO depletion leads to a change in the relative abundances of major charge carriers \citep[e.g., H$_{3}^{+}$, N$_{2}$H$^{+}$ and HCO$^{+}$;][]{ber07,cas11}. In addition, the abundance of the nitrogen hydrides and and deuterated molecules are strongly enhanced and these species are probing the gas where CO (and other carbon-bearing species) is depleted \citep{ber07,cas11}. Therefore, studies of CO depletion are very important for understanding the chemical processes in star formation. Through statistical studies of a sample of 674 PGCCs, \cite{liu13} found that the CO abundance is strongly (anti-)correlated to other physical parameters (e.g., dust temperature, dust emissivity spectral index, column density, volume density, and luminosity-to-mass ratio), suggesting that the gaseous CO abundance can be used as an evolutionary tracer for molecular clouds. Similarly, \cite{gian14} also found that less evolved ATLASGAL-selected high-mass clumps seem to show larger values for the CO depletion than their more evolved counterparts, and CO depletion increases for denser sources. However, both studies in \cite{liu13,gian14} used single-pointing molecular CO data, which limited the accurate determination of CO depletion in molecular clouds. The molecular CO mapping survey data in the TOP survey are more suitable for investigating how CO abundances or CO depletion varies inside individual molecular clouds and changes in different kinds of molecular clouds.

\begin{figure}
\centering
\includegraphics[angle=0,scale=0.35]{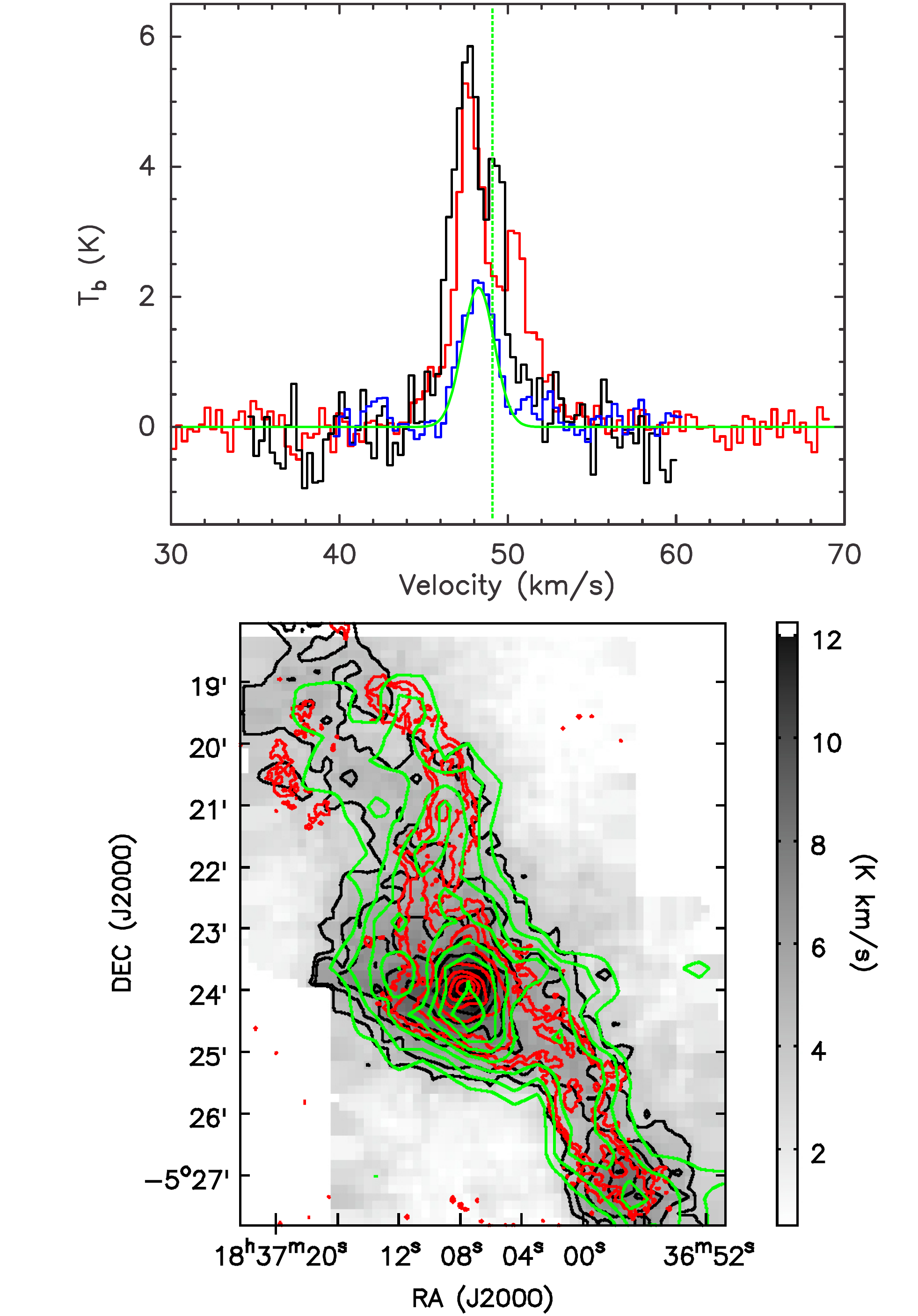}
\caption{Upper panel: The spectra of $^{13}$CO (1-0) and $^{13}$CO (2-1) at the continuum peak are shown in black and red, respectively. The spectrum of C$^{18}$O (1-0) averaged over an region of 1$\arcmin\times1\arcmin$ is shown in blue. The Gaussian fit of C$^{18}$O (1-0) is shown in green. The green dashed vertical line marks the systemic velocity of 49 km~s$^{-1}$. Lower panel: Integrated intensity (47-50 km~s$^{-1}$) map of $^{13}$CO (2-1) is shown in gray scale and black contours. The contour levels are from 30\% to 90\% in steps of 10\% of the peak value (12.9 K~km~s$^{-1}$). Integrated intensity (47-50 km~s$^{-1}$) map of C$^{18}$O (1-0) is shown in green contours. The contour levels are from 30\% to 90\% in steps of 10\% of the peak value (6.4 K~km~s$^{-1}$). The SCUBA-2 850 $\micron$ continuum emission in the \textit{\textbf{R2}} data reduction is shown in red contours. The contour levels are [0.03, 0.05, 0.1, 0.2, 0.4, 0.6, 0.8]$\times$1.40 Jy~beam$^{-1}$.\label{CO-SCUBA2}  }
\end{figure}

\begin{figure}
\centering
\includegraphics[angle=0,scale=0.45]{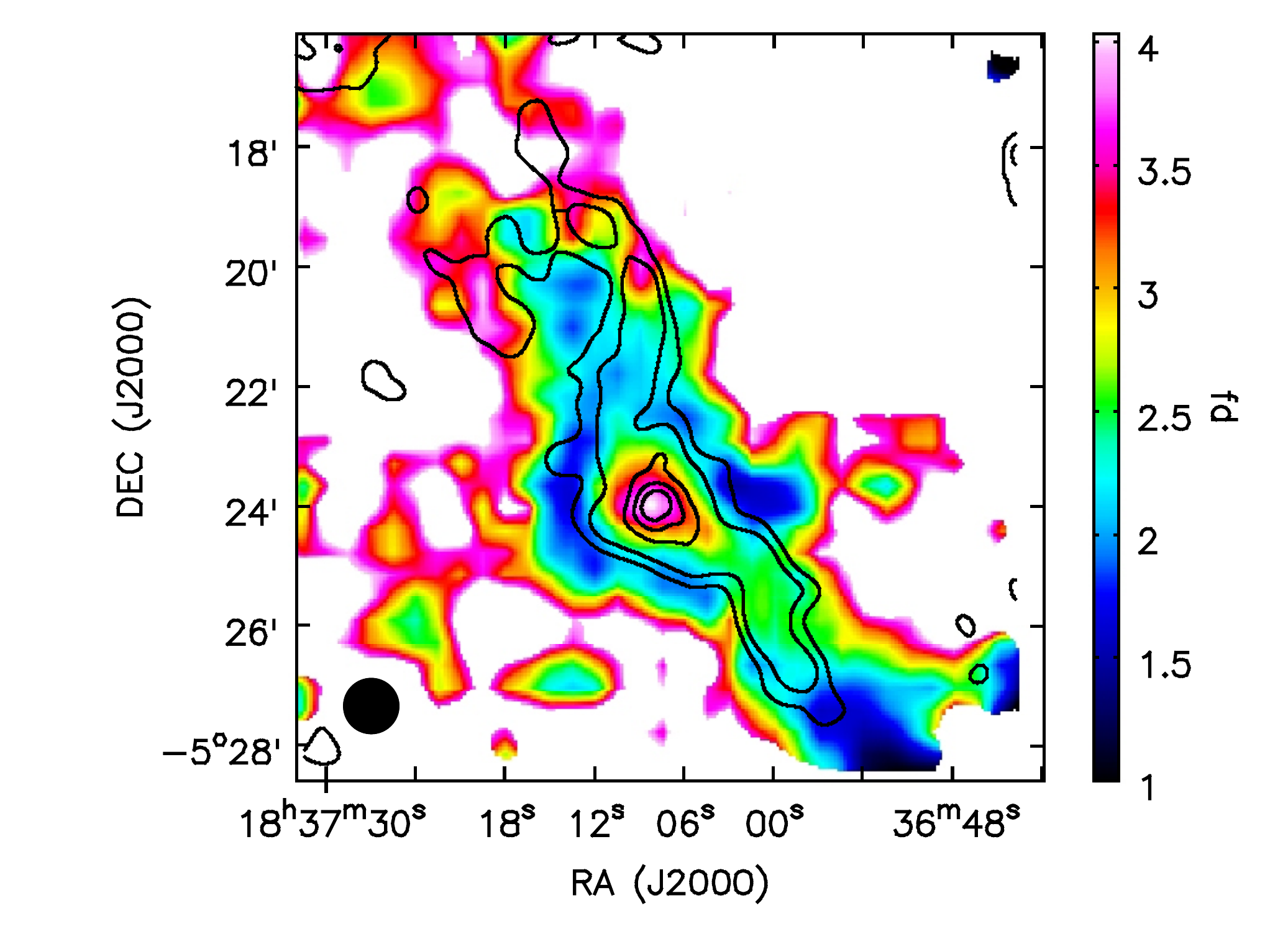}
\caption{The CO gas depletion factor is shown as color image. The column density map from SED fits with Herschel and combined Planck and SCUBA-2 850 $\micron$ data in \textit{\textbf{R2}} is shown in contours. The contour levels are [0.15, 0.2, 0.4, 0.6, 0.8]$\times$2.76$\times10^{23}$ cm$^{-2}$.\label{deple} }
\end{figure}

The upper panel of Figure \ref{CO-SCUBA2} presents the spectra of $^{13}$CO (1-0), $^{13}$CO (2-1) and C$^{18}$O (1-0) at the continuum peak for G26.  The $^{13}$CO (2-1) and C$^{18}$O (1-0) emission have a spatial distribution similar to that of the SCUBA-2 850 $\micron$ continuum as shown in the lower panel of Figure \ref{CO-SCUBA2}.

C$^{18}$O (1-0) emission is optically thin in G26 (see Table \ref{linepara}). We derived C$^{18}$O column densities following \cite{liu13} by assuming that the excitation temperature of C$^{18}$O (1-0) equals to the dust temperature. The C$^{18}$O abundance (X$_{C^{18}O}$) was derived by comparing the C$^{18}$O column densities with the H$_{2}$ column density map from the SED fits with Herschel data and Planck+SCUBA-2 data. The C$^{18}$O column density map and H$_{2}$ column density map were smoothed to the same resolution of $\sim52\arcsec$. The CO gas depletion factor (f$_{D}$) is defined as $f_{D}=\frac{max(X_{C^{18}O})}{X_{C^{18}O}}$, where max(X$_{C^{18}O}$) is the maximum C$^{18}$O abundance ($\sim1.5\times10^{-7}$) across the map with C$^{18}$O emission larger than 3 $\sigma$. The CO gas depletion factor map is presented in Figure \ref{deple}. The highest depletion ($f_{D}\sim$5) occurs at the central massive clump (``6") due to its largest column density and relatively low temperature. However, since MM6 \textbf{hosts} a relatively strong IR source (bright at 24 $\micron$ and 70 $\micron$), the higher CO depletion means that the star formation process inside MM6 is really new and most of the clump gas is still cold. Due to the large beam of TRAO observations, we see most of the bulk, cold gas rather than the small, localized hot region, further suggesting that the IRDC G26 is still very young.

The CO gas in the surroundings of the central clump MM6 is less depleted ($f_{D}\sim$1.5-2.5). The outskirts of the filament also shows higher apparent CO depletion, which reflects the lower CO abundance therein. The low CO abundance in the outskirts suggests that CO gas is released from dust grains due to external heating from interstellar radiation and then photodissociated.

\subsubsection{Molecular lines from KVN 21-m single pointing observations}

\begin{figure*}
\centering
\includegraphics[angle=-90,scale=0.65]{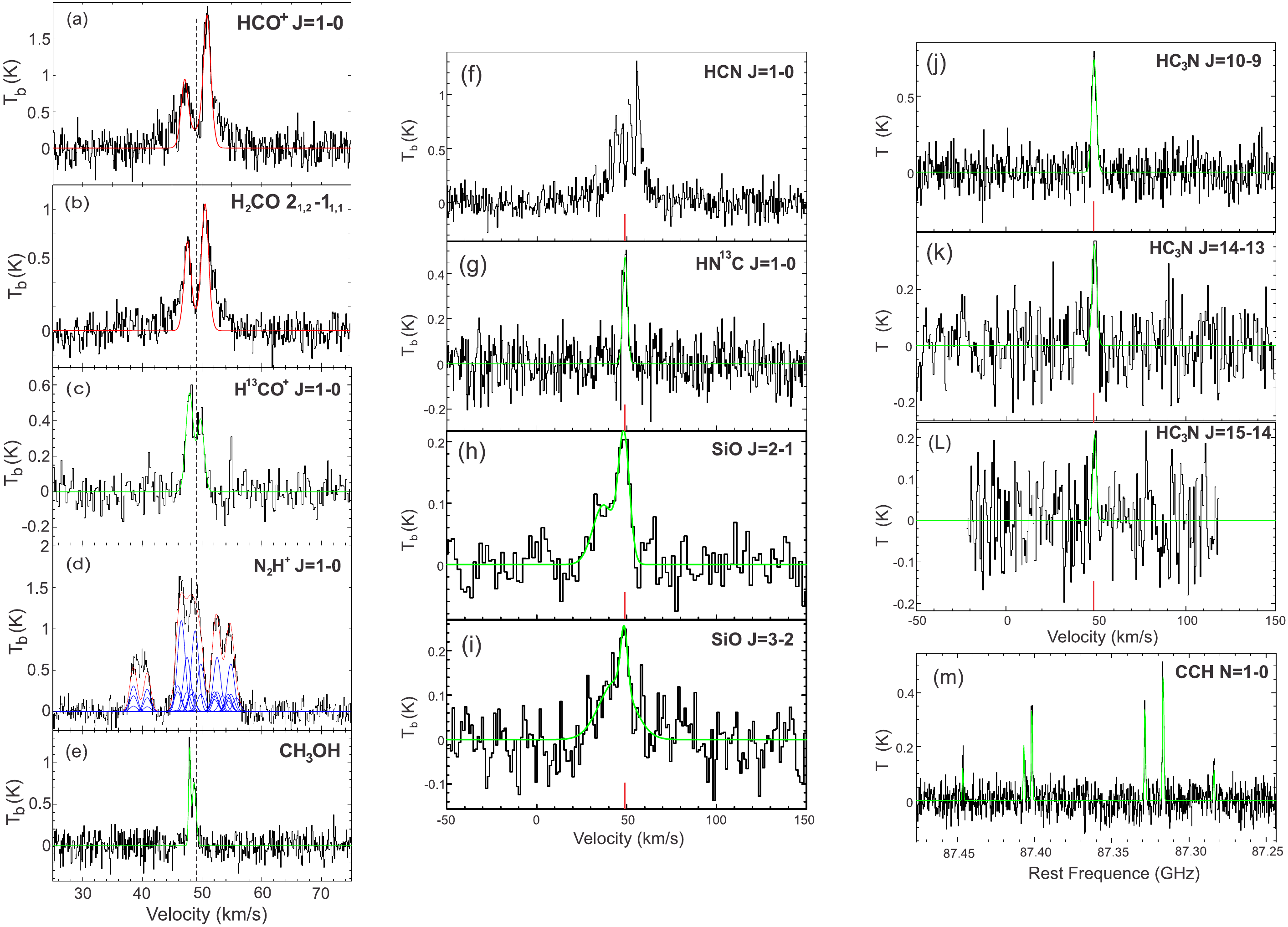}
\caption{The spectra of the central clump from KVN 21-m single-dish observations. The observed spectra are shown in black. The red lines in panels (a) and (b) show the two layer model fits. The blue lines in panel (d) show the hyperfine structure fit toward the N$_{2}$H$^{+}$ (1-0) line spectrum. The red line is the sum of the hyperfine structures. The green line in panel (m) shows the hyperfine structure fit toward the CCH (N=1-0) line. The green lines in other panels (a to L) show the Gaussian fits of the spectra. The line transition name is labeled in the upper-right corner in each panel. The black dashed vertical lines in panels (a to e) and the red vertical lines in panels (f to L) mark the systemic velocity of 49 km~s$^{-1}$.\label{KVN} }
\end{figure*}

\begin{figure}
\centering
\includegraphics[angle=0,scale=0.45]{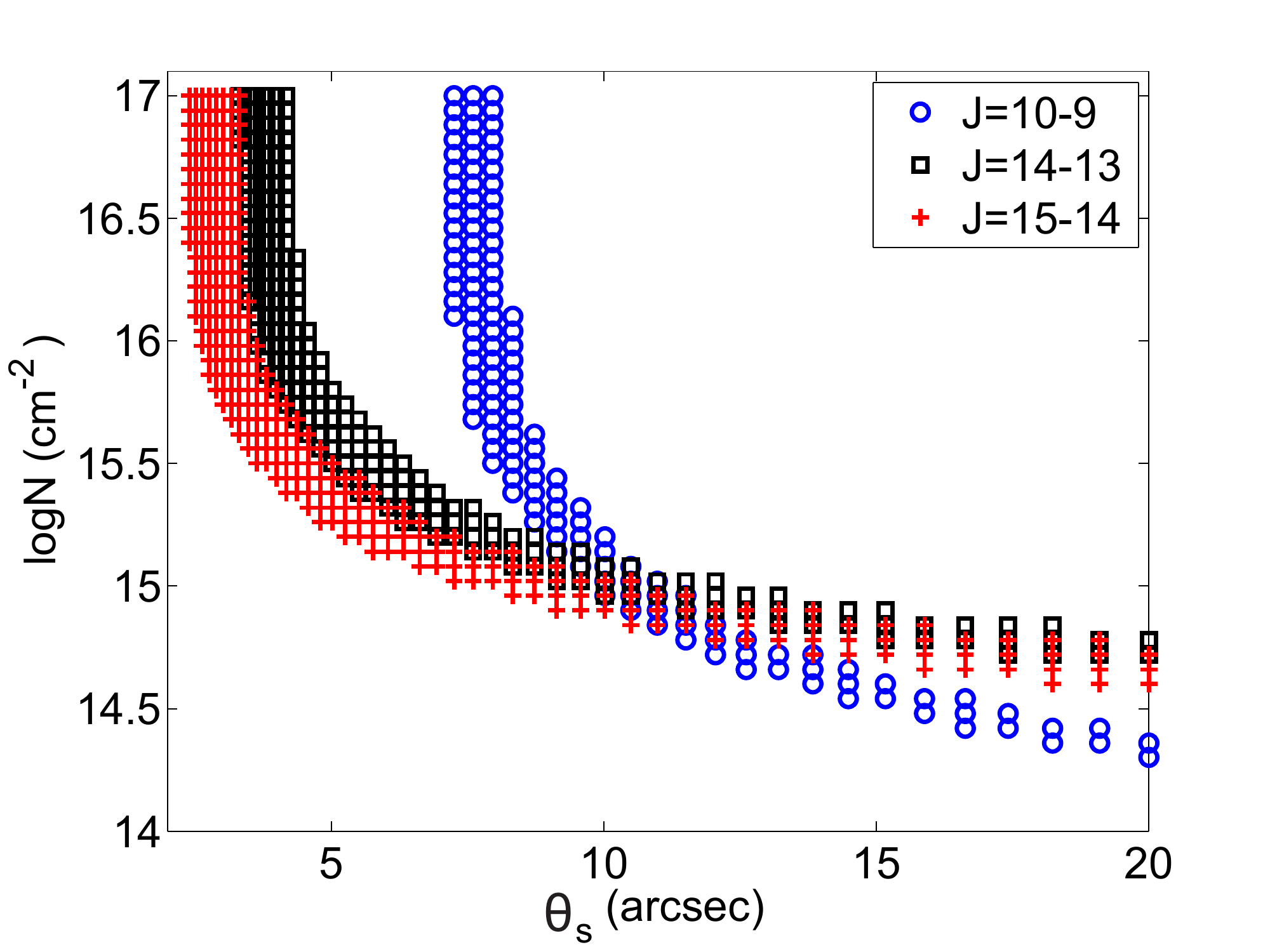}
\caption{Source size vs. column density in RADEX calculations of HC$_{3}$N transitions. The kinetic temperature and H$_{2}$ number density in RADEX calculations are fixed to 17 K and 1.8$\times10^{4}$ cm$^{-3}$. The source size and column density of HC$_{3}$N in best fit are 11$\arcsec$ ($\sim$0.22 pc) and 9.2$\times10^{14}$ cm$^{-2}$. \label{HC3N} }
\end{figure}

\begin{deluxetable*}{ccccccccc}
\centering
\tiny
\tablecolumns{9} \tablewidth{0pc}\setlength{\tabcolsep}{0.05in}
\tablecaption{Derived parameters of molecular lines \label{linepara}}\tablehead{\colhead{Line} & \colhead{$\int T_{b}dv$\tablenotemark{a}} & \colhead{V$_{LSR}$\tablenotemark{a}} & \colhead{FWHM\tablenotemark{a}} & \colhead{T$_{mb}$} & \colhead{T$_{ex}$\tablenotemark{b}} &  \colhead{$\tau$\tablenotemark{b}} &  \colhead{N\tablenotemark{b}} &  \colhead{X\tablenotemark{c}}\\
\colhead{}  & \colhead{(K~km~s$^{-1}$)} & \colhead{(km~s$^{-1}$)} & \colhead{(km~s$^{-1}$)} & \colhead{(K)} & \colhead{(K)} & \colhead{} & \colhead{(cm$^{-2}$)}}
\startdata
C$^{18}$O J=1-0                  & 5.05(0.25) & 48.23(0.05)    & 2.22(0.12)  & 2.14  & 18.2 & 0.1 & 4.6E15 & 1.0E-07\\
CH$_{3}$OH $7_0-6_1A+$           & 0.48(0.06)   & 47.90(0.02)    & 0.39(0.04)  & 1.16 \\
                                 & 0.64(0.07)   & 48.66(0.04)    & 0.79(0.11)  & 0.76 \\
HCO$^{+}$ J=1-0                  & 3.34(0.16)   & 50.95(0.04)  &   1.85(0.11) &1.70 \\
                                 & 2.51(0.20)   & 47.06(0.11)  &   3.15(0.36) &0.75 \\
H$^{13}$CO$^{+}$ J=1-0           & 0.59(0.11)   & 49.78(0.12)  &   1.37(0.28) &0.40 & 3.9  & 0.5  & 1.9E13&4.2E-10 \\
                                 & 0.86(0.12)   & 47.91(0.09)  &   1.46(0.23) &0.55 & 4.0  & 0.6  & 3.6E13&8.0E-10 \\
HC$_{3}$N J=10-9                 & 2.43(0.18)   & 49.03(0.11)  &   3.07(0.27) &0.75 & 5.4 &  0.4  & 9.2E14&2.0E-08 \\
HC$_{3}$N J=14-13                & 1.18(0.19)   & 49.12(0.25)  &   3.07(0.55) &0.36 & 6.0  & 0.1  &         &          \\
HC$_{3}$N J=15-14                & 0.52(0.13)   & 49.29(0.31)  &   2.37(0.55) &0.21 & 6.6  & 0.1  &         &          \\
SiO J=2-1                        & 1.57(0.40)   & 48.54(0.61)  &   7.28(1.23) &0.20 & 4.8  & 4.0  & 1.3E13  &2.9E-10 \\
                                 & 1.44(0.44)   & 37.16(2.29)  &  14.05(4.13) &0.10  \\
SiO J=3-2                        & 3.12(0.52)   & 44.34(1.75)  &  22.01(2.79) &0.13  \\
                                 & 0.63(0.24)   & 48.66(0.49)  &   4.44(1.45) &0.14 & 3.2  & 0.5  & 8.6E13  &1.9E-09 \\
HN$^{13}$C J=1-0                 & 1.56(0.14)   & 49.06(0.14)  &   3.08(0.30) &0.48 & 3.5  & 1.0  & 2.5E14  & 5.6E-09 \\
H$_{2}$CO $2_{1,2}-1_{1,1}$      & 2.35(0.10)   & 50.81(0.04)  &   2.38(0.13) &0.93 \\
                                 & 1.32(0.09)   & 47.42(0.06)  &   1.87(0.17) &0.66 \\
\cline{1-9}
                                 & T$_{ex}$\tablenotemark{d}     & V$_{LSR}$\tablenotemark{d}    &   FWHM\tablenotemark{d}       & $\tau$ & T$_{ex}$ & $\tau$ & \colhead{$N$}\\
\cline{1-9}
CCH N=1-0                        & 4.4(0.3)     & 49.00(0.08)  &   3.02(0.20) &0.43(0.13)  &  4.5      & 0.4       & 1.5E14 & 3.3E-09 \\
N$_{2}$H$^{+}$ J=1-0             & 5.58(0.59)   & 47.72(0.03)  &   0.58(0.03) &1.88(0.54)  &  4.1      & 2.2       & 2.4E13 & 5.3E-10 \\
                                 & 5.53(0.79)   & 50.05(0.03)  &   0.57(0.04) &1.66(0.62)  &  4.0      & 2.0       & 2.0E13 & 4.4E-10 \\
\enddata
\tablenotetext{a}{From Gaussian fits with Gildas/Class}
\tablenotetext{b}{The values for H$^{13}$CO$^{+}$, HC$_{3}$N and HN$^{13}$C are derived by assuming a source size of 11 arcsec. The others are derived by assuming a filling factor of 1.}
\tablenotetext{c}{N$_{H_{2}}=4.5\times10^{22}$ cm$^{-2}$}
\tablenotetext{d}{From hyperfine structure fits with Gildas/Class}
\end{deluxetable*}

We observed the massive clump ``6" in molecular lines that trace dense gas with the KVN 21-m single dishes to characterize its chemical properties. Figure \ref{KVN} presents the spectra from KVN 21-m single pointing observations. HCO$^{+}$ J=1-0 and H$_{2}$CO $2_{1,2}-1_{1,1}$ lines show ``red asymmetry profiles" morphology with line wings, indicating that their emission is affected by outflows. H$^{13}$CO$^{+}$ J=1-0 and N$_{2}$H$^{+}$ J=1-0 spectra show two velocity components. Since the H$^{13}$CO$^{+}$ J=1-0 and the isolated hyperfine components of N$_{2}$H$^{+}$ J=1-0 lines are usually optically thin in IRDCs \citep{sanh12}, their two velocity components should not be the result of self-absorption. Indeed, the two velocity components may indicate the existence of converging flows inside the clump as resolved in other high-mass clumps \citep[e.g., SDC335, G10.6-0.4, G33.92+0.11,  AFGL 5142;][]{pere13,liu13b,liu15,liu16b} or interacting sub-filaments \citep[e.g., G028.23-00.19;][]{sanh13}.

We did not detect the 22 GHz water maser but detected the 44 GHz methanol maser. The 44 GHz methanol maser has two velocity components peaked at $\sim$47.9 km~s$^{-1}$ and $\sim$48.7 km~s$^{-1}$, respectively. The 44 GHz methanol maser is blueshifted with respect to the systemic velocity ($\sim$49 km~s$^{-1}$) and has much smaller line widths than other lines that trace denser gas.

SiO emission shows broad lines and two velocity components. The blueshifted component in SiO emission has much broader line widths than the central component, which may be caused by outflow shocks. CCH N=1-0 and N$_{2}$H$^{+}$ (1-0) each have hyperfine structures. Their hyperfine lines were fitted assuming LTE to obtain their line widths and optical depths. HCN (1-0) shows very complicated line profile, which is likely caused by its hyperfine structures and self-absorption. We did not fit the HCN (1-0) spectrum. The other lines were fitted with a Gaussian function. The line parameters are summarized in Table \ref{linepara}. The systemic velocity ($\sim$49 km~s$^{-1}$) is obtained from averaging the peak velocities of the optically thin lines (C$^{18}$O, HC$_{3}$N and HN$^{13}$C). The HDCO 2$_{0,2}-1_{0,1}$ line was not detected at an rms level of 0.06 K.

We estimated the column densities from molecular lines with the RADEX\footnote{RADEX is a one-dimensional non-LTE radiative transfer code, that uses the escape probability formulation assuming an isothermal and homogeneous medium without large-scale velocity fields \citep{van07}.} radiation transfer code by fixing the kinetic temperature to 17 K, i.e., the dust temperature. For C$^{18}$O, C$_{2}$H, and N$_{2}$H$^{+}$, the H$_{2}$ volume density of clump 6 is set to 6$\times10^{3}$ cm$^{-3}$ from the \textit{\textbf{R4}} imaging scheme. For the other lines (H$^{13}$CO$^{+}$, HN$^{13}$C and HC$_{3}$N), the H$_{2}$ volume density of clump 6 is set as 1.8$\times10^{4}$ cm$^{-3}$ from the \textit{\textbf{R2}} imaging scheme since their effective excitation densities are about one order of magnitude higher than that of N$_{2}$H$^{+}$ \citep{shir15}.

As shown in Figure \ref{HC3N}, with three transitions, the source size ($\theta_{S}$) of the HC$_{3}$N emitting area can be well determined, i.e., 11$\arcsec$. We applied the same filling factor ($f=\frac{\theta_{S}^2}{\theta_{S}^2+\theta_{beam}^2}$) for the H$^{13}$CO$^{+}$ and HN$^{13}$C calculations. We assume a filling factor of 1 in the RADEX calculations for the other lines. The RADEX-derived excitation temperature, optical depth, column density, and abundance from each transition are listed in the last four columns of Table \ref{linepara}.

As a major destroyer of N$_{2}$H$^{+}$, the absence of CO in the gas phase could enhance the abundance of N$_{2}$H$^{+}$. In warm regions, however, CO evaporation
could destroy N$_{2}$H$^{+}$ and enhance the abundance of HCO$^{+}$ by the following reaction \citep{lee04,bus11}:
\begin{equation}\label{eq_lacc}
N_{2}H^{+}+CO \rightarrow HCO^{+}+N_{2}
\end{equation}
Therefore, a small N$_{2}$H$^{+}$ to H$^{13}$CO$^{+}$ abundance ratio may indicate that clumps are warm and thus chemically evolved. \cite{Feng16} recently observed a sample of four
70 $\micron$ dark IRDC clumps, whose CO seems to be heavily depleted with depletion factors in the range of 14-50, i.e., about 3-10 times larger than that of G26 clump 6.
They also derived high [N$_{2}$H$^{+}$]/[H$^{13}$CO$^{+}$] abundance ratios ($\sim$400-10000) in the 70 $\micron$ dark young IRDC clumps, which are more than two orders of magnitudes larger than that of clump 6 in G26 ($\sim$0.8). Therefore, clump 6 in G26 is more chemically evolved than those IRDC clumps. In addition, the abundance of HC$_{3}$N, a hot core tracer, is about two orders of magnitude larger in clump 6 than in the 70 $\micron$ dark IRDC clumps, indicating that G26 clump 6 may be in a more chemically evolved phase. Indeed, protostars have already formed in G26 clump 6.

\subsection{Dynamical properties of dense clumps}

The CO molecular line data from the ``TOP" and ``SAMPLING" mapping surveys can be used to study large scale kinematics (e.g., collisions, filament accretion, outflows). The HCO$^{+}$ J=1-0 and H$_{2}$CO $2_{1,2}-1_{1,1}$ lines in the KVN survey are also good tracers of kinematics like infall and outflows associated with dense clumps/cores. Below, we study the kinematics of G26 based on the $^{13}$CO (2-1) line emission from the ``SAMPLING" survey and the HCO$^{+}$ J=1-0 and H$_{2}$CO $2_{1,2}-1_{1,1}$ lines from KVN observations.

\subsubsection{High-velocity Outflows}

\begin{figure}
\centering
\includegraphics[angle=0,scale=0.45]{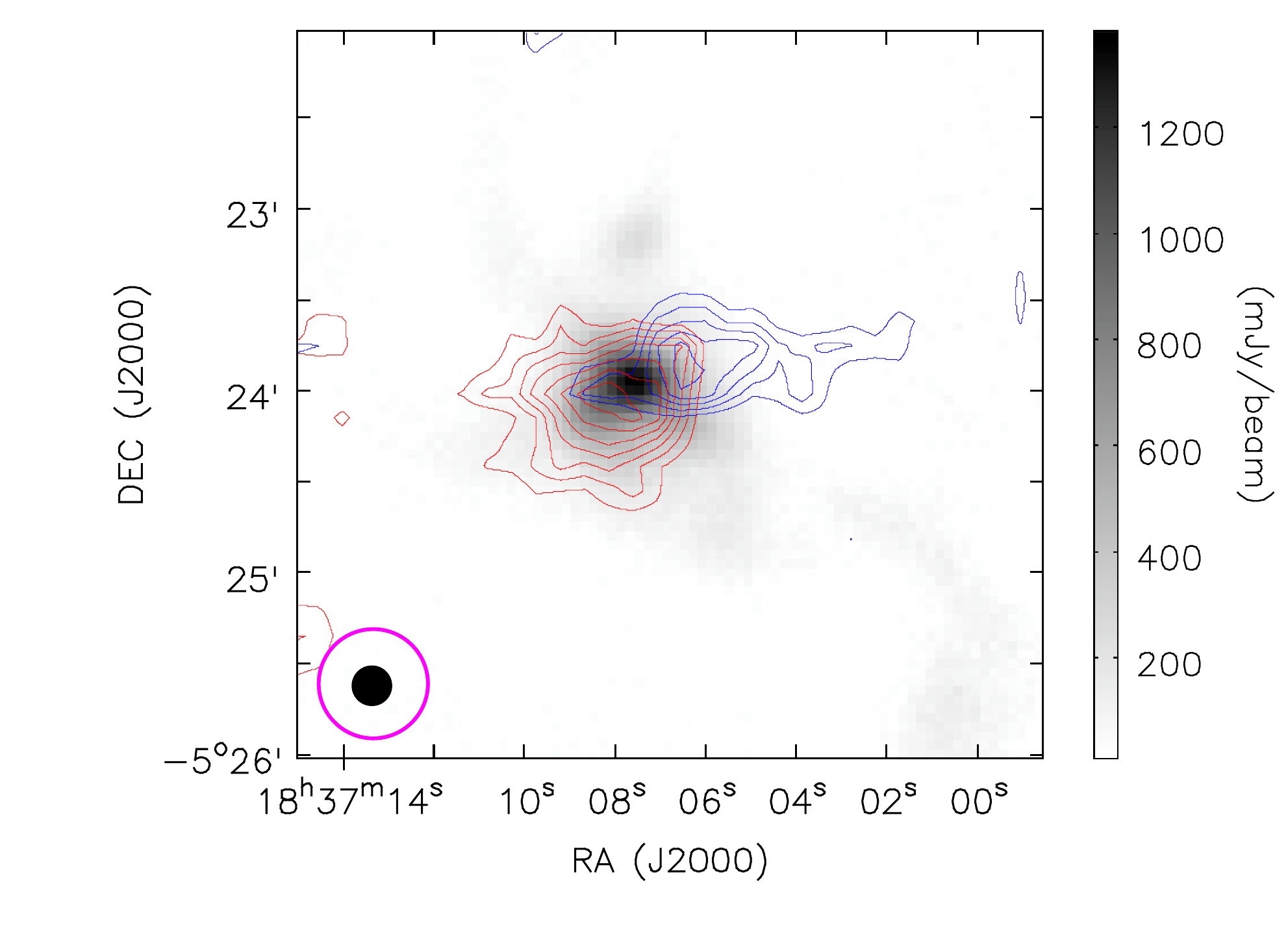}
\caption{The blueshifted (43-46.5 km~s$^{-1}$) and redshifted (51-55 km~s$^{-1}$) $^{13}$CO (2-1) outflow emission in clump 6 are shown in blue and red contours, respectively. The contour levels of blueshifted outflow emission are [0.5, 0.6, 0.7, 0.8, 0.9]$\times$3.06 K~km~s$^{-1}$. The contour levels of redshifted outflow emission are [0.3, 0.4, 0.5, 0.6, 0.7, 0.8, 0.9]$\times$5.65 K~km~s$^{-1}$. The SCUBA-2 850 $\micron$ continuum emission is shown in gray scale image.\label{outflow} }
\end{figure}

Outflow wings can be easily identified in the PV diagrams in Figure \ref{PV}. Figure \ref{outflow} shows the integrated intensity maps of the outflow emission. Since the blue and red outflow lobes are well separated and have very high flow velocities, neither the very low (pole-on) inclination nor the very high (edge-on) inclination is likely. Therefore, we take 45$\arcdeg$ as the inclination angle for the outflows with respect to the line of sight. The radii of the blueshifted and redshifted lobes are $\sim$1.6 pc and $\sim$1.1 pc, respectively. The characteristic outflow velocities of the blueshifted and redshifted lobes are $\sim$5.1 km~s$^{-1}$ and $\sim$3.3 km~s$^{-1}$, respectively. Assuming the excitation temperature of $^{13}$CO (2-1) line wing emission is 17 K, i.e., the dust temperature, the integrated intensity can be converted to the outflow masses following \cite{gard91}. The $^{13}$CO abundance (7$\times10^{-7}$) is converted from C$^{18}$O abundance (1$\times10^{-7}$) with a [$^{13}$CO/C$^{18}$O] abundance ratio ($\sim$7) derived from empirical relations between molecular abundances and the Galactocentric distance \citep{wils94,gian14}. The masses of the blueshifted and redshifted lobes are $\sim$67 M$_{\sun}$ and $\sim$97 M$_{\sun}$, respectively. The momentum of the blueshifted and redshifted lobes is $\sim$342 M$_{\sun}$~km~s$^{-1}$ and $\sim$314 M$_{\sun}$~km~s$^{-1}$, respectively. The energy of the blueshifted and redshifted lobes are $\sim$1.7$\times10^{46}$ and $\sim$1.1$\times10^{46}$ erg, respectively. The dynamical time scales ($t_{dyn}=\frac{2R}{V_{flow}}$) of the blueshifted and redshifted lobes are 6.2$\times10^5$ yr and 6.8$\times10^5$ yr, respectively. The total outflow mass loss rate is $\sim2.5\times10^{-4}$ M$_{\sun}$~yr$^{-1}$.

\subsubsection{Outflow driven expansion}

\begin{deluxetable*}{ccccccccc}
\centering
\tiny
\tablecolumns{9} \tablewidth{0pc}\setlength{\tabcolsep}{0.05in}
\tablecaption{Parameters of the two layer model \label{modelpara}}\tablehead{\colhead{Line} & \colhead{$\tau_0$} & \colhead{$\Phi$} & \colhead{J$_{c}$} & \colhead{J$_{f}$} & \colhead{J$_{r}$} &  \colhead{V$_{cont}$} &  \colhead{$\sigma$} &  \colhead{V$_{out}$}\\
\colhead{}  & \colhead{} & \colhead{}  & \colhead{(K)} & \colhead{(K)} & \colhead{(K)} & \colhead{(km~s$^{-1}$)} & \colhead{(km~s$^{-1}$)} & \colhead{(km~s$^{-1}$)}}
\startdata
HCO$^{+}$ J=1-0   & 3.56 & 0.74 & 15.70 & 11.69 & 19.60 & 48.98 & 1.04 & 0.15\\
H$_{2}$CO $2_{1,2}-1_{1,1}$ & 2.21 & 0.79 & 15.44 & 12.03 & 19.54 & 49.07 & 0.93 & 0.06\\
\enddata
\end{deluxetable*}

The HCO$^{+}$ J=1-0 and H$_{2}$CO $2_{1,2}-1_{1,1}$ lines show ``red asymmetric profiles" morphologies, suggesting expansion. We apply a simple ``two layer model" to fit the lines. The model is described in APPENDIX B. The results are presented in Table \ref{modelpara}. The expansion velocities inferred from the HCO$^{+}$ J=1-0 and H$_{2}$CO $2_{1,2}-1_{1,1}$ lines are $\sim$0.15 km~s$^{-1}$ and $\sim$0.06 km~s$^{-1}$, respectively. Assuming the whole clump (with M$\sim$2100 M$_{\sun}$) is in expanding with a velocity of 0.1 km~s$^{-1}$, the total momentum and energy of the expanding clump are 210 M$_{\sun}$~km~s$^{-1}$ and $\sim$2.1$\times10^{44}$ erg, respectively. Since it is unlikely that the whole clump is in expansion, these momentum and energy values should be treated as upper limits. Even in this extreme case, however, both the momentum and energy are smaller than those of the low-velocity outflows, indicating that the outflows have enough energy themselves to drive the envelope expansion.

The virial mass of the clump MM6 considering turbulent support can be derived as $\frac{M_{vir}}{M_{\sun}}=210(\frac{R}{pc})(\frac{\Delta V}{km~s^{-1}})^2$ \citep{mac88,zhang15}, where $R$ is the radius (1.15 pc) and $\Delta V$ is the linewidth (2.22 km~s$^{-1}$) of C$^{18}$O (1-0). The virial mass is $\sim$1200 M$_{\sun}$, which is much smaller than the clump mass ($\sim$2100 M$_{sun}$), indicating that the whole clump may be still in collapse. Therefore, the expansion revealed by HCO$^{+}$ J=1-0 and H$_{2}$CO $2_{1,2}-1_{1,1}$ lines should be very localized.

\begin{figure}
\centering
\includegraphics[angle=0,scale=0.35]{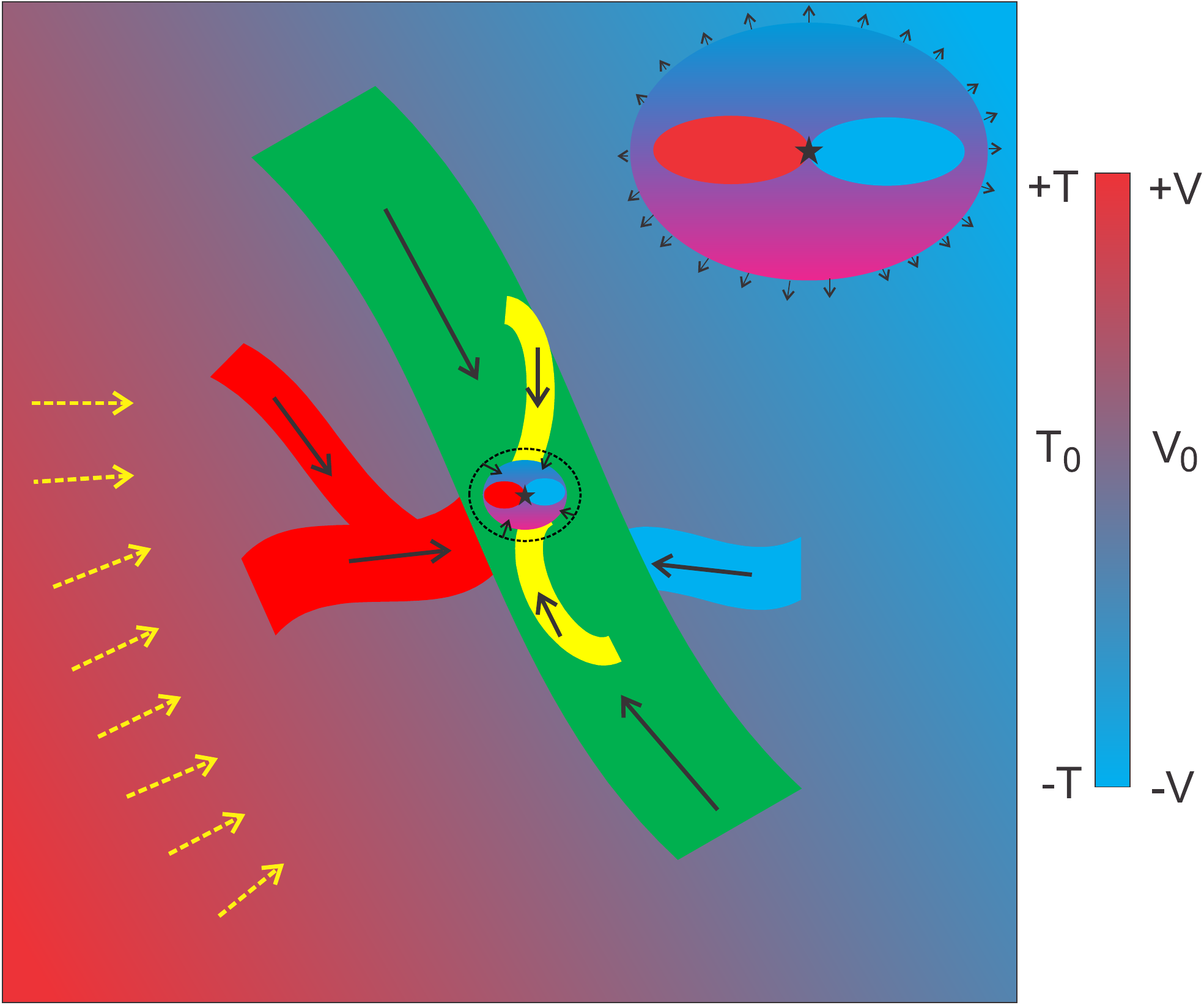}
\caption{This cartoon shows the environment of PGCC G26.53+0.17. The main gas filament is shown in green. The SCUBA-2 filament is shown in yellow. Other small sub-filaments are shown in red, green, and blue. PGCC G26.53+0.17 is located in a parent cloud with velocity and temperature gradients along the east-west direction. The central clump MM6 (dashed ellipse) may be still in collapse while its inner part (color filled ellipse) is undergoing expansion due to outflow feedback from the protostars. The zoomed-in picture of the central clump is shown in the upper-right corner. The red and blue ellipses represent the outflow lobes. The arrows show the directions of gas flows. \label{model} }
\end{figure}

Figure \ref{model} shows a cartoon to describe the PGCC G26 based on the results from the present data. Our findings of G26 are: (i) The G26 filament may be formed due to large-scale compression flows evidenced by the temperature and velocity gradients across its natal cloud; (ii) Sub-filaments in G26 have different velocities and may interact with each other; (iii) The central massive clump where massive stars are forming may still accrete gas along sub-filaments; (iv) The massive clump in whole may be still in global collapse while its inner part seems to be undergoing expansion due to outflow feedback.

\section{Summary}

We introduce here the survey strategy, observations, data reduction and example science for a joint survey program ``TOP-SCOPE", targeting 1000-2000 Planck Galactic Cold Clumps at TRAO 14-m and JCMT 15-m telescopes. ``TOP," standing for ``TRAO Observations of Planck cold clumps," is a $^{12}$CO/$^{13}$CO J=1-0 line survey of 2000 PGCCs with the Taeduk Radio Astronomy Observatory 14-meter telescope. ``SCOPE," standing for ``SCUBA-2 Continuum Observations of Pre-protostellar Evolution," is a legacy survey using SCUBA-2 on the James Clerk Maxwell Telescope (JCMT) at the East Asia Observatory (EAO) to survey 1000 PGCCs at 850 $\micron$ continuum. Follow-up observations toward SCUBA-2 detected dense clumps/cores with other ground-based telescopes (NRO 45-m, SMT, KVN, SMA, ALMA) are on-going.

We introduced the example science of the ``TOP-SCOPE" survey and other joint survey programs with an exemplar source, PGCC
G26.53+0.17 (i.e., G26). The main findings for G26 are:

(1). The total mass, the length and the mean line-mass (M/L) of the G26 filament revealed are $\sim$6200 M$_{\sun}$, $\sim$12 and $\sim$500 M$_{\sun}$~pc$^{-1}$, respectively. Both the length and line-mass of G26 are comparable to those of the integral shaped filament in the Orion A cloud \citep{kai17}.

(2). Ten clumps are found along the filament with a mean separation of $\sim$1.3 pc. Only two of them (MM6, MM10) are associated with protostars. The others seem to be starless. The clumps in G26 especially the dense and massive ones are very likely bounded by gravity and have the ability to form high-mass stars. The typical spacing and mass of clumps in G26 are comparable to the predictions in gravitational fragmentation of an isothermal, non-magnetized, and turbulent supported cylinder, suggesting that the fragmentation in G26 from cloud (at $\sim$10 pc scale) to clump (at $\sim$1 pc scale) is very likely dominated by turbulence.

(3). Evidence for grain growth along the filament is found. The histograms of dust emissivity spectral index $\beta$ can be well fitted with two normal distributions regardless of whether or not 850 $\micron$ data were included in the SED fits. Such a bimodal behavior in $\beta$ distribution suggests grain growth along the filament.

(4). The N-PDF and two-point correlation functions (2PT) of the natal cloud harboring G26 filament are investigated. The slope (-3.9) of the power-law tail is comparable to those of IRDCs G28.34+0.06 (-3.9) and G14.225-0.506 (-4.1). The 2PT
function shows a smooth decay of correlation strengths over all spatial scales up to $\sim$15 pc. The flat 2PT correlation indicates that the column density distribution of G26 is very homogenous over all spatial scales and the mass is less concentrated overall than in high-mass protoclusters.

(5). From comparing the SCUBA-2 images with Planck+SCUBA-2 combined images, we found that only 30\%-40\% of the cloud gas in G26 is dense gas.

(6). The G26 filament may be formed due to large-scale compression flows evidenced by the temperature and velocity gradients across its natal cloud. Sub-filaments in G26 have different velocities and may interact with each other. The central massive clump where massive stars are forming may still accrete gas along sub-filaments with a natal mass inflow rate of $\sim1\times10^{-3}$ M$_{\sun}$~yr$^{-1}$.

(7). The mass of the most massive clump MM6 is much larger than its virial mass, suggesting that it may be still in global collapse. Its inner part traced by HCO$^{+}$ J=1-0 and H$_{2}$CO $2_{1,2}-1_{1,1}$ lines, however, seems to be undergoing expansion due to outflow feedback. The total mass and total outflow mass loss rate of the outflows are $\sim$164 M$_{\sun}$ and $\sim2.5\times10^{-4}$ M$_{\sun}$~yr$^{-1}$, respectively.

(8). The most massive clump MM6 has much smaller CO gas depletion factor, much smaller [N$_{2}$H$^{+}$]/[H$^{13}$CO$^{+}$] abundance ratio, and much larger HC$_{3}$N abundance than 70 $\micron$ dark young IRDC clumps, indicating that clump 6 containing protostars is more chemically evolved and is in a more chemically evolved phase.

All the example science discussed for G26 will be fully explored through more statistical studies with the ``TOP-SCOPE" data and other follow-up survey data toward larger samples.

\renewcommand{\thefigure}{A\arabic{figure}}

\setcounter{figure}{0}

\renewcommand{\theequation}{A\arabic{equation}}

\setcounter{equation}{0}

\renewcommand{\thetable}{A\arabic{table}}

\setcounter{table}{0}

\section{APPENDIX A\\
Observations and Data reduction for PGCC G26.53+0.17}

The JCMT/SCUBA-2, TRAO 14-m, SMT 10-m and KVN 21-m observations of PGCC G26.53+0.17 are summarized in Table \ref{obsG26}.

\begin{deluxetable*}{cccccccc}
\centering
\scriptsize
\tablecolumns{8} \tablewidth{0pc} \setlength{\tabcolsep}{0.04in}
\tablecaption{Observations of PGCC G26.53+0.17 \label{obsG26}}\tablehead{
\multicolumn{8}{c}{JCMT/SCUBA-2 850 $\micron$ continuum observations}\\
\cline{1-8}
\colhead{Date}  &    \colhead{Mode}     & \colhead{$\tau_{225}$} & & \colhead{$\theta_{B} (\arcsec)$} & \colhead{FCF} & \colhead{rms (mJy~beam$^{-1}$)} & \colhead{Tracer}}
\startdata
\cline{1-8}
2016-04-21     &    CV Daisy & 0.057  & & 14.1  & 543.3 & 8.3  & dense clumps/filaments\\
2017-05-13     &    Pong1800 & 0.090  & & 14.1  & 513.2 & 18.7 & dense clumps/filaments\\
2017-05-14     &    Pong1800 & 0.115  & & 14.1  & 556.7 & 17.5 & dense clumps/filaments\\
2017-05-16     &    Pong1800 & 0.071  & & 14.1  & 547.0 & 25.7 & dense clumps/filaments\\
2017-05-23     &    Pong1800 & 0.122  & & 14.1  & 561.6 & 30.7 & dense clumps/filaments\\
2017-05        &    Pong1800 & ---    & & 14.1  &  & 10.7 & dense clumps/filaments\\
\cline{1-8}
\multicolumn{8}{c}{TRAO 14-m OTF observations on 2017-01-09\tablenotemark{a}}\\
\cline{1-8}
 Line           & Frequency (GHz)  & T$_{sys}$ (K)  & $\Delta$v (km~s$^{-1}$) & $\theta_{B} (\arcsec)$ & B$_{eff}$ & rms (K) & Tracer\\
\cline{1-8}
 $^{12}$CO (1-0) & 115.27120  & 551            & 0.32       &  45   &0.51    & 1.1 & large scale structure and kinematics\\
 $^{13}$CO (1-0) & 110.20135  & 244            & 0.33       &  47   &0.54    & 0.4 & large scale structure and kinematics\\
 C$^{18}$O (1-0) & 109.78217  & 233            & 0.33       &  47   &0.54    & 0.2 & dense clumps/filament\\
\cline{1-8}
\multicolumn{8}{c}{SMT 10-m OTF observations\tablenotemark{a}}\\
\cline{1-8}
 Line            & Frequency (GHz) & T$_{sys}$ (K)  & $\Delta$v (km~s$^{-1}$) & $\theta_{B} (\arcsec)$ & B$_{eff}$ & rms (K) & Tracer\\
\cline{1-8}
 $^{13}$CO (2-1) & 220.39868  &    250         & 0.34      &36     &  0.70      &0.2 & large scale structure and kinematics \\
\cline{1-8}
\multicolumn{8}{c}{KVN 21-m single pointing observations on 2017-05-27\tablenotemark{a}}\\
\cline{1-8}
 Line           & Frequency (GHz)  & T$_{sys}$ (K)  & $\Delta$v (km~s$^{-1}$) & $\theta_{B} (\arcsec)$ & B$_{eff}$ & rms (K) & Tracer\\
 \cline{1-8}
H$_{2}$O $6_{1,6}-5_{2,3}$\tablenotemark{b}      &   22.23508      & 110 & 0.42  &126 &0.46 & 0.08   & maser \& shock tracer \\
CH$_{3}$OH $7_0-6_1A+$          &   44.06941      & 117 & 0.11  &63  &0.48 & 0.12   & maser \& shock tracer\\
HCN J=1-0                         & 	88.63185      & 287 & 0.42  &32  &0.41 & 0.09   & outflow \& infall\\
HCO$^{+}$ J=1-0                   & 	89.18852      & 203 & 0.21  &32  &0.41 & 0.10   & outflow \& infall \\
H$^{13}$CO$^{+}$ J=1-0            &  	86.75429      & 200 & 0.22  &32  &0.41 & 0.08   & dense gas, V$_{lsr}$ \& $\sigma$ \\
HC$_{3}$N J=10-9                  &   90.97902      & 308 & 0.41  &32  &0.41 & 0.10   & hot core  \\
HC$_{3}$N J=14-13                 &   127.36767     & 496 & 0.59  &23  &0.39 & 0.09   & hot core  \\
HC$_{3}$N J=15-14                 &   136.46441     &437  & 0.55  &23  &0.39 & 0.07   & hot core  \\
SiO J=2-1                         &  	86.84696      & 240 & 1.73  &32  &0.41 & 0.03   & shock tracer  \\
SiO J=3-2                         & 	130.26861     & 345 & 1.15  &23  &0.39 & 0.04   & shock tracer  \\
HN$^{13}$C J=1-0                  &   87.09085      & 240 & 0.86  &32  &0.41 & 0.05   & dense gas, V$_{lsr}$ \& $\sigma$ \\
CCH N=1-0                         &   87.31692      & 206 & 0.43  &32  &0.41 & 0.07   & photodissociation tracer \\
N$_{2}$H$^{+}$ J=1-0              &   93.17340      & 236 & 0.10  &32  &0.41 & 0.09   & dense gas, V$_{lsr}$ \& $\sigma$  \\
H$_{2}$CO $2_{1,2}-1_{1,1}$       & 	140.83950     & 282 & 0.13  &23  &0.39 & 0.09   & outflow \& infall  \\
HDCO 2$_{0,2}-1_{0,1}$\tablenotemark{b}            &   128.81286     & 477 & 1.15  &23  &0.39 & 0.06   & deuterium fraction  \\
\enddata
\tablenotetext{a}{For molecular line observations, the columns are line name, line frequency, system temperature, velocity resolution, beam size, main beam efficiency, rms level in brightness temperature
and tracers. The spectra are usually slightly smoothed to get a better S/N level. The velocity resolutions are thus different from original ones. The rms levels were estimated from smoothed spectra.}
\tablenotetext{b}{These lines were not detected toward PGCC G26.53+0.17.}
\end{deluxetable*}

\subsection{Herschel archival data}

G26 was mapped with Herschel as part of the Hi-GAL project
\citep{Molinari2010}. The observations include 70 $\micron$, 160 $\micron$, 250 $\micron$, 350 $\micron$, and
500 $\micron$ bands that were observed in parallel mode with 60$\arcsec$/s
scanning speed.  We use the level 2.5 maps available in the Herschel
Science Archive\footnote{http://archives.esac.esa.int/hsa}, SPIRE data
(250-500\,$\mu$m) using extended source calibration and the PACS maps
(70, 160\,$\mu$m) made with the UNIMAP method \citep{Piazzo2015}. The
resolutions of the original maps are approximately 12$\arcsec$,
15$\arcsec$,  18.3$\arcsec$, 24.9$\arcsec$,  and 36.3$\arcsec$, respectively. Given the fast scanning speed, the effective PACS beams are
elongated.

\begin{figure}
\centering
\includegraphics[angle=0,scale=0.45]{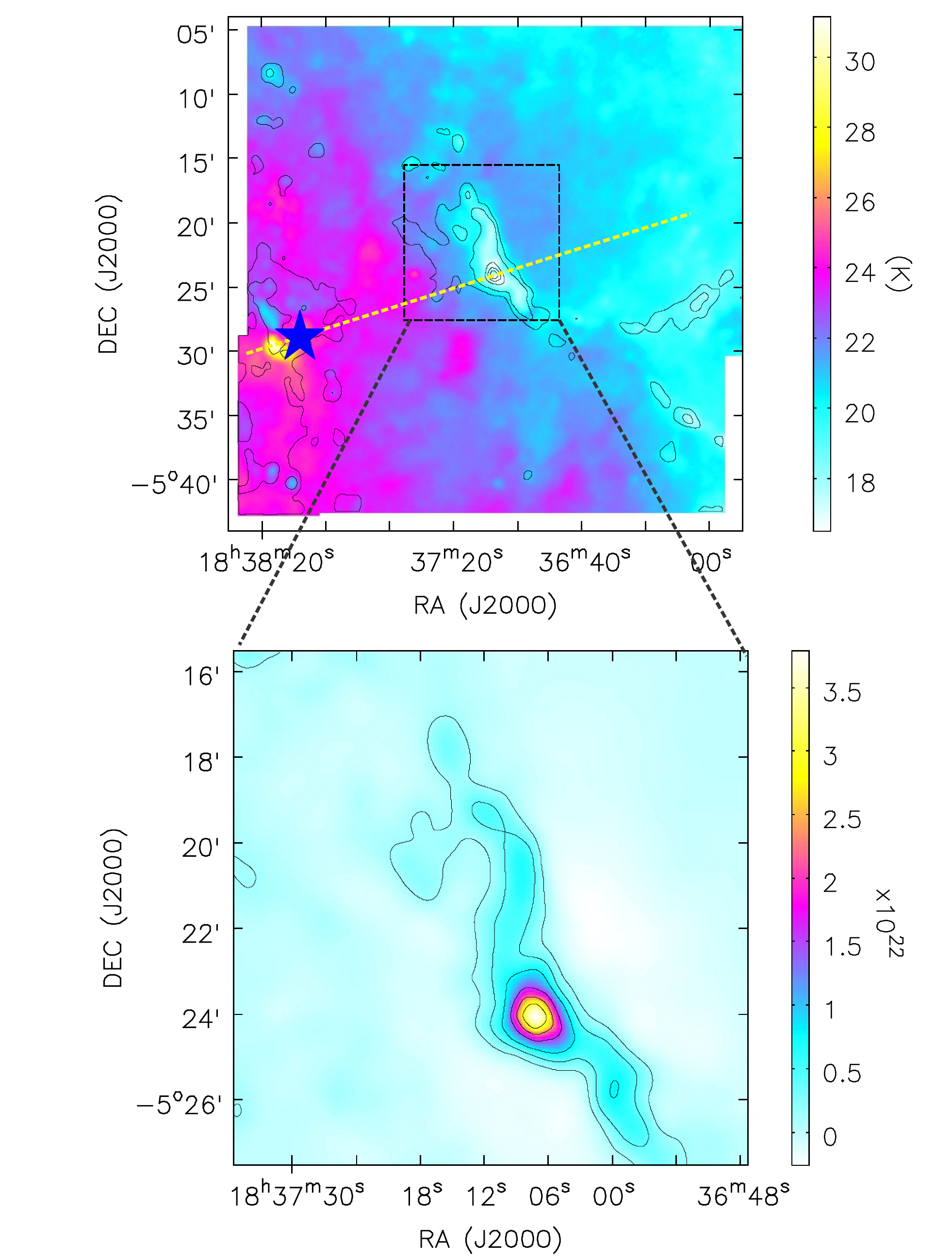}
\caption{Upper panel: From Herschel data, column density contours overlaid on a dust temperature color image. The contour levels are [0.15, 0.2, 0.4, 0.6, 0.8]$\times4.8\times10^{22}$ cm$^{-2}$. The blue star marks the position of the infrared bubble shown in Figure \ref{HII}. The yellow dashed line marks the direction along which the cuts in Figure \ref{compress} were made. Lower panel: Close-up of column density map from Herschel data over the SCUBA-2 mapped region. The large scale emission in Herschel bands was filtered out. The contour levels are [0.03, 0.1, 0.2, 0.4, 0.6, 0.8]$\times3.8\times10^{22}$ cm$^{-2}$.\label{Herschel}}
\end{figure}

Figure \ref{Herschel} presents the dust temperature and column density maps of PGCC G26.53+0.17 derived from Herschel/SPIRE data. The column density maps in Figure \ref{Herschel} are used as external masks in the below JCMT/SCUBA-2 data reduction. We fit the SEDs pixel-by-pixel with a modified blackbody function using the three Herschel/SPIRE bands. The modified blackbody function is:
\begin{equation}
F(\nu)=F(\nu_0)\frac{B(\nu,T)}{B(\nu_0,T)}(\frac{\nu}{\nu_0})^{\beta}=\kappa(\nu)B(T)M/d^2,
\end{equation}
where $F(\nu)$ is the flux density at frequency $\nu$, $\nu_0$ is a selected reference frequency, $B$
stands for the Planck law, $\beta$ is the opacity spectra index, $M$ is the clump mass and $d$ is the
cloud distance. The Herschel data were fitted with modified blackbody spectra
with $\beta=1.8$ to estimate colour correction factors and to derive
estimates of the dust optical depth at a common spatial resolution of
40$\arcsec$. The values were also converted to estimates of hydrogen
column density assuming a dust opacity of
$\kappa$=0.1($\nu$/1000 GHz)$^\beta$ cm$^2$/g \citep{Beckwith1990}.

The upper panel in Figure \ref{Herschel} presents the dust temperature and column density maps derived from Herschel/SPIRE data without applying any local spatial filtering. A temperature gradient along the SE-NW direction is seen in the dust temperature map. The lower panel in Figure \ref{Herschel} presents the column density map of the G26 filament derived from Herschel/SPIRE data with the extended emission larger than 300$\arcsec$ filtered out. The last closed contours in the column density maps outline the external masks used in the JCMT/SCUBA-2 data reduction in section 6.2.

\subsection{JCMT/SCUBA-2 data reduction}

The CV Daisy observations of G26 were conducted on 21 April 2016. To test how well CV Daisy observations can recover faint and extended emission, we also conducted Pong1800 maps from 13 May to 23 May 2017 for comparison. In contrast to the CV Daisy observations, the Pong1800 maps, which are designed for mapping larger fields, cover a circular area of $\sim$30$\arcmin$ diameter with a uniform sensitivity and thus better to recover extended emission.

The JCMT data reduction methods are similar to those used in the JCMT Gould Belt Survey and JCMT transient survey \citep[Herczeg et al. submitted]{ward07,mairs15,mairs17}. The data reduction procedure was performed using the iterative map-making technique makemap \citep{chap13} in the SMURF package of the Starlink software \citep{curr14,jenn13}. The user can also supply an external mask which surrounds the astronomical signal to constrain the solution derived by makemap, which together with a larger spatial filter, can better recover faint and extended structure. Generally, the largest recoverable scales are $\sim600\arcsec$ before atmospheric signal becomes significant. The details of JCMT data reduction can be found in \cite{mairs15,mairs17}. In this paper, we performed four individual data reductions labeled \textit{\textbf{R1, R2, R3, and R4}} with different external masks and spatial filters for comparison. The parameters of makemap used in the data reductions are the same as in \cite{mairs15}.

\begin{figure}
\centering
\includegraphics[angle=0,scale=0.45]{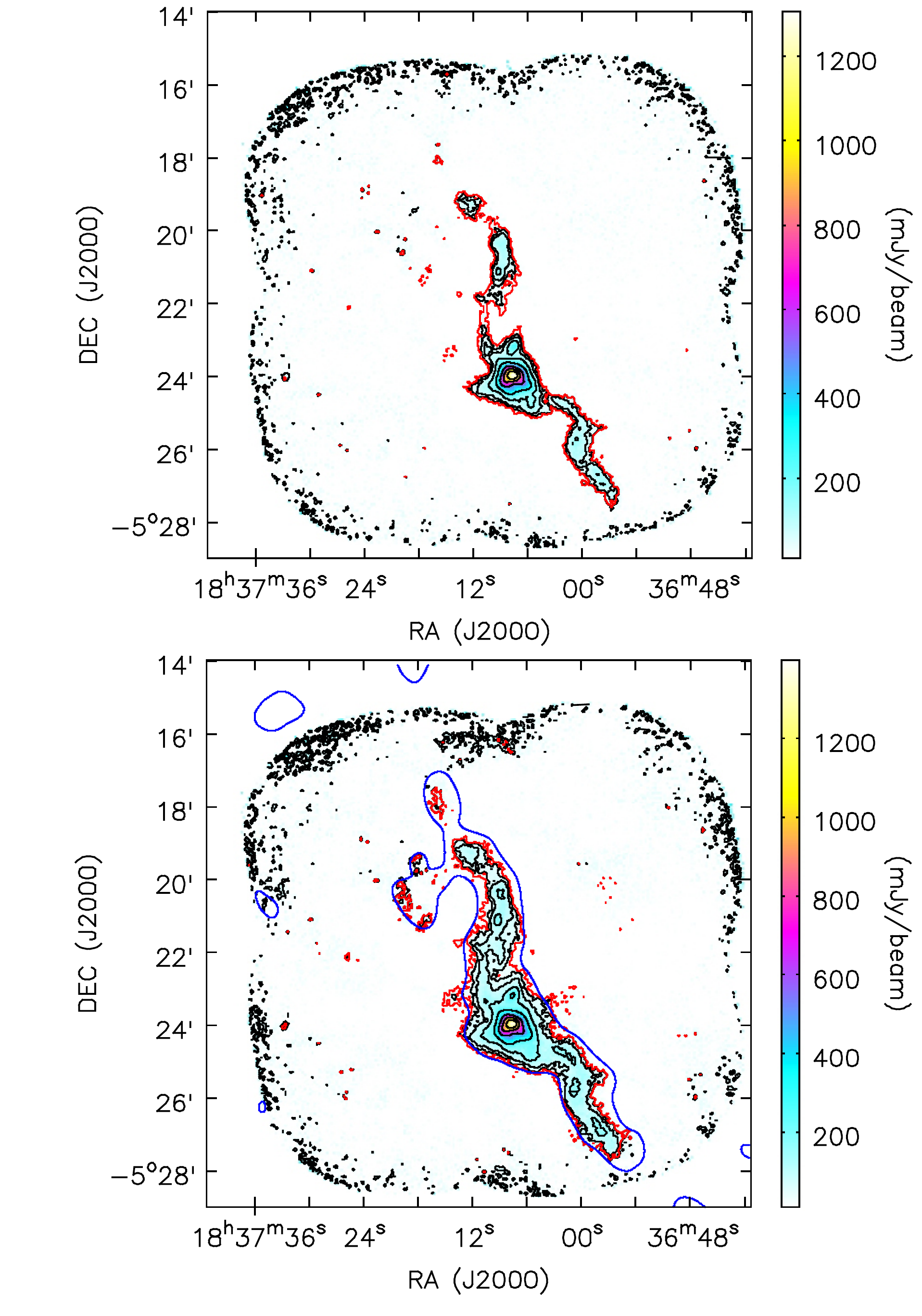}
\caption{Upper panel: The SCUBA-2 850 $\micron$ continuum emission in the \textit{\textbf{R1}} data reduction for G26. The contour levels are [0.03, 0.05, 0.1, 0.2, 0.4, 0.6, 0.8]$\times$1.35 Jy~beam$^{-1}$. Lower panel: The SCUBA-2 850 $\micron$ continuum emission in the \textit{\textbf{R2}} data reduction. The contour levels are [0.03, 0.05, 0.1, 0.2, 0.4, 0.6, 0.8]$\times$1.40 Jy~beam$^{-1}$. The red contour in both panels shows the S/N level of 3.\label{SCUBA-2} }
\end{figure}

\textit{\textbf{(1) R1:}} We first run makemap with an effective spatial filter of 200$\arcsec$ and no external mask. Then an external mask was constructed from the final map with S/N$>$3. The external mask was used to constrain the solution derived by makemap in a second run. Therefore, the structure was filtered to 200$\arcsec$ in the \textit{\textbf{R1}} data reduction. This data reduction is efficient to detect dense clumps and is used in the first data release of the ``SCOPE" project. The upper panel of Figure \ref{SCUBA-2} shows the SCUBA-2 850 $\micron$ continuum emission in the \textit{\textbf{R1}} data reduction.

\textit{\textbf{(2) R2:}} An external mask from the Herschel data (the last closed contour in the lower panel of Figure \ref{Herschel}) was applied in the data reduction. The external mask corresponds to the contour of 3\% of the peak column density value, which is the largest astronomical structure (see the blue contour in the lower panel of Figure \ref{SCUBA-2}) that could be recovered in the CV Daisy map. Besides the external mask, an effective spatial filter of 300$\arcsec$ was applied in the data reduction with makemap. The lower panel of Figure \ref{SCUBA-2} shows the SCUBA-2 850 $\micron$ continuum emission in the \textit{\textbf{R2}} data reduction. Compared with \textit{\textbf{R1}}, \textit{\textbf{R2}} better reveals not only dense clumps but also extended structures between them. The peak flux density (1.40 Jy~beam$^{-1}$) in \textit{\textbf{R2}} is slightly larger than the value (1.35 Jy~beam$^{-1}$) in \textit{\textbf{R1}}.

\begin{figure}
\centering
\includegraphics[angle=0,scale=0.6]{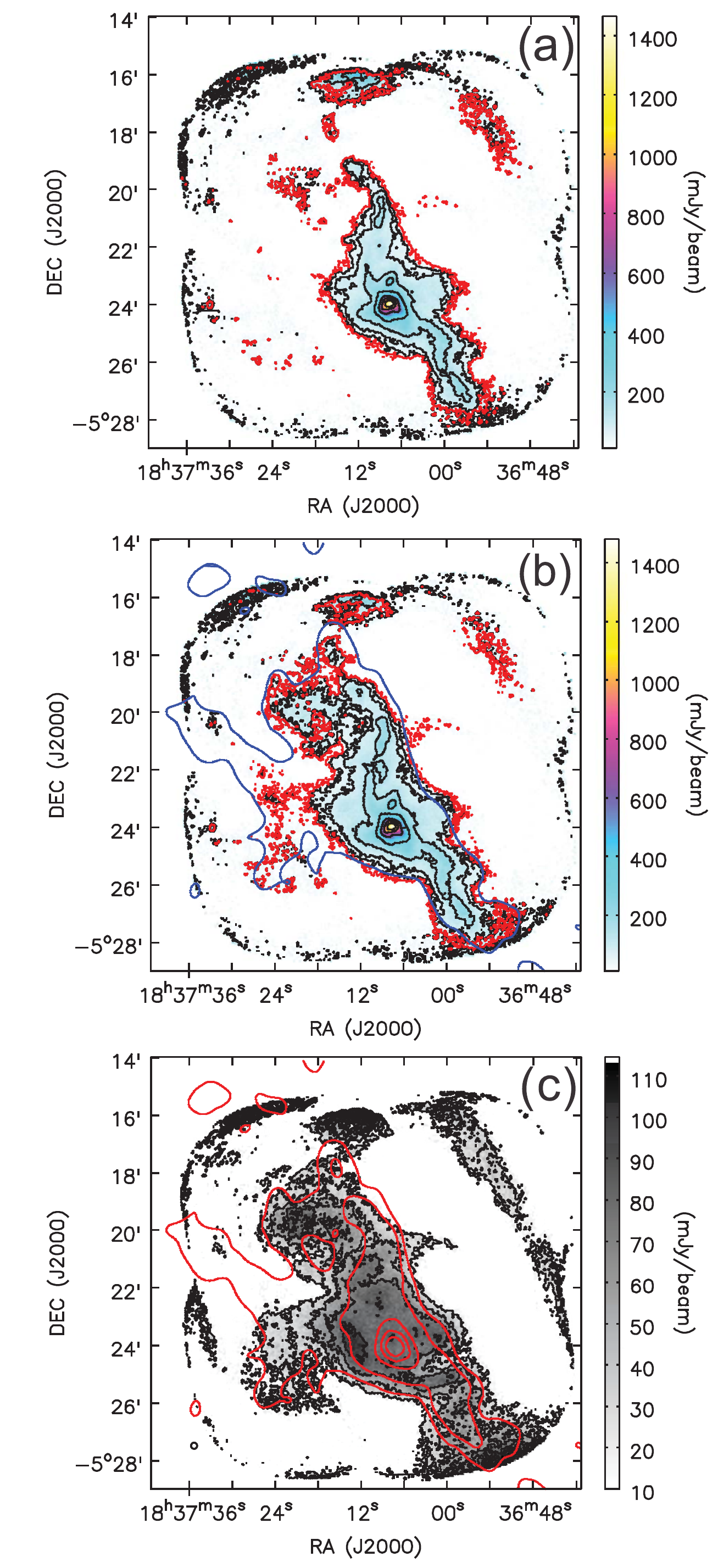}
\caption{(a). The SCUBA-2 850 $\micron$ continuum emission in the \textit{\textbf{R3}} data reduction is shown in color-scale image and black contours. The contour levels are [0.03, 0.05, 0.1, 0.2, 0.4, 0.6, 0.8]$\times$1.47 Jy~beam$^{-1}$. The red contour shows the S/N level of 3. (b). The SCUBA-2 850 $\micron$ continuum emission in the \textit{\textbf{R4}} data reduction is shown in color-scale image and black contours. The contour levels are [0.03, 0.05, 0.1, 0.2, 0.4, 0.6, 0.8]$\times$1.48 Jy~beam$^{-1}$. The red contour shows the S/N level of 3. The blue contour outlines the external mask from the Herschel column density map. (c). The extended SCUBA-2 850 $\micron$ continuum emission (\textit{\textbf{R4}}-\textit{\textbf{R2}}) is shown in gray-scale image and black contours. The contour levels are [0.2, 0.4, 0.6, 0.8]$\times$0.11 Jy~beam$^{-1}$. The red contours show the column density derived from Herschel/SPIRE data. The contour levels are [0.15, 0.2, 0.4, 0.6, 0.8]$\times4.8\times10^{22}$ cm$^{-2}$.\label{SCUBA-Lab} }
\end{figure}

\textit{\textbf{(3).R3:}} Same as \textit{\textbf{R1}} but with a larger effective spatial filter of 600$\arcsec$. Figure \ref{SCUBA-Lab}a shows the SCUBA-2 850 $\micron$ continuum emission in the \textit{\textbf{R3}} data reduction. In contrast to \textit{\textbf{R1}} and \textit{\textbf{R2}}, \textit{\textbf{R3}} recovers more extended emission.

\textit{\textbf{(4).R4:}} Same as \textit{\textbf{R2}} but with a larger effective spatial filter of 600$\arcsec$. The external mask was constructed from the Herschel data without any spatial filtering (the last closed contour in the upper panel of Figure \ref{Herschel}). The external mask corresponds to the contour of 15\% of the peak column density value, which is the largest astronomical structure (see the blue contour in Figure \ref{SCUBA-Lab}b) could be recovered in the CV Daisy map. Figure \ref{SCUBA-Lab}b shows the SCUBA-2 850 $\micron$ continuum emission in the \textit{\textbf{R4}} data reduction. Only the central coherent structure in the masked region is reliable. In general, \textit{\textbf{R4}} and \textit{\textbf{R3}} reveal similar structure. With the external mask from Herschel, makemap, however, recovers more extended and faint structures in \textit{\textbf{R4}} than in \textit{\textbf{R3}}. Figure \ref{SCUBA-Lab}c reveals the extended emission in the difference image between \textit{\textbf{R4}} and \textit{\textbf{R2}}. \textit{\textbf{R4}} recovers a more flattened structure with extended emission ($\sim$ 50 mJy~beam$^{-1}$) than \textit{\textbf{R2}}.

In general, maps with external masks recover more flux. Therefore, we suggest to apply external masks in SCUBA-2 data reduction if Herschel data are available. Although \textit{\textbf{R1}} was used in the first data release of the SCOPE survey, \textit{\textbf{R1}} cannot recover the flux of the dense clumps as well as the other data reductions (\textit{\textbf{R2}}, \textit{\textbf{R3}} and \textit{\textbf{R4}}). \textit{\textbf{R3}} can recover most flux but detect the least number of dense clumps. Therefore, in this paper (except in section 4.2), we mainly use the data reductions \textit{\textbf{R2}} and \textit{\textbf{R4}} in the further analysis. \textit{\textbf{R2}} is very efficient to reveal dense structures (e.g., dense clumps and filaments). In contrast, \textit{\textbf{R4}} recovers more extended structures surrounding the dense clumps.

\subsection{Comparing Daisy map with Pong1800 map}

\begin{figure}
\centering
\includegraphics[angle=0,scale=0.6]{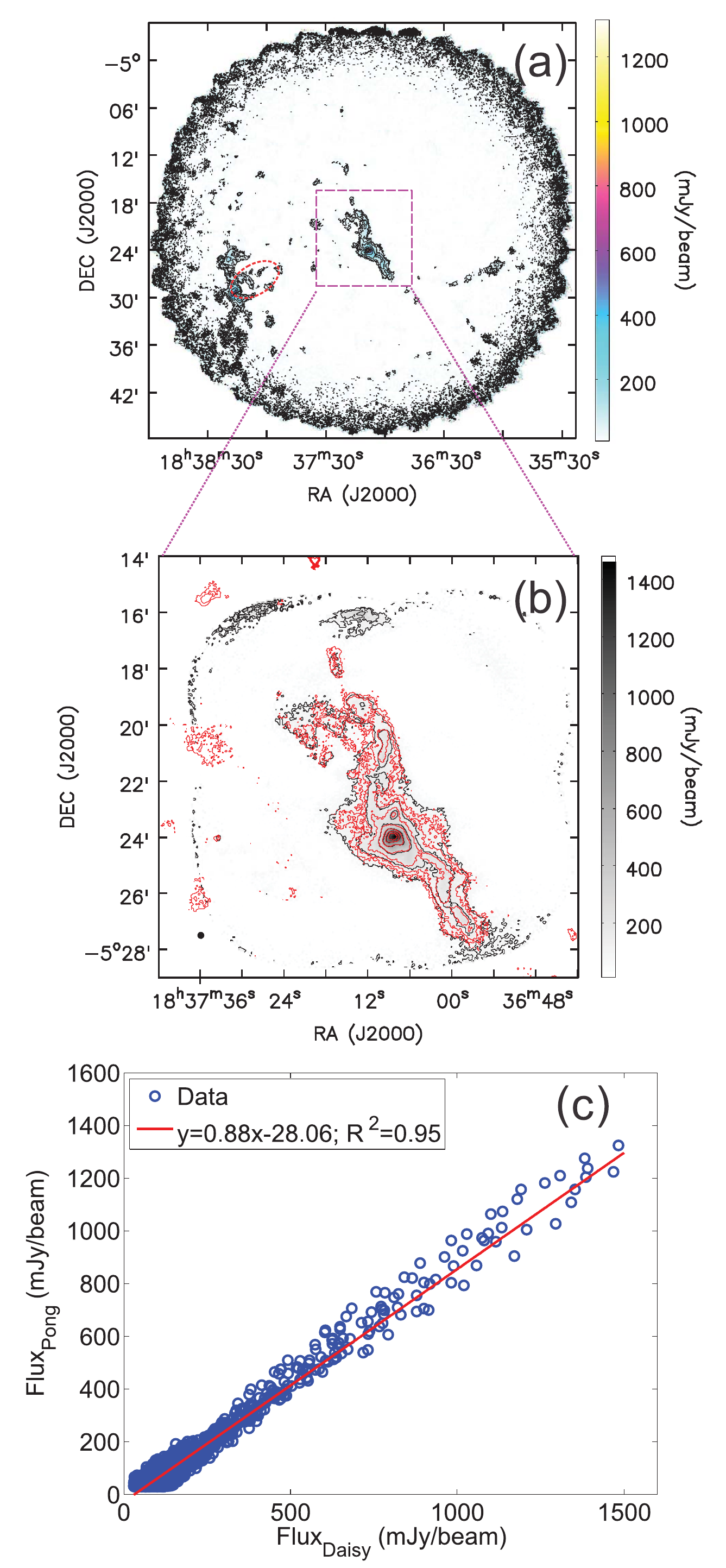}
\caption{(a). The Pong1800 SCUBA-2 850 $\micron$ map. The contours are [0.03, 0.05, 0.1, 0.2, 0.4, 0.6, 0.8]$\times$1.32 Jy~beam$^{-1}$. The red dashed ellipse marks the position of the infrared bubble shown in Figure \ref{HII}. (b). Close-up of the Pong1800 SCUBA-2 850 $\micron$ map in the Daisy map region is shown in red contours. The gray scale image and black contours show the CV Daisy map. The contour levels are [0.05, 0.1, 0.2, 0.4, 0.6, 0.8]$\times$1.48 Jy~beam$^{-1}$. (c). Comparison of flux in the Daisy map with flux in the Pong map for pixels with S/N$>$3. The red line shows the linear fit.\label{Pong}  }
\end{figure}

To verify the detection of faint and extended structure in CV Daisy maps, we compare the Daisy and Pong1800 maps, both of which are obtained with the \textit{\textbf{R4}} imaging scheme.
Figure \ref{Pong}a shows the Pong1800 map. In Figure \ref{Pong}b, we compare Daisy and Pong1800 maps. In general, the Daisy map and the Pong1800 map show very consistent morphology, indicating that the Daisy map can recover large scale structures as well as the Pong map. The flux recovered in the Daisy map is also linearly correlated with the flux in the Pong map as shown in Figure \ref{Pong}c: $Flux_{Pong}=(0.883\pm0.003)Flux_{Daisy}-(28.062\pm0.532); R^2=0.95$. The Pong1800 map will be discussed in detail in another paper.

\section{APPENDIX B\\
``Two layer" models for fitting HCO$^{+}$ J=1-0 and H$_{2}$CO $2_{1,2}-1_{1,1}$ lines}

To investigate the properties of the expanding gas, the profiles of the HCO$^{+}$ J=1-0 and H$_{2}$CO $2_{1,2}-1_{1,1}$ lines were modeled with an enhanced version of the "two-layer" model, which is similar to the infall models used in \cite{mye96,fra02}. In this model, a continuum source is located in between the two layers. Each layer has peak optical depth $\tau_{0}$, velocity dispersion $\sigma$,
and expanding speed $V_{exp}$ with respect to the continuum source. The gas is moving away from the continuum source if $V_{out}$ is positive. The line brightness temperature
at velocity V is:
\begin{eqnarray*}
\Delta T_{B}=(J_{f}-J_{cr})[1-exp(-\tau_{f})]\\
+(1-\Phi)(J_{r}-J_{b})\times[1-exp(-\tau_{r}-\tau_{f})]
\end{eqnarray*}
where
\begin{equation*}
J_{cr}=\Phi J_{c}+(1-\Phi)J_{r}
\end{equation*}
and
\begin{equation*}
\tau_{f}=\tau_{0}exp[\frac{-(V+V_{out}-V_{cont})^{2}}{2\sigma^{2}}]
\end{equation*}
\begin{equation*}
\tau_{r}=\tau_{0}exp[\frac{-(V-V_{out}-V_{cont})^{2}}{2\sigma^{2}}]
\end{equation*}
$J_{c}$, $J_{f}$, $J_{r}$, $J_{b}$ are the Planck temperatures of the continuum source, the "front" layer, the "rear" layer
and the cosmic background radiation, respectively. $J=\frac{h\nu}{k}\frac{1}{exp (T_{0}/T)-1}$ is related to the blackbody temperature T at frequency $\nu$, where h is Planck's constant, and k is Boltzmann's constant. $\Phi$ and $V_{cont}$ are the filling factors and the systemic velocities of the continuum sources, respectively.

\section*{Acknowledgment}
\begin{acknowledgements}

Tie Liu is supported by KASI fellowship and EACOA fellowship. Ke Wang is supported by grant WA3628-1/1 of the German Research Foundation (DFG) through the priority program 1573 (``Physics of the Interstellar Medium''). S.-L. Qin is supported by  the Joint Research Fund
in Astronomy (U1631237) under cooperative
agreement between the National Natural Science Foundation of
China (NSFC) and Chinese Academy of Sciences (CAS), by
the Top Talents Program of Yunnan Province (2015HA030). MJ acknowledges the support of the Academy of Finland Grant No. 285769. JMa acknowledges the support of ERC-2015-STG No. 679852 RADFEEDBACK. CWL was supported by the Basic Science Research Program though the National Research Foundation of Korea (NRF) funded by the Ministry of Education, Science, and Technology (NRF-2016R1A2B4012593). Miju Kang was supported by Basic Science Research Program through the National Research Foundation of Korea(NRF) funded by the Ministry of Science, ICT \& Future Planning (No. NRF-2015R1C1A1A01052160). The James Clerk Maxwell Telescope is operated by the East Asian Observatory on behalf of The National Astronomical Observatory of Japan, Academia Sinica Institute of Astronomy and Astrophysics, the Korea Astronomy and Space Science Institute, the National Astronomical Observatories of China and the Chinese Academy of Sciences (Grant No. XDB09000000), with additional funding support from the Science and Technology Facilities Council of the United Kingdom and participating universities in the United Kingdom and Canada. The data presented in this paper are based on the ESO-ARO programme ID
196.C-0999(A). DM, GG and LB acknowledge support from CONICYT Project PFB-06.

\software{Starlink software \citep{curr14,jenn13,chap13}, GILDAS (http://www.iram.fr/IRAMFR/GILDAS/), RADEX \citep{van07}, MIRIAD (https://www.cfa.harvard.edu/sma/miriad/) and CASA (https://casa.nrao.edu/).}

\end{acknowledgements}


\begin{thebibliography}

\bibitem[Alves et al. (2007)]{alves07}Alves, J., Lombardi, M., \& Lada, C. J., 2007, \aap, 462L, 17

\bibitem[Andr\'{e} et al. (2010)]{and10}Andr\'{e}, Ph., Men'shchikov, A., Bontemps, S., et al. 2010, \aap, 518, 102

\bibitem[Andr\'{e} et al. (2014)]{and14}Andr\'{e}, P., Di Francesco, J., Ward-Thompson, D., et al. 2014, Protostars and Planets VI, Henrik Beuther, Ralf S. Klessen, Cornelis P. Dullemond, and Thomas Henning (eds.), University of Arizona Press, Tucson, 914 pp., p.27-51

\bibitem[Bacmann et al. (2002)]{bac02}Bacmann A., Lefloch B., Ceccarelli C., et al. 2002. \aap, 389, L6

\bibitem[Bally et al. (1987)]{bally87}Bally, J., Langer, W. D., Stark, A. A., Wilson, R. W., 1987, \apj, 312, 45

\bibitem[Bergin \& Tafalla (2007)]{ber07}Bergin, E. A. \& Tafalla, M., 2007, AR\aap, 45, 339

\bibitem[Berry (2015)]{berry15}Berry D. S., 2015, Astron. Comput., 10, 22

\bibitem[Berry (2007)]{berry07}Berry D. S., Reinhold K., Jenness T., Economou F., 2007, in Shaw R. A.,
Hill F., Bell D. J., eds, ASP Conf. Ser. Vol. 376, Astronomical Data
Analysis Software and Systems XVI. Astron. Soc. Pac., San Francisco,
p. 425

\bibitem[Beckwith et al. (1990)]{Beckwith1990}Beckwith, S. V. W., Sargent, A. I., Chini, R. S., Guesten, R., 1990, \aj, 99, 924

\bibitem[Bintley et al. (2014)]{bint14}Bintley, D., Holland, W. S., MacIntosh, M., J., et al. ,2014, SPIE, 9153, 3

\bibitem[Busquet et al.\ (2011)]{bus11}Busquet, G., Estalella, R., Zhang, Q., et al.\, 2011, \aap, 525, 141

\bibitem[Caselli (2011)]{cas11}Caselli, P., 2011, IAUS, 280, 19

\bibitem[Chandrasekhar \& Fermi (1953)]{Chand53}Chandrasekhar S. \& Fermi E., 1953, \apj, 118, 116

\bibitem[Chapin et al. (2013)]{chap13}Chapin, E. L., Berry, D. S., Gibb, A. G., et al. 2013, \mnras, 430, 2545

\bibitem[Chen et al. (2016)]{chen16}Chen, M. C.-Y., Di Francesco, J., Johnstone, D., et al. 2016, \apj, 826, 95

\bibitem[Contreras et al. (2013)]{cont13}Contreras, Y., Schuller, F., Urquhart, J. S., et al. 2013, \aap, 549, 45

\bibitem[Contreras et al. (2016)]{cont16}Contreras, Y., Garay, G., Rathborne, J. M., Sanhueza, P., 2016, \mnras, 456, 2041

\bibitem[Csengeri et al. (2016)]{cseng16}Csengeri, T., Weiss, A., Wyrowski, F., et al. 2016, \aap, 585, 104

\bibitem[Cox et al. (2010)]{cox10}Cox, N. L. J., Arzoumanian, D., Andr\'{e}, Ph., et al. \aap, 590, 110

\bibitem[Currie et al. (2014)]{curr14}Currie, M. J., Berry, D. S., Jenness, T., et al. 2014, in Astronomical Society of the Pacific Conference Series, Vol. 485, Astronomical Data Analysis Software and Systems XXIII, ed. N. Manset \& P. Forshay, 391

\bibitem[Dempsey et al. (2013)]{demp13}Dempsey, J. T., Friberg, P., Jenness, T., et al., 2013, \mnras, 430, 2534

\bibitem[Di Francesco et al. (2002)]{fra02}Di Francesco, J., Myers, P. C., Wilner, D. J., Ohashi, N., Mardones, D., 2001, \apj, 562, 770

\bibitem[Eden et al. (2017)]{Eden17}Eden, D. J., Moore, T. J. T., Plume, R., et al. 2017, \mnras, 469, 2163

\bibitem[Federrath \& Klessen(2013)]{Fed13} Federrath, C., \& Klessen, R.~S.\ 2013, \apj, 763, 51

\bibitem[Federrath (2016)]{Fed16}Federrath, C., 2016, \mnras, 457, 375

\bibitem[Feng et al. (2016)]{Feng16}Feng, S., Beuther, H., Zhang, Q., et al. 2016, \aap, 592, 21

\bibitem[Fiege \& Pudritz (2000)]{fiege00}Fiege, Jason D. \& Pudritz, Ralph E., 2000, \mnras, 311, 105

\bibitem[Fischera \& Martin (2012)]{fisch12}Fischera J., \& Martin P. G., 2012, \aap, 542, A77

\bibitem[Goldsmith et al. (2016)]{gold16}Goldsmith, P. F., Pineda, J. L., Langer, W. D., et al. 2016, \apj, 824, 141

\bibitem[Garden et al. (1991)]{gard91}Garden, R. P., Hayashi, M., Hasegawa, T., Gatley, I., Kaifu, N., 1991, \apj, 374, 540

\bibitem[Gao \& Solomon\ (2004)]{gao04}Gao, Y., \& Solomon, P. M. 2004, ApJ, 606, 271

\bibitem[Giannetti et al. (2014)]{gian14}Giannetti, A., Wyrowski, F., Brand, J., et al. 2014, \aap, 570, 65

\bibitem[Holland, et al. (2013)]{holl13}Holland, W. S., Bintley, D., Chapin, E. L., et al., 2013, \mnras, 430, 2513

\bibitem[Inutsuka \& Miyama (1992)]{inut92}Inutsuka S.-I., \& Miyama S. M., 1992, \apj, 388, 392

\bibitem[Inutsuka et al. (2015)]{Inu15}Inutsuka, S., Inoue, T., Iwasaki, K., et al. 2015, \aap, 580, 49

\bibitem[Jackson et al. (2010)]{jack10}Jackson J. M., Finn S. C., Chambers E. T., Rathborne J. M., Simon R., 2010, \apj, 719, L185

\bibitem[Jenness et al. (2013)]{jenn13}Jenness, T., Chapin, E. L., Berry, D. S., et al. 2013, SMURF: SubMillimeter User Reduction Facility, Astrophysics Source Code Library., , , ascl:1310.007:

\bibitem[Jim{\'e}nez-Serra et al.(2010)]{jim10} Jim{\'e}nez-Serra, I., Caselli, P., Tan, J.~C., et al.\ 2010, \mnras, 406, 187

\bibitem[Johnstone et al. (2000)]{john00}Johnstone, D., Wilson, C. D., Moriarty-Schieven, et al. 2000, \apj, 545, 327

\bibitem[Johnstone, Di Francesco, \& Kirk (2004)]{john04}Johnstone, D., Di Francesco, J., \& Kirk, H., 2004, \apj, 611L, 45

\bibitem[Juvela et al. (2011)]{juve11}Juvela, M., Ristorcelli, I., Pelkonen, V.-M., et al. 2011, \aap, 527, 111

\bibitem[Juvela et al. (2012)]{GCC-III}Juvela, M., Ristorcelli, I., Pagani, L., et al. 2012, \aap, 541, 12

\bibitem[Juvela et al. (2015a)]{GCC-V}Juvela, M., Ristorcelli, I., Marshall, D. J., et al. 2015a, \aap, 584, 93

\bibitem[Juvela et al. (2015b)]{GCC-VI}Juvela, M., Demyk, K., Doi, Y., et al. 2015b, \aap, 584, 94

\bibitem[Juvela et al. (2017)]{juvela17}Juvela, M., He, J., Pattle, K., et al., 2017, \aap, in press, arXiv:1711.09425

\bibitem[Kang et al. (2015)]{kang15}Kang, M., Choi, M., Stutz, A. M., Tatematsu, K., 2015, \apj, 814, 31

\bibitem[Kainulainen et al. (2017)]{kai17}Kainulainen, J., Stutz, A. M., Stanke, T., et al. 2017, \aap, 600, 141

\bibitem[Kauffmann \& Pillai (2010)]{kauff10}Kauffmann J., \& Pillai T., 2010, \apj, 723, L7

\bibitem[Kennicutt et al.\ (1998)]{ken98}Kennicutt, R. C., Jr. 1998, \apj, 498, 541

\bibitem[Kirk, Johnstone, \& Di Francesco (2006)]{kirk06}Kirk, H., Johnstone, D., \& Di Francesco, J., 2006, \apj, 646, 1009

\bibitem[Kirk et al. (2013)]{kirk13}Kirk, H., Myers, P. C., Bourke, T. L., et al. 2013, \apj, 766, 115

\bibitem[Klessen(2000)]{Klessen00} Klessen, R.~S.\ 2000, \apj, 535, 869

\bibitem[K\"{o}nyves et al. (2010)]{kon10}K\"{o}nyves, V., Andr\'{e}, Ph., Men'shchikov, A., et al. 2010, \aap, 518, 106

\bibitem[K\"{o}nyves et al. (2015)]{kon15}K\"{o}nyves, V., Andr\'{e}, Ph., Men'shchikov, A., et al. 2015, \aap, 584, 91

\bibitem[Kritsuk et al.(2011)]{Kritsuk11} Kritsuk, A.~G., Norman, M.~L., \&
Wagner, R.\ 2011, \apjl, 727, L20

\bibitem[Lada, Lombardi, \& Alves\ (2010)]{lada10}Lada, C. J., Lombardi, M., \& Alves, J. F. 2010, ApJ, 724, 687

\bibitem[Lee, Bergin \& Evans (2004)]{lee04}Lee, J.-E., Bergin, E. A., \& Evans, N. J., II, 2004, \apj, 617, 360

\bibitem[Li et al. (2016)]{li16}Li, G.-X., Urquhart, J. S., Leurini, S., et al. 2016, \aap, 591, 5

\bibitem[Li (2017)]{Li17}Li G.-X., 2017, \mnras, 465, 667

\bibitem[Li, Klein \& McKee (2017)]{LiP17}Li, P. S., Klein, R. I., McKee, C. F., et al. 2018, \mnras, 473, 4220

\bibitem[Lin et al. (2016)]{lin16}Lin, Y., Liu, H. B., Li, D., et al. 2016, \apj, 828, 32

\bibitem[Lin et al. (2017)]{lin17}Lin, Y., Liu, H. B., Dale, J. E., et al., 2017, \apj, 840, 22

\bibitem[Liu et al. (2012)]{liu12}Liu, T., Wu, Y., Zhang, H,. 2012, \apjs, 202, 4

\bibitem[Liu et al. (2013)]{liu13}Liu, T., Wu, Y., Zhang, H.,. 2013, \apj, 775, L2

\bibitem[Liu et al. (2013b)]{liu13b}Liu, T., Wu, Y., Wu, J., Qin, S.-L., Zhang, H., 2013b, \mnras, 836, 1335

\bibitem[Liu et al. (2014)]{liu14}Liu, T., Wu, Y., Mardones, D., et al. 2014, PKAS, 30, 79L

\bibitem[Liu et al. (2015)]{liu15}Liu, H. B., Galv\'{a}n-Madrid, R., Jim\'{e}nez-Serra, I., et al. 2015, \apj, 804, 37

\bibitem[Liu et al. (2016)]{liu16}Liu, T., Zhang, Q., Kim, K. -T., et al. 2016, \apjs, 222, 7

\bibitem[Liu et al. (2016b)]{liu16b}Liu, T., Zhang, Q., Kim, K.-T., et al. 2016b, \apj, 824, 31

\bibitem[Liu et al. (2016c)]{liu16c}Liu, T., Kim, K.-T., Yoo, H., et al., 2016c, \apj, 829, 59

\bibitem[Liu et al. (2017)]{liu17}Liu, T., Lacy, J., Li, P. S., et al., 2017, \apj, 849, 25

\bibitem[MacLaren et al. (1988)]{mac88}MacLaren, I., Richardson, K. M., Wolfendale, A. W., 1988, \apj, 333, 821

\bibitem[Mairs et al. (2015)]{mairs15}Mairs, S., Johnstone, D., Kirk, H., et al. 2015, \mnras, 454, 2557

\bibitem[Mairs et al. (2017)]{mairs17}Mairs, S., Lane, J., Johnstone, D., et al., 2017, arXiv170601897M

\bibitem[Meng et al. (2013)]{meng13}Meng, F., Wu, Y., Liu, T., 2013, \apjs 209, 37

\bibitem[Montillaud et al. (2015)]{GCC-IV}Montillaud, J., Juvela, M., Rivera-Ingraham, A., et al., 2015, \aap, 584, 92

\bibitem[Molinari et al. (2010)]{Molinari2010}Molinari, S., Swinyard, B., Bally, J., et al., 2010, \aap, 518, L100

\bibitem[Montier et al. (2010)]{mont10}Montier, L. A., Pelkonen, V.-M., Juvela, M., Ristorcelli, I., Marshall, D. J., 2010, \aap, 522, 83

\bibitem[Moore et al. (2015)]{Moore15}Moore, T. J. T., Plume, R., Thompson, M. A., et al., 2015, \mnras, 453, 4264

\bibitem[Motte, Andre, \& Neri (1998)]{motte98}Motte, F., Andre, P., \& Neri, R., 1998, \aap, 336, 150

\bibitem[Myers et al. (1996)]{mye96}Myers, P. C., Mardones, D., Tafalla, M., Williams, J. P., \& Wilner, D. J. 1996, \apj, 465, L133

\bibitem[Ormel et al. (2009)]{Ormel09}Ormel, C. W., Paszun, D., Dominik, C., Tielens, A. G. G. M., 2009, \aap, 502, 8450

\bibitem[Palmeirim et al. (2013)]{pal13}Palmeirim, P., Andr\'{e}, Ph., Kirk, J., et al., 2013, \aap, 550, 38

\bibitem[Peretto \& Fuller (2009)]{pere09}Peretto, N. \& Fuller, G. A., 2009, \aap, 505, 405

\bibitem[Peretto et al. (2012)]{pere12}Peretto, N., Andr\'{e}, Ph.; K\"{o}nyves, V., et al., 2012, \aap, 541, 63

\bibitem[Peretto et al. (2013)]{pere13}Peretto, N., Fuller, G. A., Duarte-Cabral, A., et al. 2013, \aap, 555, 112

\bibitem[Piazzo et al. (2015)]{Piazzo2015}Piazzo, L., Calzoletti, L., Faustini, F., et al. 2015, \mnras, 447, 1471

\bibitem[Van der Tak et al. (2007)]{van07}Van der Tak, F.F.S., Black, J.H., Sch\"{o}ier, F.L., Jansen, D.J., van Dishoeck, E.F. 2007, \aap, 468, 627

\bibitem[Rivera-Ingraham et al. (2016)]{GCC-VII}Rivera-Ingraham, A., Ristorcelli, I., Juvela, M., et al., 2016, \aap, 591, 90

\bibitem[Rivera-Ingraham et al. (2017)]{GCC-VIII}Rivera-Ingraham, A., Ristorcelli, I., Juvela, M., et al., 2017, \aap, 601, 94

\bibitem[Ostriker et al. (1964)]{ostr64}Ostriker J., 1964, \apj, 140, 1056

\bibitem[Pagani et al. (2015)]{paga15}Pagani, L., Lef\`{e}vre, C., Juvela, M., Pelkonen, V.-M., Schuller, F., 2015, \aap, 574, L5

\bibitem[Planck Collaboration XXIII (2011)]{planck11}Planck Collaboration, Ade, P. A. R., Aghanim, N., et al. 2011, \aap, 536, 23

\bibitem[Planck Collaboration XXII (2011b)]{PlanckII}Planck Collaboration, Ade, P. A. R., Aghanim, N., et al. 2011b, \aap, 536, 22

\bibitem[Planck Collaboration XXVIII (2016)]{planck16}Planck Collaboration, Ade, P. A. R., Aghanim, N., et al. 2016, \aap, 594, A28

\bibitem[Sadavoy et al. (2013)]{sad13}Sadavoy, S. I., Di Francesco, J., Johnstone, D., et al. 2013, \apj, 767, 126

\bibitem[Schmidt\ (1959)]{sch59}Schmidt, M. 1959, \apj, 129, 243

\bibitem[Sanhueza et al.(2012)]{sanh12}Sanhueza, P., Jackson, J. M., Foster, J. B., et al. 2012, \apj, 756, 60

\bibitem[Sanhueza et al.(2013)]{sanh13}Sanhueza, P., Jackson, J.~M., Foster, J.~B., et al.\ 2013, \apj, 773, 123

\bibitem[Schneider et al.(2012)]{Schneider12} Schneider, N.,
Csengeri, T., Hennemann, M., et al.\ 2012, \aap, 540, L11

\bibitem[Schneider et al.(2015)]{Schneider15}Schneider, N., Bontemps, S.,
Girichidis, P., et al.\ 2015, \mnras, 453, 41

\bibitem[Shirley (2015)]{shir15}Shirley, Y. L., 2015, \pasp, 127, 299

\bibitem[Stephens et al. (2016)]{step16}Stephens, I. W., Jackson, J. M., Whitaker, J. S., et al. 2016, \apj, 824, 29

\bibitem[Stod\'{o}lkiewicz (1963)]{stod63}Stod\'{o}lkiewicz, J. S., 1963, Acta Astron., Volume 13, p. 30-54

\bibitem[Tatematsu et al. (2017)]{tat17}Tatematsu, K., Liu, T., Ohashi, S., et al. 2017, \apjs, 228, 12

\bibitem[V\'{a}zquez-Semadeni (1994)]{vaz94}V\'{a}zquez-Semadeni, E. 1994, \apj, 423, 681

\bibitem[Wang et al. (2011)]{wang11}Wang, K., Zhang, Q., Wu, Y., Zhang, H., 2011, \apj, 735, 64

\bibitem[Wang et al. (2014)]{wang14}Wang, K., Zhang, Q., Testi, L., et al. 2014, \mnras, 439, 3275

\bibitem[Wang et al. (2015)]{wang15}Wang, K., Testi, L., Ginsburg, A., Walmsley, C. M., Molinari, S., Schisano, E., 2015, \mnras, 450, 4043

\bibitem[Wang et al. (2016)]{wang16}Wang, K., Testi, L., Burkert, A., et al. 2016, \apjs, 226, 9

\bibitem[Ward-Thompson et al. (2007)]{ward07}Ward-Thompson D., Di Francesco, J., Hatchell, J., et al., 2007, \pasp, 119, 855

\bibitem[Wilson \& Rood et al. (1994)]{wils94}Wilson, T. L., \& Rood, R. 1994, AR\aap, 32, 191

\bibitem[Wu et al. (2012)]{wu12}Wu, Y., Liu, T., Meng, F., et al. 2012, \apj, 756, 76

\bibitem[Wu et al.\ (2005)]{wu05}Wu, J., Evans, N. J., II, Gao, Y., et al. 2005, \apj, 635, L173

\bibitem[Wu et al.\ (2010)]{wu10}Wu, J., Evans, N. J., II, Shirley, Y. L., et al. 2010, \apjs, 188, 313

\bibitem[Yuan et al. (2016)]{yuan16}Yuan, J., Wu, Y., Liu, T., et al. 2016, \apj, 820, 37

\bibitem[Yuan et al. (2017)]{yuan17}Yuan, J., Li, J.-Z., Wu, Y., et al. 2017, ApJ in press, arXiv:1711.08951

\bibitem[Zhang et al.\ (2014)]{zhang14}Zhang, Z.-Y., Gao, Y., Henkel, C., et al. 2014, \apjl, 784, L31

\bibitem[Zhang et al.\ (2015)]{zhang15}Zhang, Q., Wang, K., Lu, X., et al.\ 2015, \apj, 804, 141

\bibitem[Zhang et al. (2016)]{zhang16}Zhang, T., Wu, Y., Liu, T., et al. 2016, \apjs, 224,43


\end{thebibliography}
\end{document}